\documentclass{aastex63}

\shorttitle{TeV cosmic ray acceleration}
\shortauthors{Zeng et al.}
\graphicspath{{./}{figures/}}

\begin{document}

\title{TeV cosmic ray nuclei acceleration in shell-type supernova remnants with hard $\gamma$-ray spectra}

\correspondingauthor{Siming Liu}
\email{liusm@pmo.ac.cn, zhd@pmo.ac.cn}

\author[0000-0001-8500-0541]{Houdun Zeng}
\affiliation{Key Laboratory of Dark Matter and Space Astronomy, Purple Mountain Observatory, Chinese Academy of Sciences Nanjing 210034, People's Republic of China}

\author{Yuliang Xin}
\affiliation{School of Physical Science and Technology, Southwest Jiaotong University, Chengdu 610031, People's Republic of China}

\author{Shuinai Zhang}
\author[0000-0003-1039-9521]{Siming Liu}
\affiliation{Key Laboratory of Dark Matter and Space Astronomy, Purple Mountain Observatory, Chinese Academy of Sciences Nanjing 210034, People's Republic of China}




\begin{abstract}
The emission mechanism for hard $\gamma$-ray spectra from supernova remnants (SNRs) is still a matter of debate. Recent multi-wavelength observations of TeV source HESS J1912+101 show that it is associated with an SNR with an age of $\sim 100$ kyrs, making it unlikely produce the TeV $\gamma$-ray emission via leptonic processes. We analyzed Fermi observations of it and found an extended source with a hard spectrum. HESS J1912+101 may represent a peculiar stage of SNR evolution that dominates the acceleration of TeV cosmic rays. By fitting the multi-wavelength spectra of 13 SNRs with hard GeV $\gamma$-ray spectra with simple emission models with a density ratio of GeV electrons to protons of $\sim 10^{-2}$, we obtain reasonable mean densities and magnetic fields with a total energy of $\sim 10^{50}$ ergs for relativistic ions in each SNR. Among these sources, only two of them, namely SN 1006 and RCW 86, favor a leptonic origin for the $\gamma$-ray emission. The magnetic field energy is found to be comparable to that of the accelerated relativistic ions and their ratio has a tendency of increase with the age of SNRs. These results suggest that TeV cosmic rays mainly originate from SNRs with hard $\gamma$-ray spectra.
\end{abstract}

\keywords{Galactic cosmic rays (567); Gamma-ray sources (633); Non-thermal radiation sources (1119); Gamma-ray astronomy (628); Supernova remnants (1667)}

\section{Introduction}
\label{sec:intro}

Although it is generally accepted that soft $\gamma$-ray spectra of supernova remnants interacting with molecular clouds (SNRs) result from decay of $\pi^0$ produced via inelastic collisions of high-energy ions with nuclei in the background due to evolution of SNR shocks in a high-density environment \citep{2010Sci...327.1103A, Giuliani_2011, 2019ApJ...874...50Z}, the nature of hard $\gamma$-ray spectra has been a subject of extensive investigations \citep{2012ApJ...761..133Y, 2014MNRAS.445L..70G, 2016ApJ...821...43Z, 2018A&A...612A...6H, 2019MNRAS.487.3199C}. In the leptonic scenario for the $\gamma$-ray emission, the model parameters are well constrained and appear to be consistent with expectation of diffusive shock particle acceleration mechanism \citep{2019ApJ...876...24Z}. Hadronic models require stronger magnetic fields and less efficient electron acceleration \citep{2008MNRAS.386L..20B}. These results have profound implications on the origin of cosmic rays (CRs), especially those with energies lower than the CR spectral knee energy of $\sim 1$ PeV \citep{2020ChA&A..44....1Z}. 

Considering the anomalous CR spectra discovered with space measurements \citep{PhysRevLett.114.171103,PhysRevLett.115.211101}, \cite{2017ApJ...844L...3Z} proposes that GeV CRs are mostly accelerated in SNRs interacting with molecular clouds with relatively lower shock speeds, giving rise to softer spectra, while higher energy CRs may be attributed to particle acceleration in relatively younger SNRs with higher shock speeds, leading to slightly harder high-energy spectra. 
In this scenario, high-energy electron acceleration efficiency in young SNRs can indeed be lower than that for GeV electrons in old SNRs \citep{2019MNRAS.482.5268Z}, reminiscence of the hadronic scenario for hard $\gamma$-ray spectra  \citep{2008MNRAS.386L..20B}. 
One of the challenges facing hadronic models is that ion spectra need to cut off at tens of TeVs to account for the observed $\gamma$-ray spectral cutoffs, suggesting that SNRs are not PeVatrons. Recent CR proton spectral measurement by the DAMPE shows that there appears to be a spectral hump at tens of TeVs \citep{eaax3793}, which has been attributed to a nearby SNR, such as Geminga \citep{2019JCAP...12..007Q}, implying indeed that shocks of SNRs can only accelerate protons up to a few tens of TeV \citep{1983A&A...125..249L, 2013MNRAS.431..415B}. Observations of young SNR Cas A and $\gamma$-Cygni SNR also imply a high-energy cutoff of the ion distribution in the TeV energy range \citep{2019ApJ...874...98Z, 2020ApJ...894...51A, 2020arXiv201015854M}.   

HESS J1912+101 was first discovered in 2008 by the H.E.S.S. collaboration with a shell structure \citep{2008A&A...484..435A}. However its radio counterpart had not been identified until very recently with polarization measurement \citep{2019RAA....19...45R}, implying presence of large scale strong magnetic field.
Interestingly, observations of molecular clouds in the direction of HESS J1912+101 suggest that it is associated with an SNR with an age of $70-200$ kilo-years (kyrs) \citep{Su_2017}, which is consistent with the characteristic age of 170 kyrs for pulsar J1913+1011 inside it \citep{2002MNRAS.335..275M}. SNR G279.0+1.1 has similar properties with $\gamma$-ray emission up to 0.5 TeV detected recently \citep{2020MNRAS.492.5980A}. A recent population study of SNRs shows that higher energy particles are preferentially accelerated in younger SNRs with higher shock speeds \citep{2019ApJ...874...50Z}. In the absence of continuous TeV particle acceleration in old SNRs, it is very challenging to produce the TeV emission from HESS J1912+101 via leptonic processes.

In section \ref{sec:1912}, we show challenges to reproduce $\gamma$-ray emission from HESS J1912+101 and SNR G279.0+1.1 via leptonic processes. Section \ref{sec:296} is dedicated to the study of G296.5+10.0 for its hard $\gamma$-ray spectrum similar to HESS J1912+101. 
In section \ref{sec:sample} we fit the multi-wavelength spectra of another 10 SNRs with hard $\gamma$-ray spectra in the hadronic and leptonic scenarios and discuss the model implications. Our conclusions are drawn in section \ref{sec:conclusion}.

\section{HESS J1912+101} \label{sec:1912}

Recent radio observations of HESS J1912+101 revealed strongly polarized emission at 6 cm from the northeast half of the shell \citep{2019RAA....19...45R}. Due to strong radio emission from the surrounding, a shell structure cannot be identified in the total intensity maps. The total polarized flux density at 6 cm is about 0.5$\pm$0.2 Jy, which can be considered as a lower limit to the total flux density. Assuming a polarization fraction of 20\%, \citet{2019RAA....19...45R} obtained a total flux density of 2.5$\pm$ 1.0 Jy, which will be treated as an upper limit in the following. The strong polarization of radio emission implies the presence of large scale magnetic fields, which should also be stronger than the typical value of $3\mu$G for the interstellar medium.

Via CO and HI observations, \citet{Su_2017} showed that HESS J1912+101 is likely associated with an old SNR with an age of $0.7-2.0\times 10^5$ years at a distance of $\sim$ 4.1 kpc. The good correlation between the TeV emission and the disturbed gas revealed by these observations makes them suggest a hadronic origin for the $\gamma$-ray emission. 

\begin{figure}[ht!]
\plotone{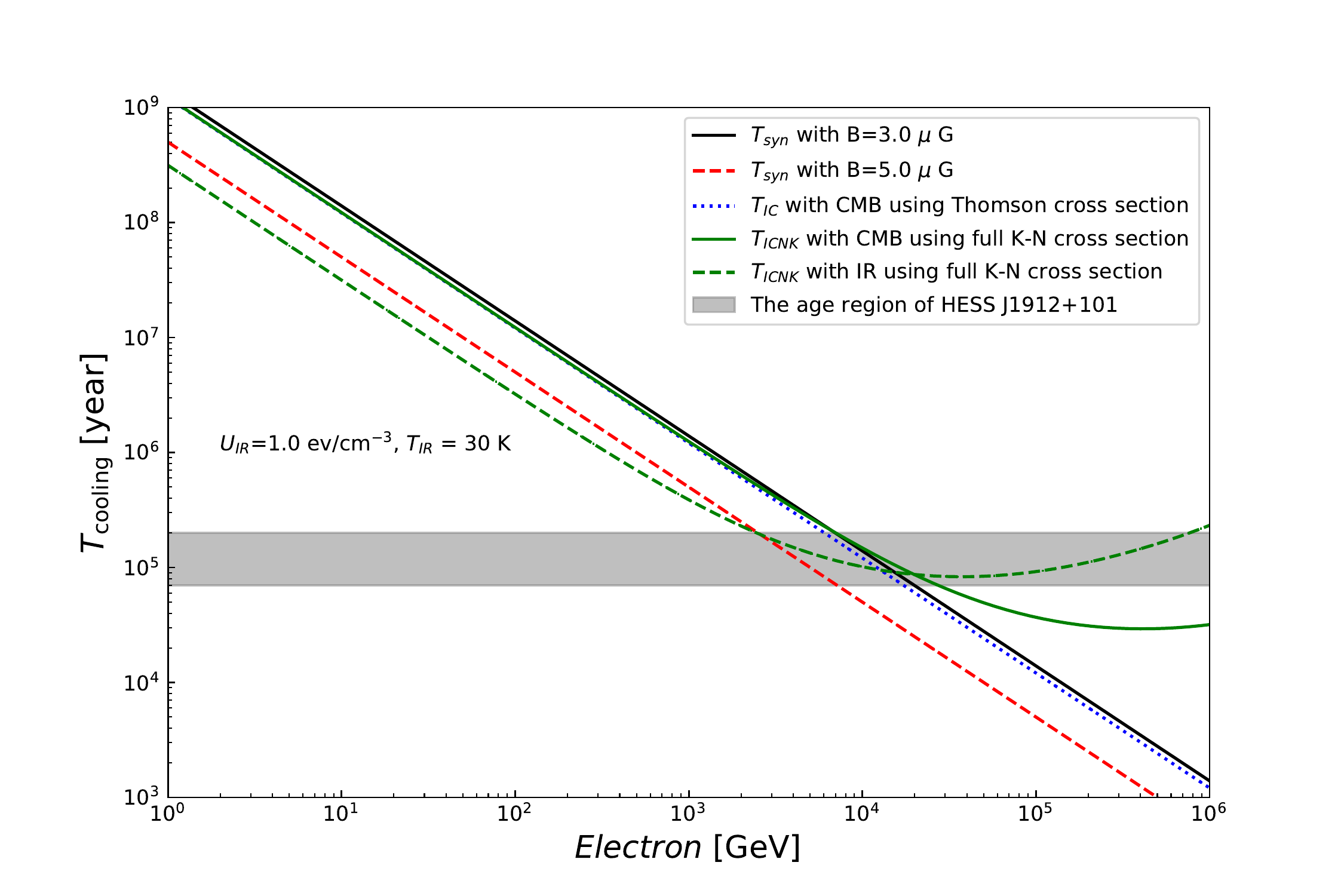}
\caption{Electron cooling time due to synchrotron and inverse Compton emission. The magnetic fields and properties of the background soft photons are indicated. The grey band indicates the age range obtained via molecular cloud observations \citep{Su_2017}. \label{fig:times}}
\end{figure}

Recently \cite{2019ApJ...874...50Z} studied a sample of $\gamma$-ray SNRs and found that in general high-energy particles in these sources have a broken power-law distribution with the break energy and the low-energy spectral index decreasing with the increase of SNR age. These results imply that higher energy particles are mostly accelerated in younger SNRs with relatively higher shock speeds. The acceleration of TeV particles quenches for SNR with an age greater than 10 kyrs. This challenges a leptonic origin for the TeV emission from HESS J1912+101.   

\begin{figure}[ht!]
\plottwo{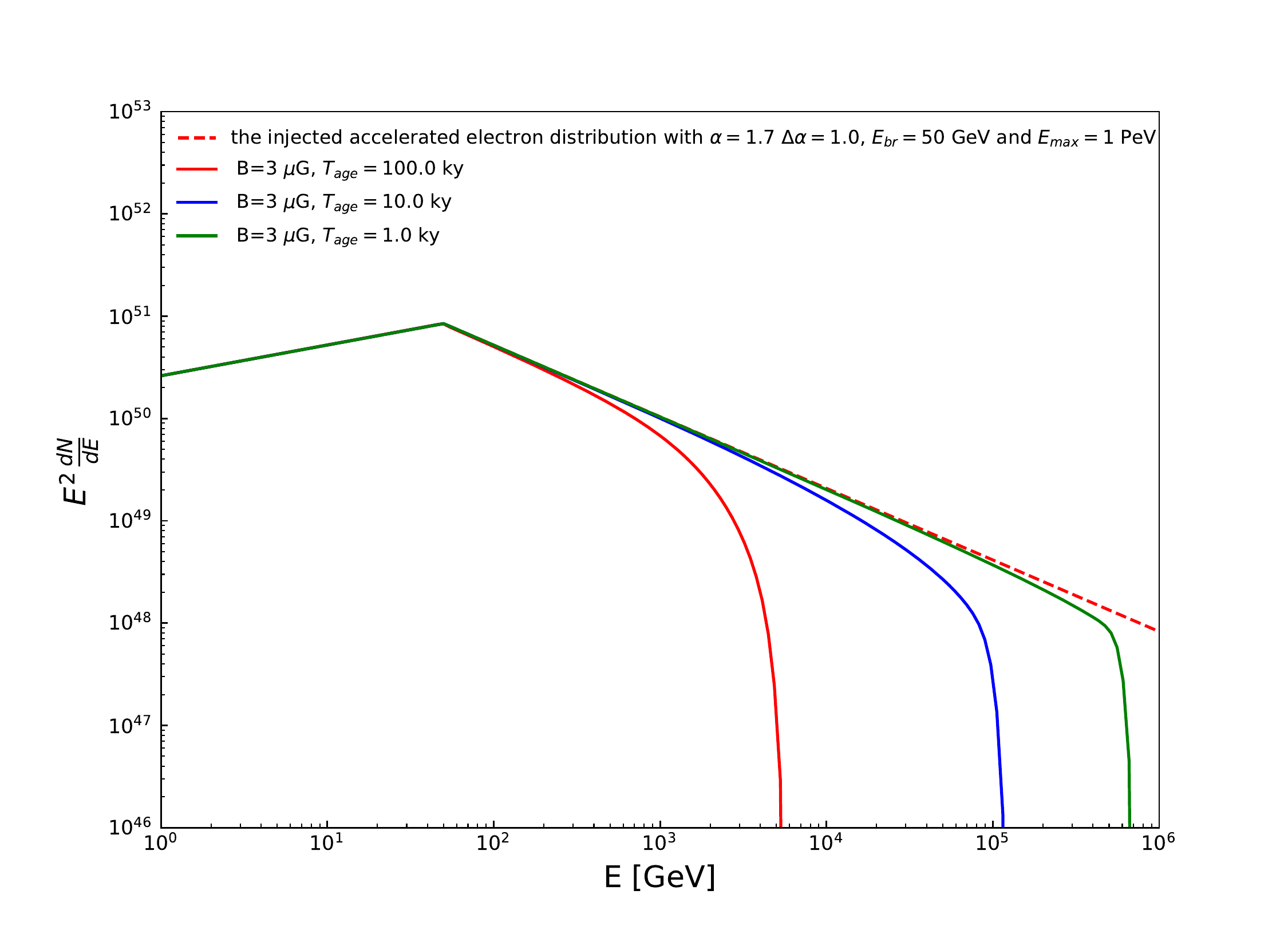}{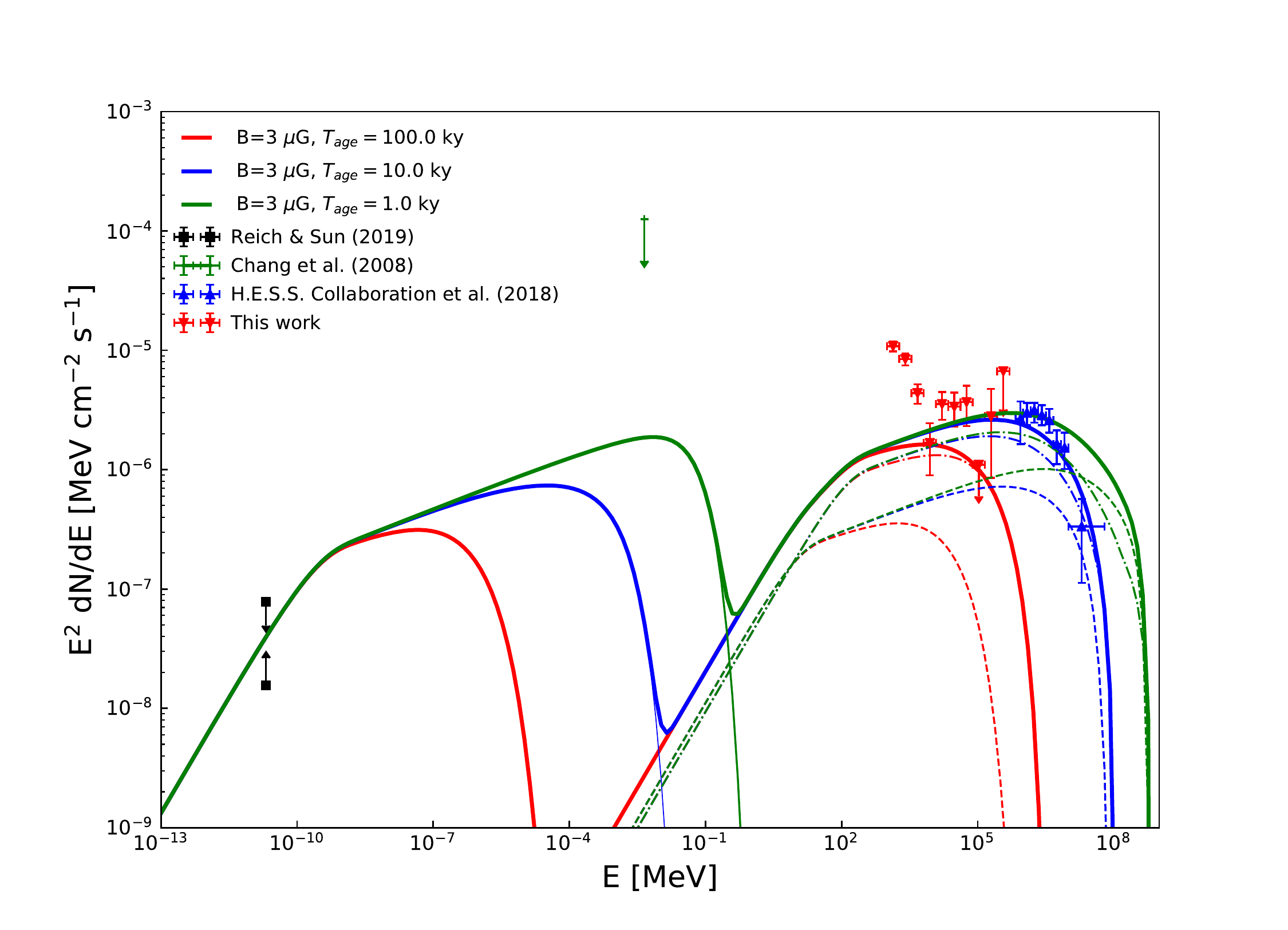}
\caption{Left: evolution of electron distribution due to synchrotron and IC losses. The initial distribution is a broken power-law with a high-energy cutoff with parameters indicated in the figure. 
Right: evolution of the corresponding emission spectra compared with the spectral energy distribution (SED) of HESS J1912+101. The radio
data are taken from \cite{2019RAA....19...45R}. The lower limit is for the polarized emission. The X-ray upper limit is from \cite{2008ApJ...682.1177C},
and the blue data points are from the H.E.S.S. collaboration \citep{2018A&A...612A...8H}. \label{fig:lep}}
\end{figure}

Figure \ref{fig:times} shows the energy loss timescale of electrons due to the synchrotron and inverse Compton (IC) processes. It can be seen that for a typical value of the interstellar magnetic field, the maximum energy of electrons in HESS J1912+101 should be about 10 TeV. The $\gamma$-ray produced via IC by such electrons should cut off below 10 TeV. The left panel of Figure \ref{fig:lep} shows the evolution of electron distribution under the influence of energy loss due to synchrotron and IC processes for an injected broken power-law spectrum with a maximum energy of 1 PeV at the beginning. Compared to the age of this SNR, TeV and higher energy particles are assumed to be accelerated instantaneously in the early stage of the SNR evolution \citep{2013MNRAS.431..415B}.

\begin{figure*}[!htb]
	\centering
	\includegraphics[width=3.5in]{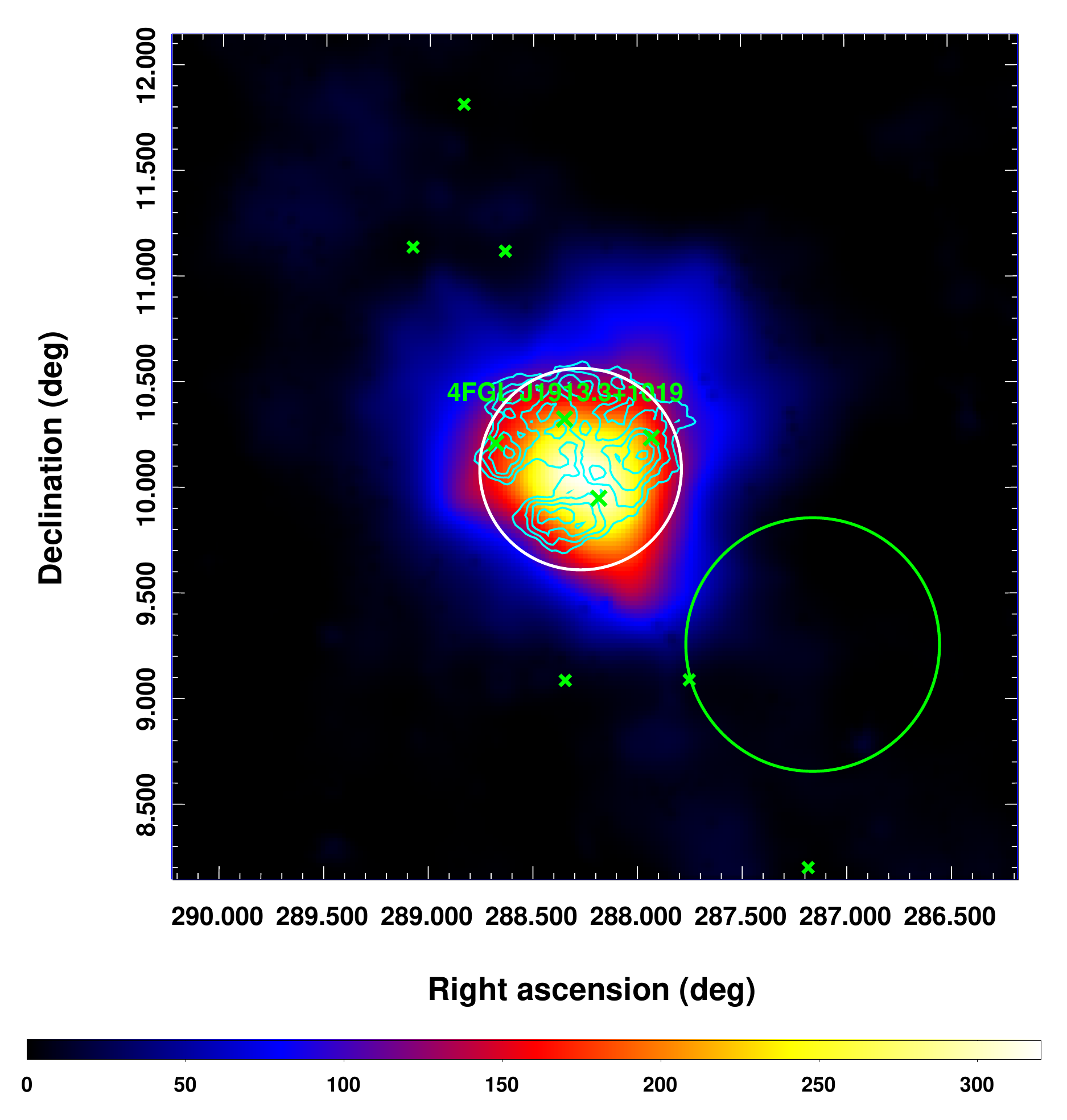}
	\includegraphics[width=3.5in]{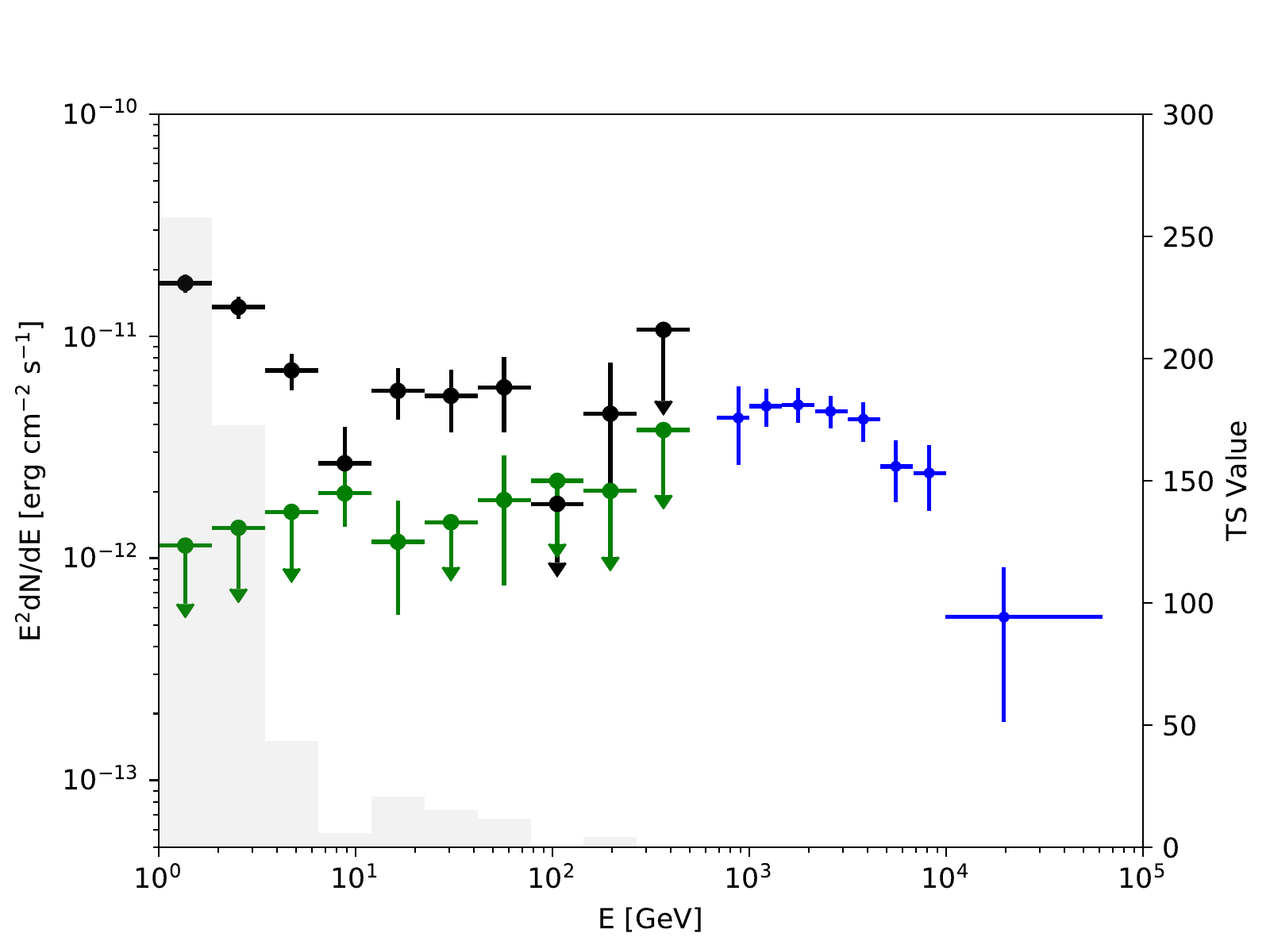}
	\caption{Left: $4^\circ \times 4^\circ$ TS map of photons above 1 GeV around HESS J1912+101 after subtracting $\gamma$-ray emission from 4FGL J1913.3+1019. The green crosses and circle represent the 4FGL sources and the best-fit radius of the uniform disk is marked by the white circle. The cyan contours show the TeV $\gamma$-ray emission of HESS J1912+101 \citep{2018A&A...612A...8H}. 
		Right: The $\gamma$-ray SED of HESS J1912+101 (black dots) and 4FGL J1913.3+1019 (green dots). The blue dots are the HESS data of HESS J1912+101 \citep{2018A&A...612A...8H}.The gray histogram denotes the TS value of HESS J1912+101 for each energy bin and the arrows indicate the upper limits with 95\% significance level.}
	\label{fig:tsmap-sed-j1912}
\end{figure*}

Recent analyses of Fermi data by \citet{2020ApJ...889...12Z} uncovered a compact GeV source with a soft spectrum within the shell of HESS J1912+101, and there appears to be extended diffuse emission. 
Here we re-analyzed the {\em Fermi}-LAT data around HESS J1912+101. 
In the region of HESS J1912+101, there are four 4FGL sources (4FGL J1913.3+1019, 4FGL J1914.7+1012, 4FGL J1911.7+1014, 4FGL J1912.7+0957). 
Among them, 4FGL J1913.3+1019 is associated with PSR J1913+1011 \citep{2002MNRAS.335..275M, 2019ApJ...871...78S, 2020ApJ...889...12Z} and others are unidentified sources with soft spectra.
We removed these three unidentified 4FGL sources from the model file and treated them as parts of the diffuse emission around HESS J1912+101.
Using the data above 1 GeV with the {\em Fermipy} package version 0.19.0 \citep{2017ICRC...35..824W}, we tested several spatial templates for HESS J1912+101, including an uniform disk, a 2D Gaussian model, and the HESS image.
The uniform disk model (R.A., decl.=$288.262^{\circ} \pm 0.033^{\circ}$, $10.075^{\circ} \pm 0.037^{\circ}$ with $\sigma = 0.496^{\circ} \pm 0.035^{\circ}$) is favored, while the 2D Gaussian template (R.A., decl.=$288.277^{\circ} \pm 0.034^{\circ}$, $10.135^{\circ} \pm 0.041^{\circ}$ with $\sigma = 0.413^{\circ} \pm 0.036^{\circ}$) gives an equally good representations of the data. 
We produced a TS map above 1 GeV after subtracting the $\gamma$-ray emission from 4FGL J1913.3+1019, which is shown in the left panel of Figure \ref{fig:tsmap-sed-j1912}.


In the energy range of 1 GeV - 500 GeV, the TS value of HESS J1912+101 is fitted to be 469.0, and the spectral index of a power-law model is 2.54$\pm$0.07. The integral photon flux from 1 GeV to 500 GeV is $(7.72\pm0.44)\times10^{-9}$ photon cm$^{-2}$ s$^{-1}$.
For the point source 4FGL J1913.3+1019, the TS value and the power-law spectral index are fitted to be 24.2 and 1.90$\pm$0.17.
The $\gamma$-ray SEDs of HESS J1912+101 and 4FGL J1913.3+1019 shown in the right panel of Figure \ref{fig:tsmap-sed-j1912} were produced by dividing all data between 1 GeV and 500 GeV into 10 bins with identical width on the logarithmic of energy. And the 95\% upper limits are for energy bins with the TS value of HESS J1912+101 smaller than 4.0. 
The $\gamma$-ray SED shows a new spectral component below 10 GeV, which may be attributed to shock interaction with molecular clouds. The nature of this soft extended component will be explored in a separate paper. Here we focus on spectral modeling above 10 GeV.

The right-panel of Figure \ref{fig:lep} shows that even with a magnetic field of $3\ \mu$G, the age of HESS J1912+101 needs to be less than 10 kyrs to reproduce the $\gamma$-ray spectrum, which contradicts molecular cloud observations \citep{Su_2017} and properties of the associated pulsar \citep{2019ApJ...871...78S}. Here the $\gamma$-ray is produced via IC processes \citep{1968PhRv..167.1159J}, and besides the cosmic microwave background radiation (CMB), we assume an infrared photon background with $T=30$ K and an energy density of 1 eV cm$^{-3}$ \citep{2006ApJ...648L..29P}. To facilitate model comparison, these values for background photons will be used in this paper unless specified otherwise. Although the $\gamma$-ray spectrum is flat, it implies an electron distribution with an index of $\sim$3 in the leptonic scenario. Considering the low radio flux density, it is evident that energetic electrons need to have a broken power law distribution with a hard low-energy spectrum. Here the spectral index $\alpha=1.7$ is indicated in the left panel of Figure \ref{fig:lep}. Beyond the break energy of $E_{\rm br}=50$ GeV, the spectral index is 2.7.

\begin{figure}[ht!]
\plottwo{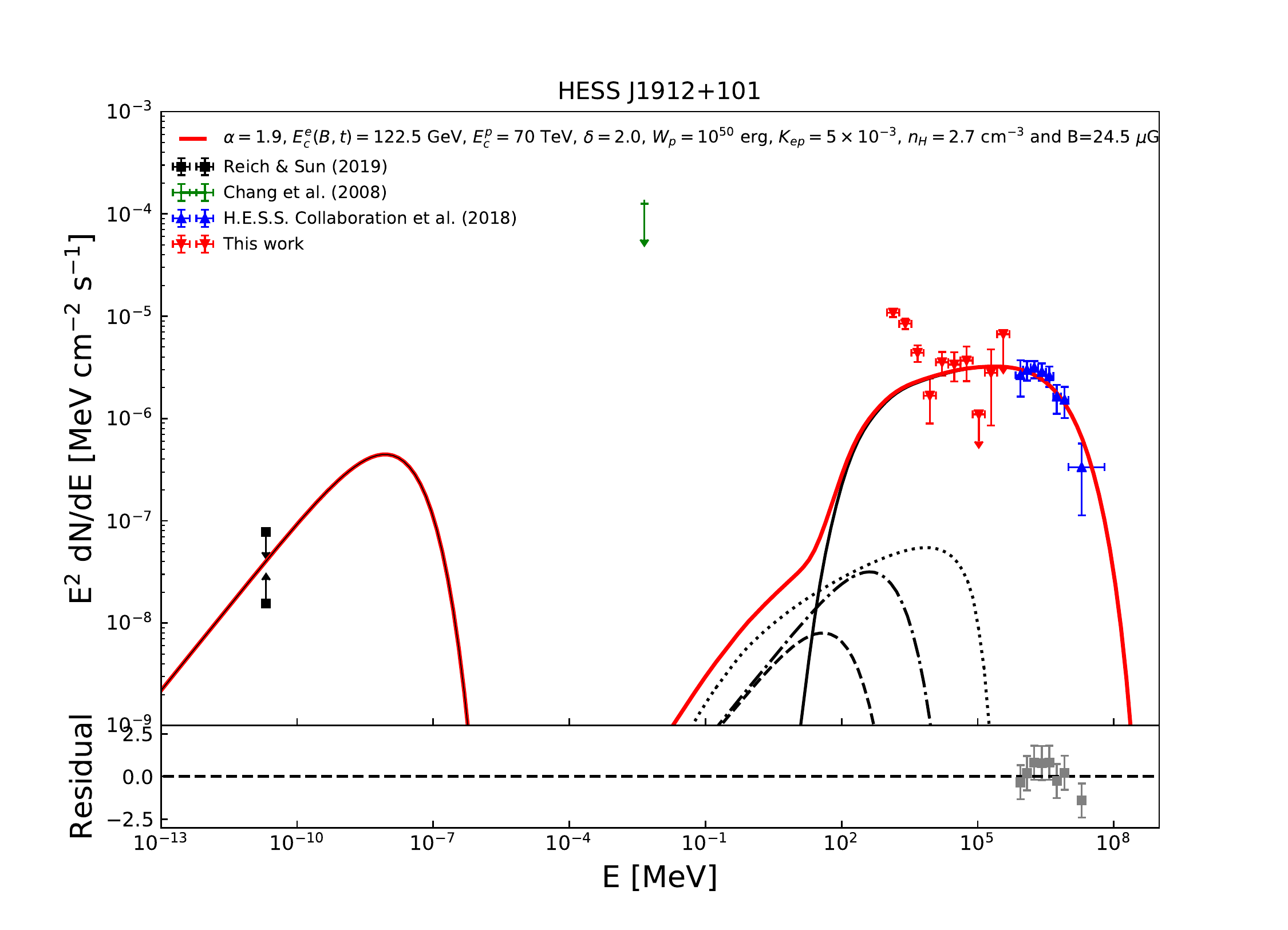}{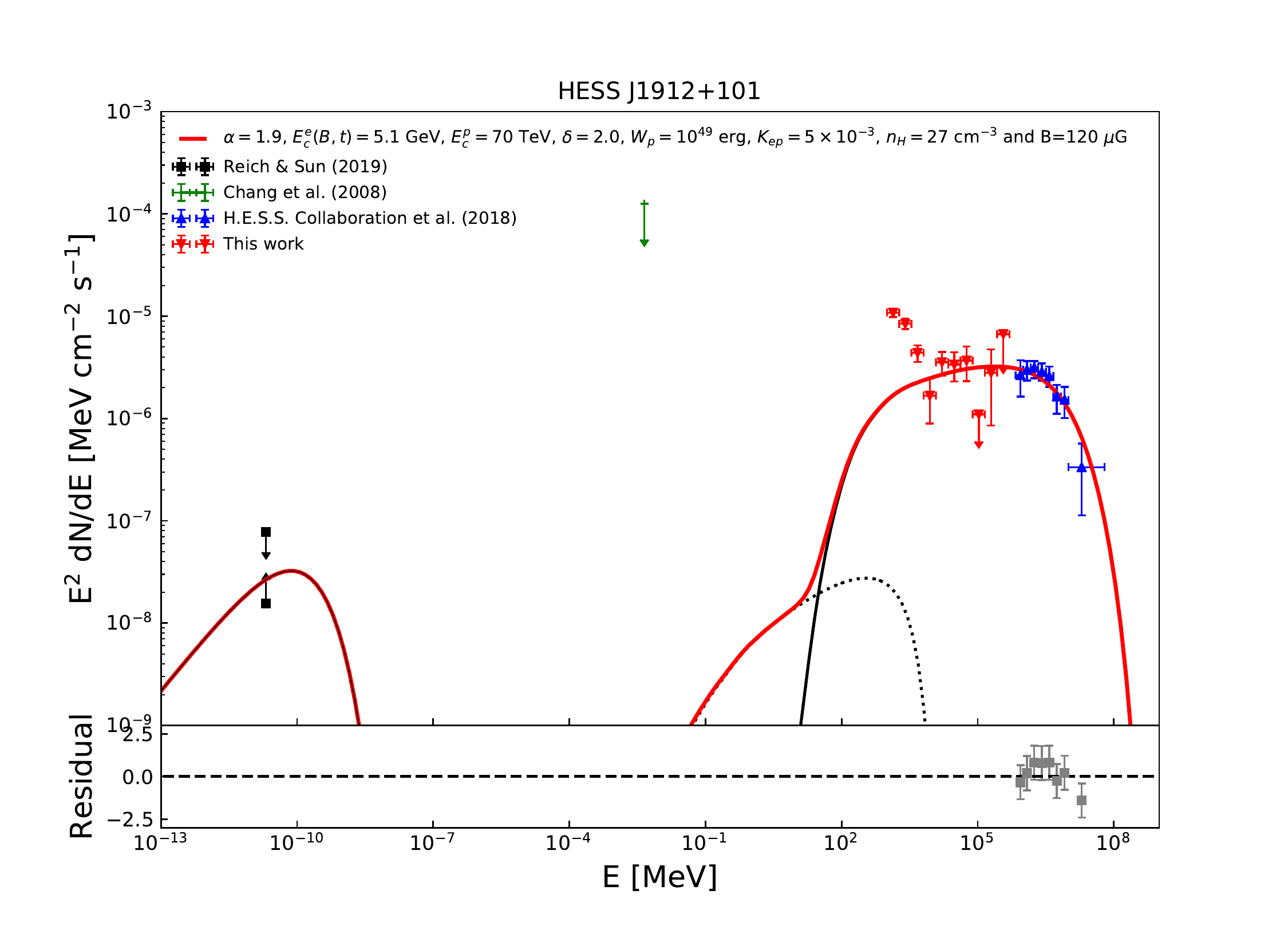}
\caption{Fit to the multi-wavelength SED of HESS J1912+101 in the hadronic scenario for the $\gamma$-ray emission. The total energy of protons above 1 GeV is $10^{50}$ erg (Left) and $10^{49}$ erg (Right). The model parameters are indicated on the Figures. The solid line is for $\gamma$-ray emission via hadronic processes, while the dotted, dotted-dashed, and dashed lines are for electron bremsstrahlung and IC of Infrared and CMB photons, respectively. \label{fig:1912}}
\end{figure}

On the other hand, the $\gamma$-ray spectrum can be readily fitted in the hadronic scenario. For the sake of simplicity, we assume a power-law distribution with an identical index for electrons and ions. 
$N(R_i) = N_{0,i} R_i^{-\alpha} $exp$[-(E_i/E^i_{\rm cut})^{\delta}]$ 
where $R_i=p_i/q_i$ is the rigidity of the particle, $E$, $p$ and $q$ are the particle energy, momentum and charge, respectively, and $"i"$ represents different particle species. 
Considering that the flux ratio of TeV CR electrons and protons is less than $0.1\%$ and high-energy electrons are subject to radiative energy loss as they propagate in the Milky Way galaxy, we fix the density ratio of electrons and protons at 1 GeV $K_{\rm ep}=N_{0,\rm e}/N_{0,\rm p}=5\times 10^{-3}$. 
The total energy content of protons above 1 GeV ($W_{\rm p}$) determines the normalization of the particle distributions. We will usually consider two cases: $W_{\rm p}=10^{49}$ and $W_{\rm p}=10^{50}$.
The mean background density $n_{\rm H}$ and magnetic field $B$, and the spectral index $\alpha$ and cutoff energy of protons $E_{\rm cut}^p$ can be adjusted to fit the multi-wavelength SED.
The cutoff energy of electrons $E_{\rm cut}^e$ is obtained by requiring the corresponding synchrotron energy loss time be equal to the age of the remnant and we assume a super exponential high-energy cutoff for the electron distribution with $\delta =2$ unless specified otherwise. The high-energy cutoff of ion distribution is always exponential with $\delta =1$. When calculating the $\gamma$-ray emission via the hadronic processes, we only consider protons and contributions from other ions are approximated by multiplying the proton produced $\gamma$-ray flux by a factor of 1.84 \citep{2009APh....31..341M}. 

Figure \ref{fig:1912} shows the results of the spectral fit and the model parameters are listed in Table \ref{tab:fitpatameters}. When calculating the total energy of the magnetic field $W_{B}$, we assume a uniform magnetic field with a volume filling factor of 1. This magnetic field energy therefore should be considered as an upper limit. Although both set of parameters give good fit to the radio and $\gamma$-ray spectra, the model with a weaker magnetic field and therefore more energy in energetic particles (left panel of Figure \ref{fig:1912}) is favored for the more reasonable value of the magnetic field energy. Moreover, the case with a strong magnetic field has a synchrotron spectrum cutting off in the radio band (the right panel of Figure \ref{fig:1912}), which appears to be too low. Of course, the magnetic field is not well constrained. For the given radio emission, one can always compensate the increase in the magnetic field with a decrease in the energetic particle energy.

\begin{figure}
\plottwo{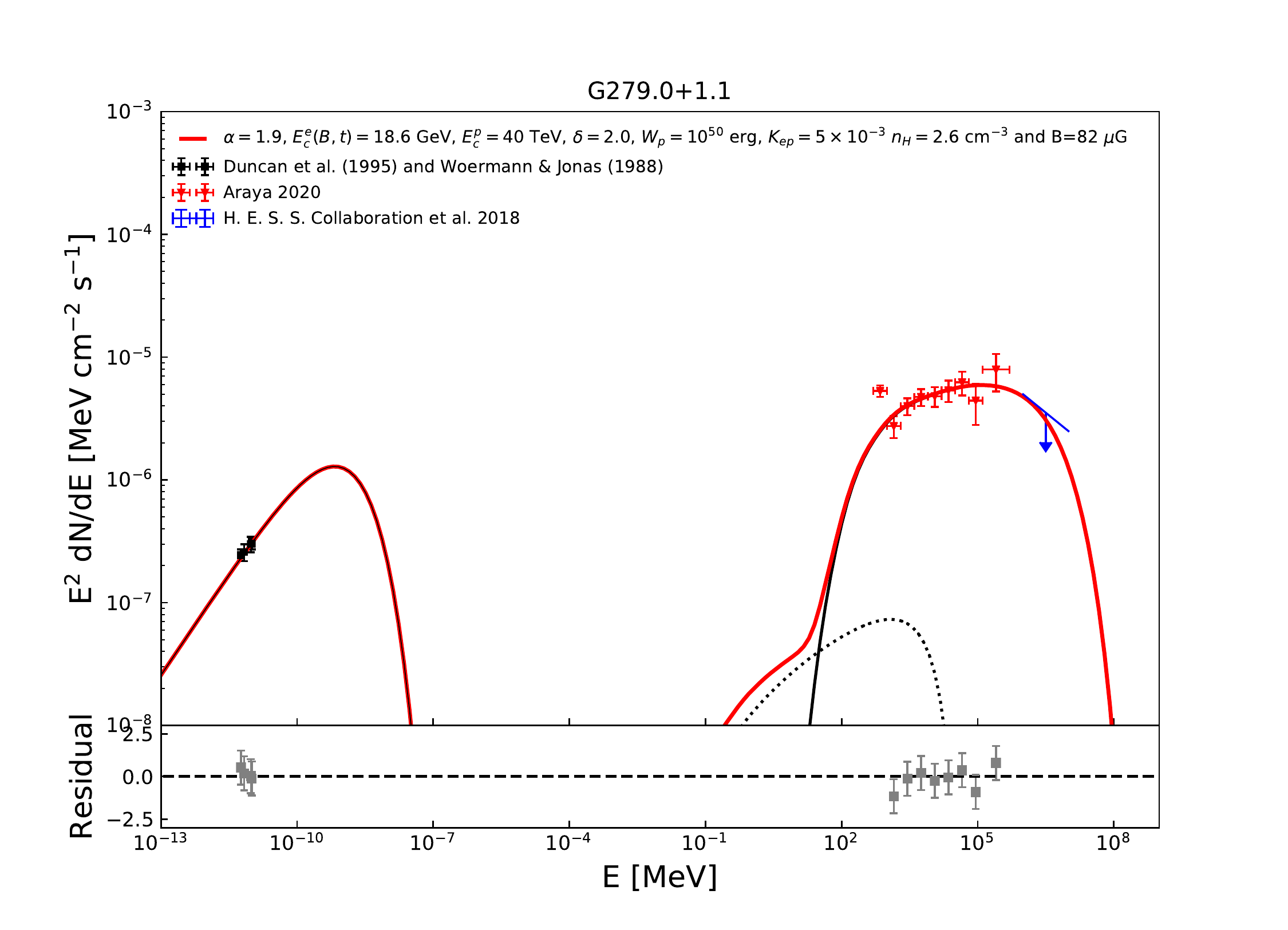}{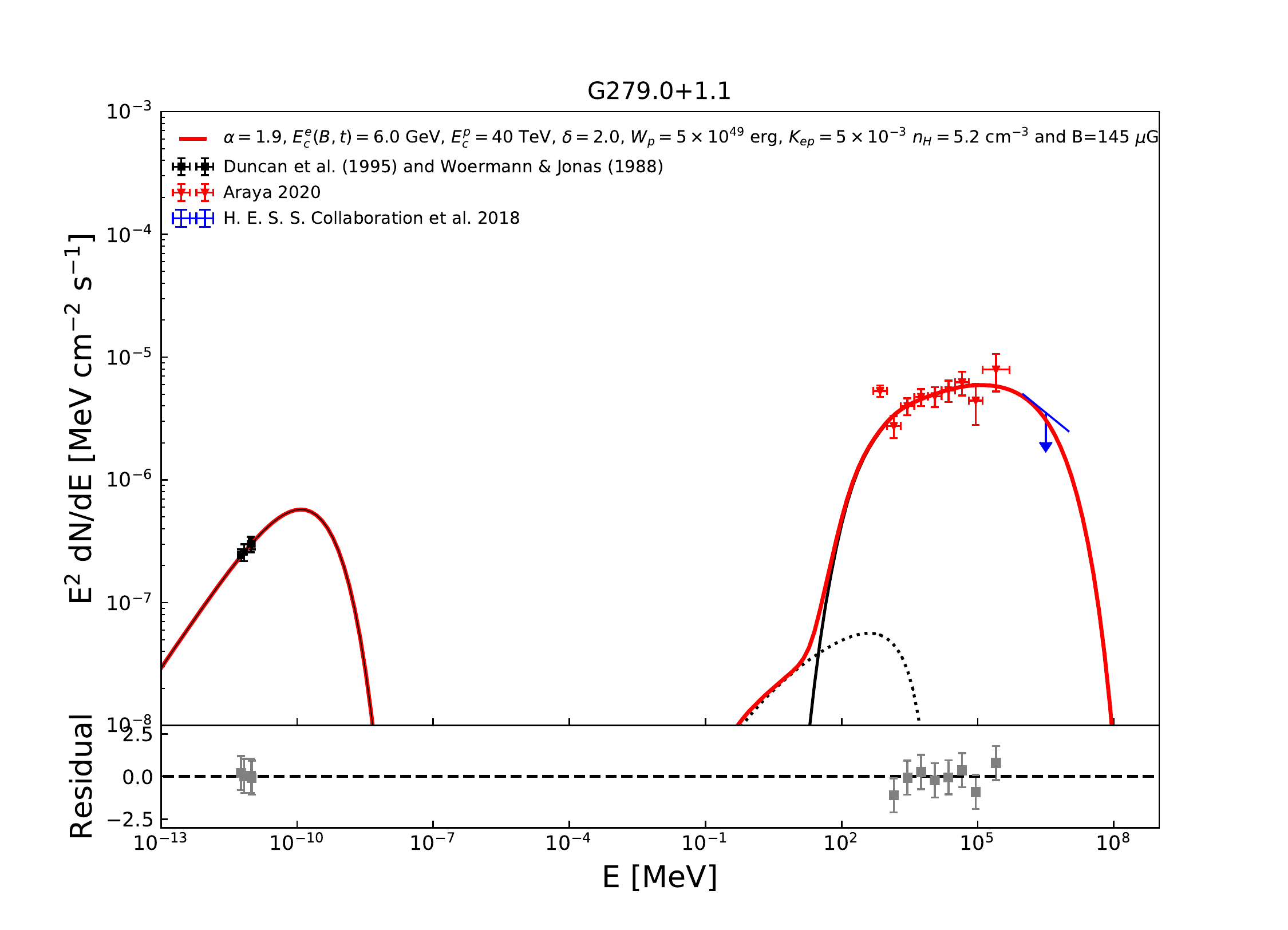}
\caption{Same as Figure \ref{fig:1912} but for G279.0$+$1.1.  The total energy of protons above 1 GeV is $10^{50}$ erg (Left) and $5\times 10^{49}$ erg (Right).
\label{fig:intermediate4}}
\end{figure}

\citet{2020MNRAS.492.5980A} recently carried out a detailed analyses of G279.0$+$1.1, a huge SNR with a radius of $\sim 40$ pc.
It has a very hard $\gamma$-ray spectrum with a spectral index of $1.86\pm 0.09$. Although it has not been detected in the TeV range, the $\gamma$-ray spectrum obtained with the Fermi-LAT extends to $0.5$ TeV without any indication of a spectral softening at high energies. The age of this remnant is greater than $100$ kyr. Figure \ref{fig:intermediate4} shows our spectral fits in the hadronic scenario for the $\gamma$-ray emission. It is interesting to note that there appears to be a low energy spectral component near 1 GeV, reminiscence of the spectral component below 10 GeV for HESS J1912+101. Compared with HESS J1912+101, SNR G279.0$+$1.1 has a similar $\gamma$-ray luminosity, a few times higher radio luminosity and radius, leading to a higher magnetic field and a large magnetic energy. However, the small volume filling factor of the radio emission region can reduce the magnetic energy significantly.

\section{G296.5+10.0} \label{sec:296}
\def\xmm{{\sl XMM-Newton}}
\def\ergcm{\hbox{erg cm$^{-2}$ s$^{-1}$ }}
\def\cmsq{\hbox{cm$^{-2}$}}
\def\cmcu{\hbox{cm$^{-3}$}}
\def\kmps{\hbox{km $\rm{s^{-1}}$}}

G296.5$+$10.0 is a bilateral morphology SNR in radio and X-rays with an angular extension of 90 arcmin $\times$ 65 arcmin, and the distance is estimated as 2.1 kpc \citep{2000AJ....119..281G}. It has relatively bright radio emission with a typical spectral index of $-0.5$ \citep{1994MNRAS.270..106M}. Due to its large size, Chandra, XMM-Newton, and Suzaku have not mapped the entire SNR, and only ROSAT PSD data provided the thermal spectrum for the whole SNR \citep{1987MNRAS.225..199K}. Nevertheless, five \xmm~observations taken in 2016 roughly cover the bright limbs of G296.5+10.0 (PI: Brian Whillianms).

\begin{figure}[htbp] 
 \centering
       \includegraphics[angle=0,width=0.24\textwidth]{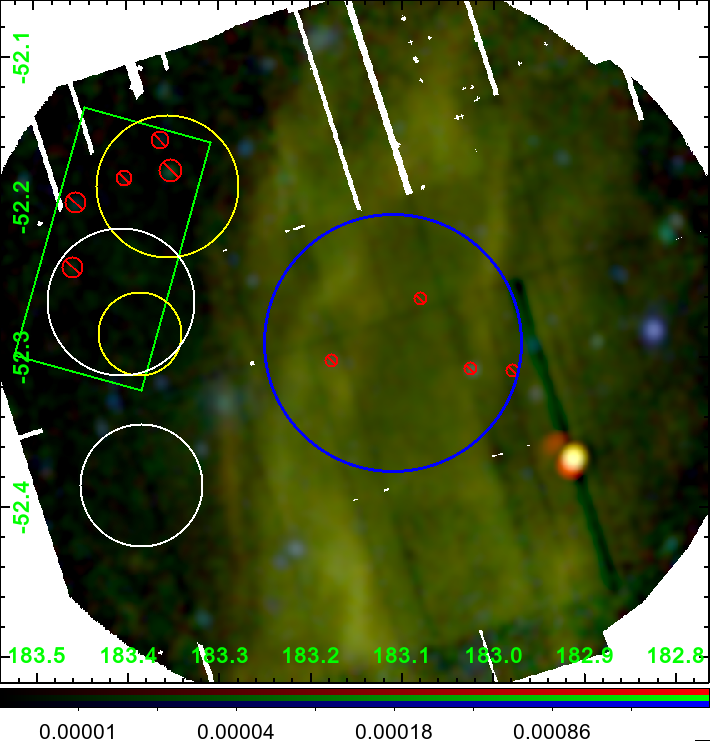} 
       \includegraphics[angle=0,width=0.35\textwidth]{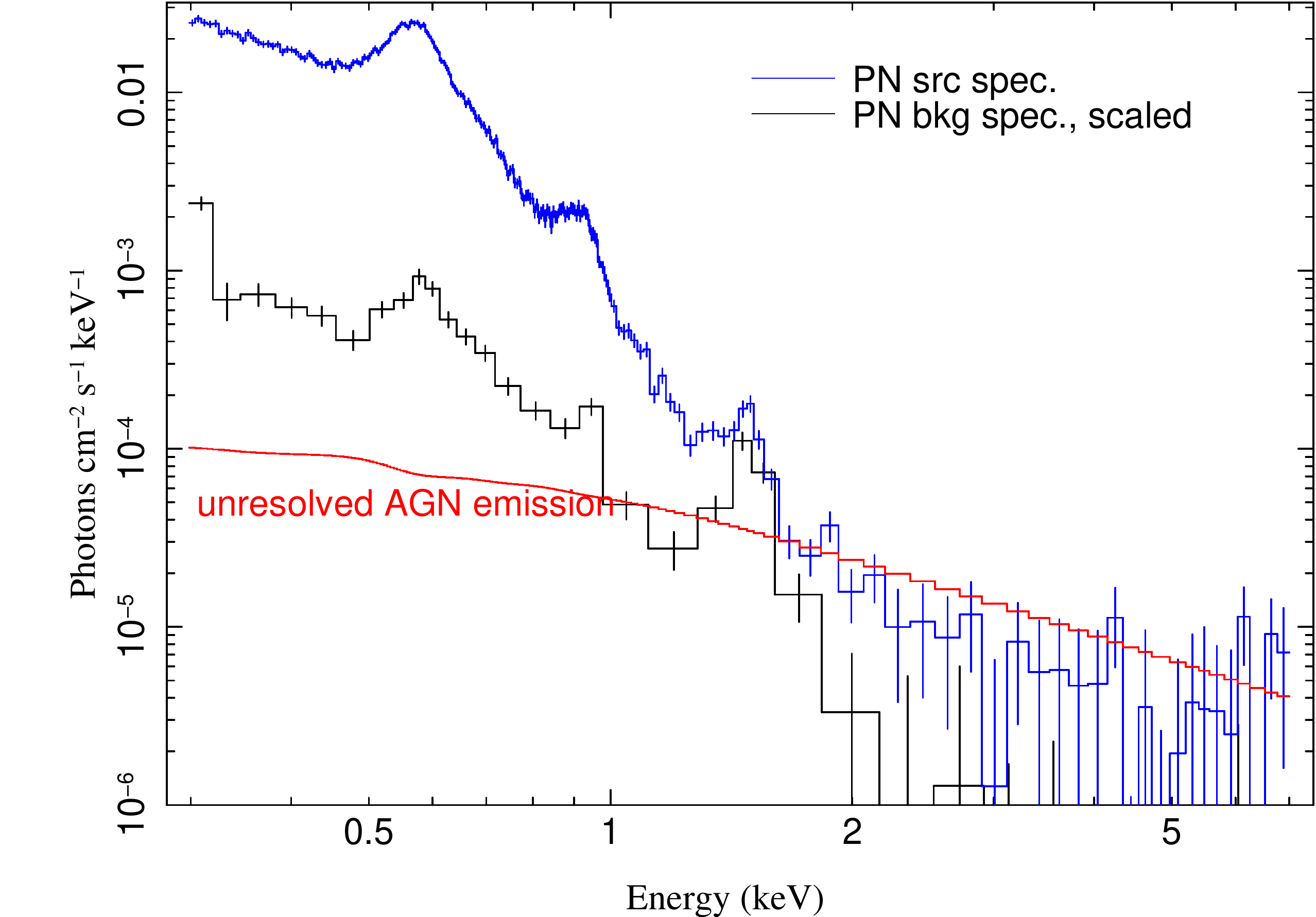} 
       \includegraphics[angle=0,width=0.36\textwidth,origin=c]{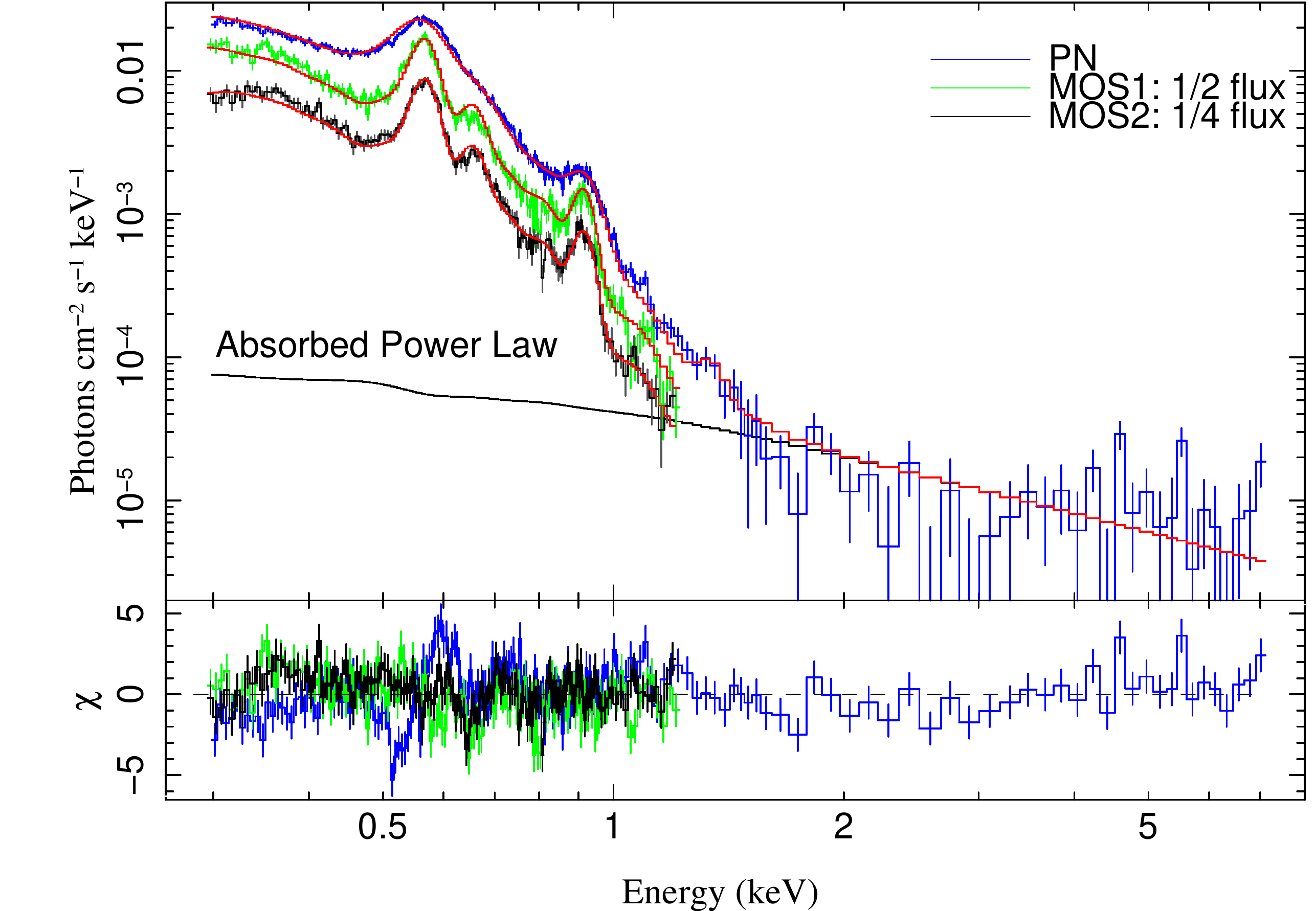} 
 \caption{The left panel shows X-ray images of the northeast part of SNR G296.5+10.0, where the red, green, and blue colors represent emission in the 0.2-0.5 keV, 0.5-1.0 keV, and 1.0-2.0 keV bands, respectively. The color bars in the bottom are in units of photons cm$^{-2}$ s$^{-1}$. The blue circle with a radius of 5.15$''$ is the region for spectral extraction, while the other circles and the square are for the `local background' regions. The green box is for the PN data, and the white and the yellow circles are for the MOS1 and MOS2 data, respectively. A few point sources are marked out with red. The middle panel shows the source-region spectrum and the scaled `local background' spectrum (according to the effective area) of the PN data, with the model of unresolved AGN emission in the cosmic X-ray background over-plotted. The right panel shows the \xmm~PN and MOS spectra with their `local background' subtracted, and the best-fit models are shown in the red color. The black curve, as modeled by an absorbed power law, represents the non-thermal emission, which is likely caused by the cosmic X-ray background.}
\label{fig:xmm} 
\end{figure}

\begin{figure}[hbt]
\centering
\includegraphics[width=3.3in]{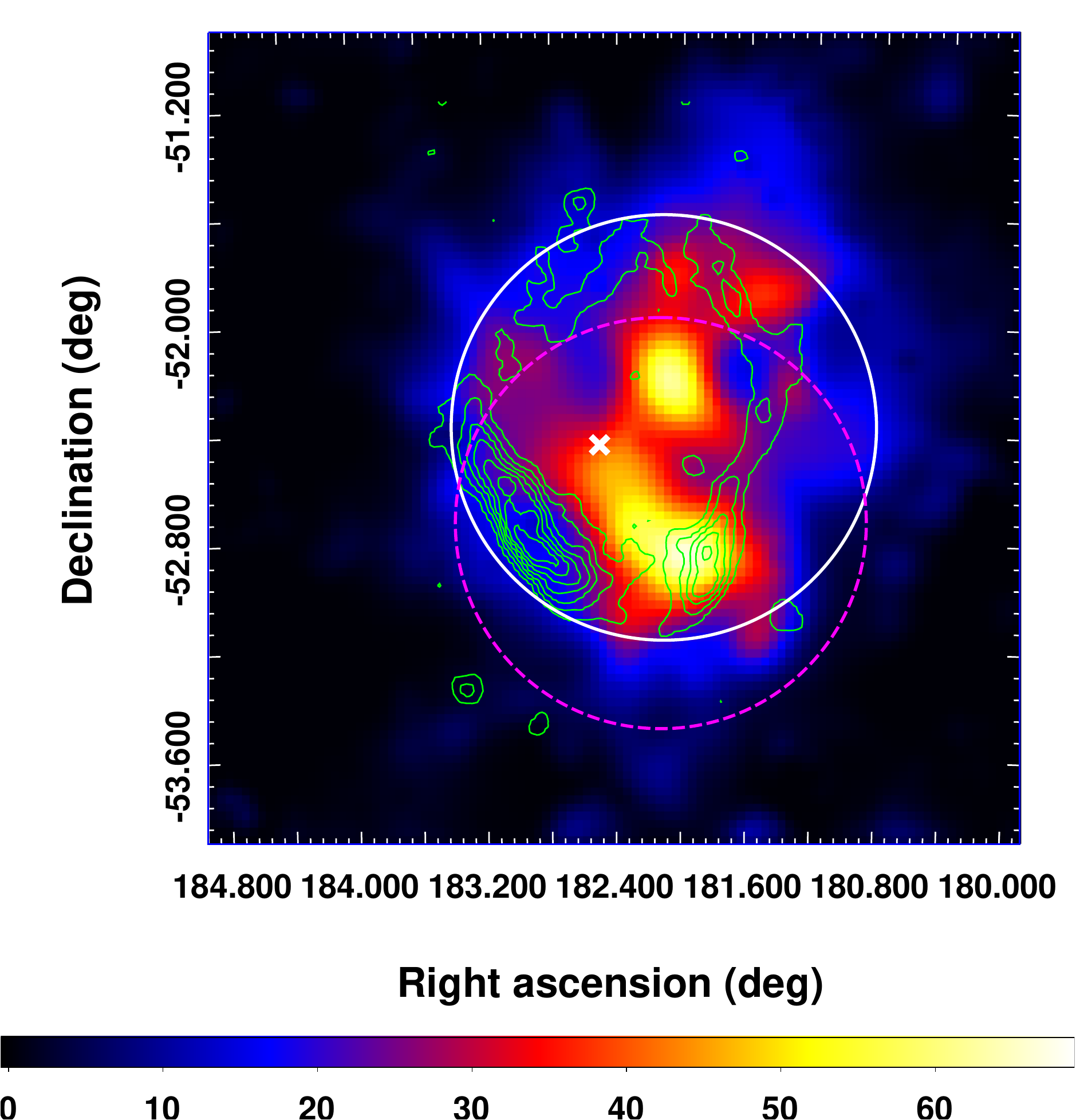}
\includegraphics[width=3.3in]{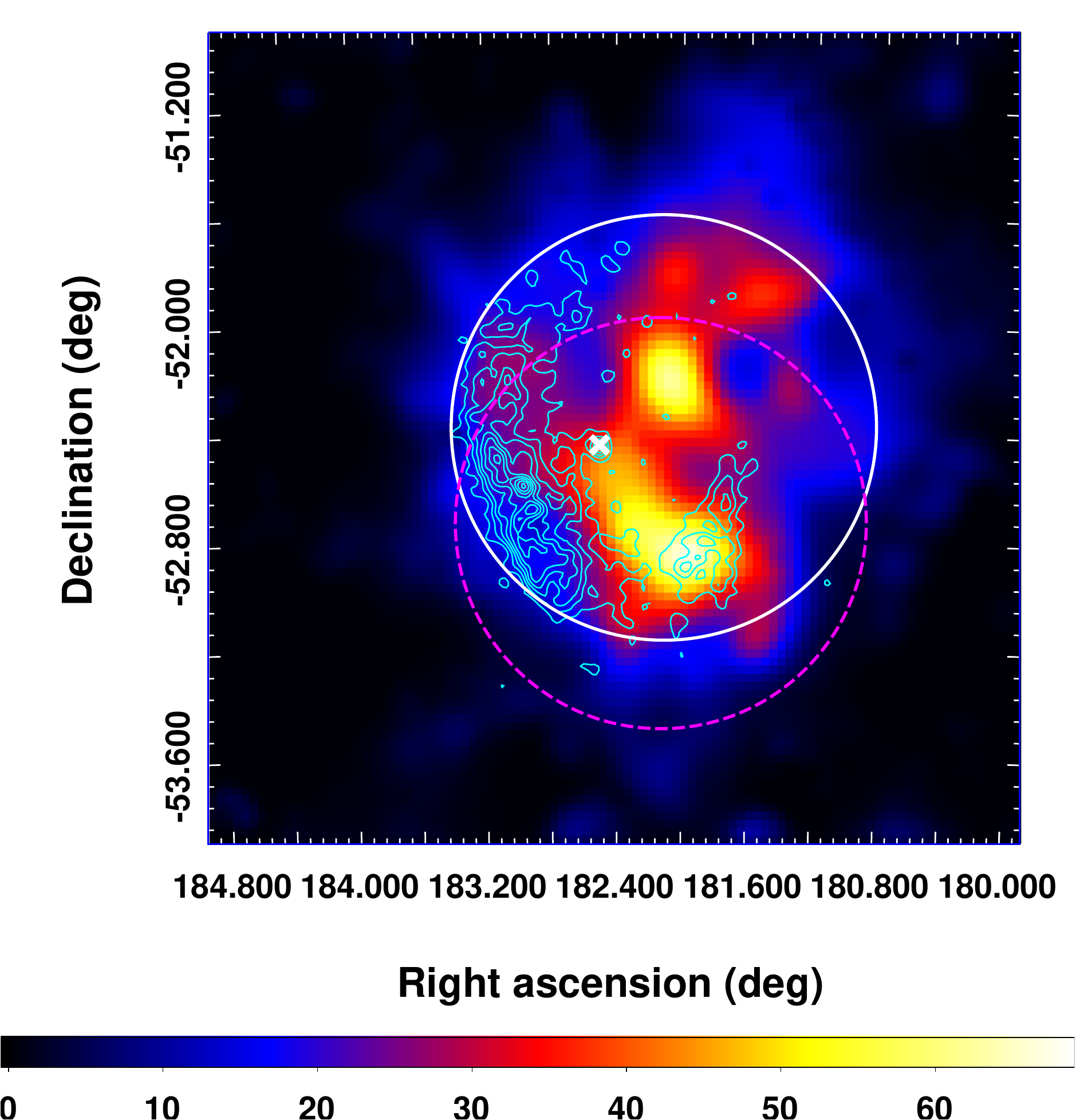}
\includegraphics[width=3.3in]{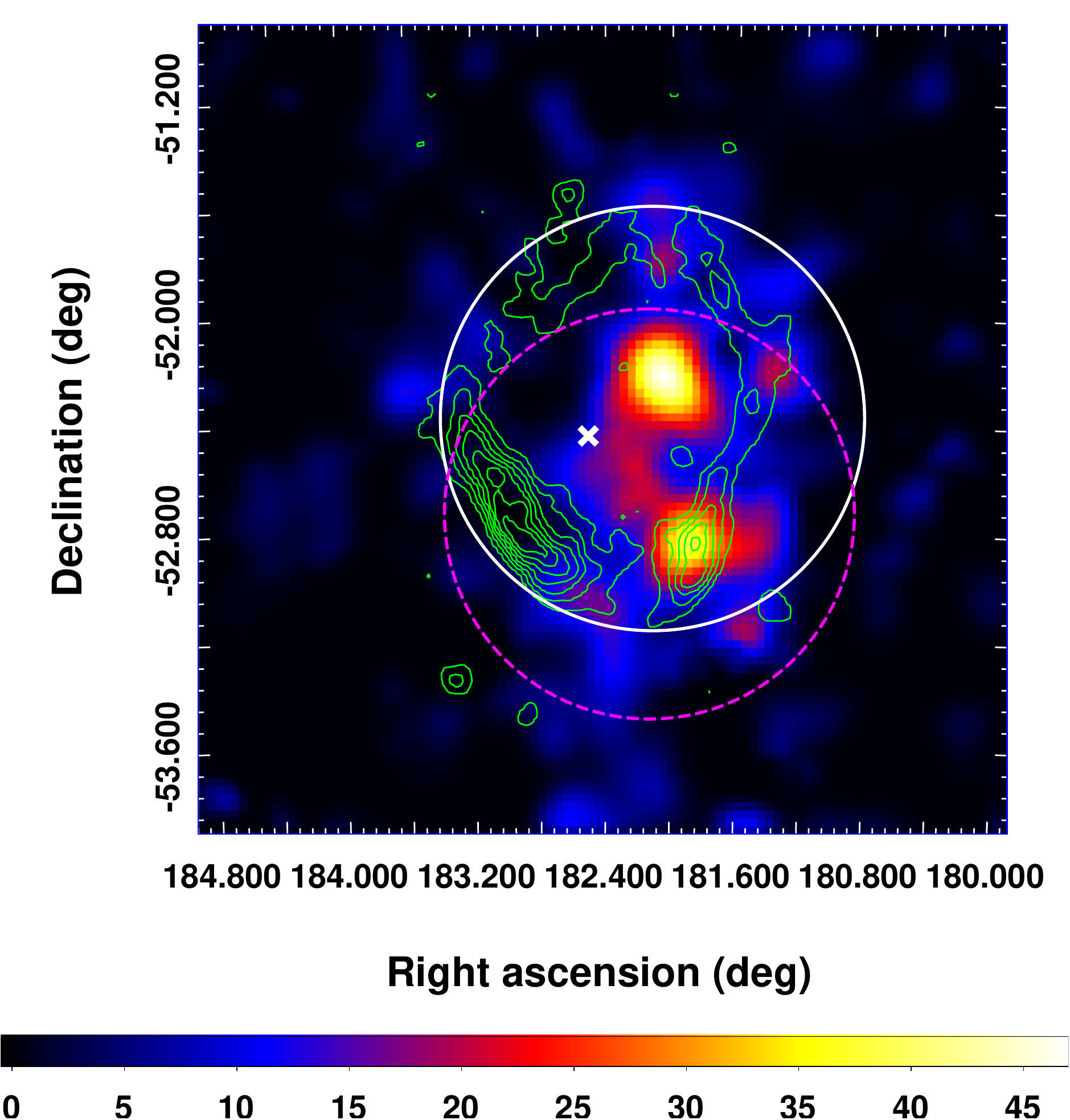}
\includegraphics[width=3.5in]{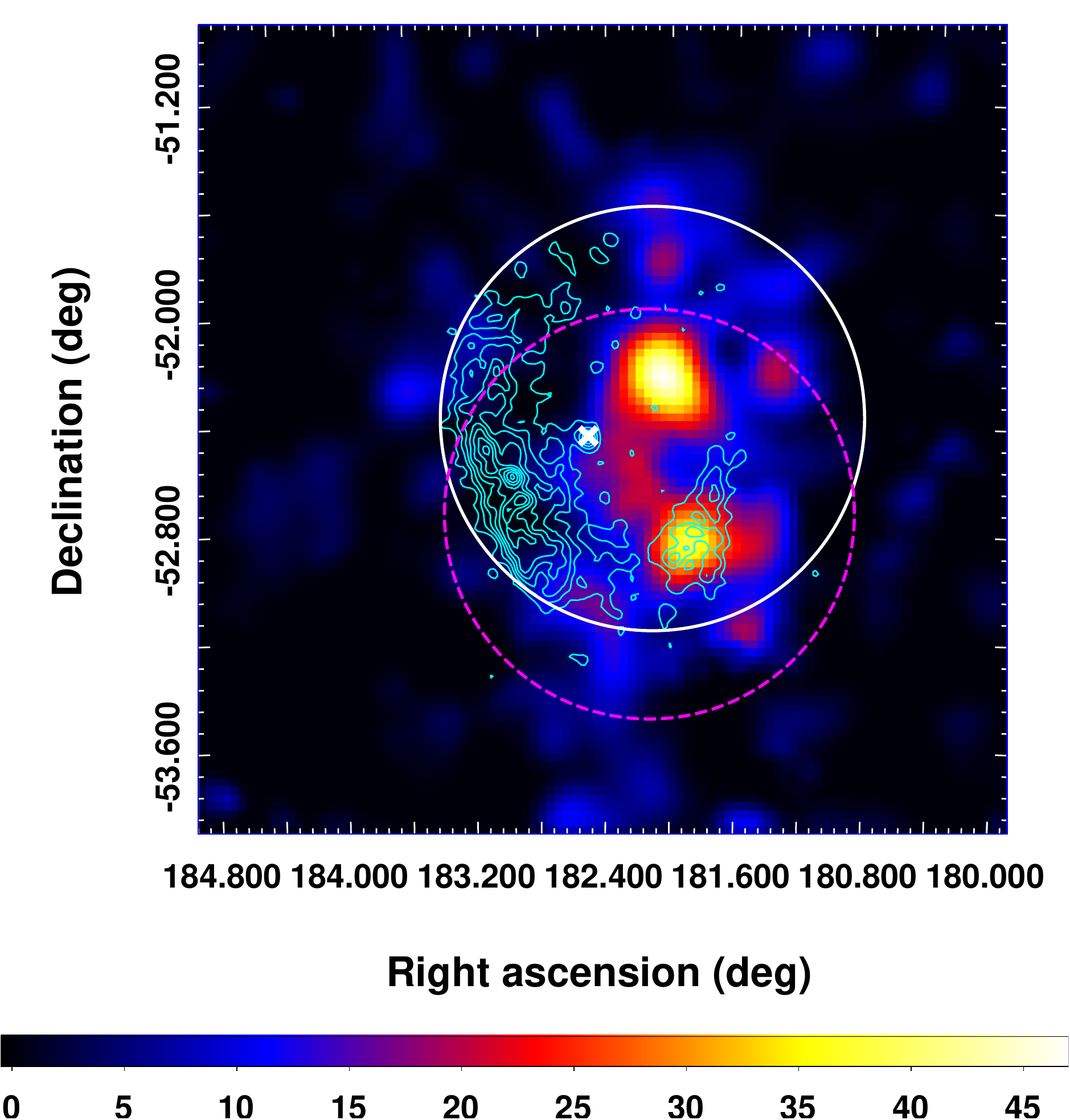}
\caption{TS maps of a $3^\circ \times 3^\circ$ region around G296.5+10.0. The top two panels is for photons in the range of 1 GeV - 1 TeV, and the bottom two panels is for photons from 5 GeV to 1 TeV. The magenta dashed circle and the white solid circle represent the size of spatial template of G296.5+10.0 in 4FGL and this work by {\em Fermi}-py, respectively. The white cross show the position of the radio-quiet X-ray-emitting neutron star (1E 1207.4-5209) in it \citep{1984Natur.307..215H, 2000ApJ...540L..25Z}. The green (left) and cyan (right) contours describe the radio and X-ray emission of G296.5+10.0, respectively.}
\label{fig:tsmap-g296}
\end{figure}

Here we analyze the \xmm~data to estimate the flux of non-thermal emission.
This emission could be fairly dim as it was not reported in the ROSAT data,  and even the residual contamination from the soft proton flares may affect the detection.
As a result, only the ID:0781720101 observation is used, which didn't suffer apparent soft proton flares during its 28 ks exposure.
The data reduction employs the standard procedure in the \xmm~Science Analysis System (SAS; version: 15.0.0) and the Extended Source Analysis Software (ESAS) package.
The diagnostic file created by the {\sl `pn-filter'} shows a steady light curve with PN count rate of $\sim$3.2 count/s in this observation, suggesting the absence of soft proton flares.
We use the {\sl `eimageget'} script to generate the PN and MOS images, and the {\sl `eimagecombine'} to combine them and produce a  background-subtracted, vignetting-corrected, and smoothed image, where the background is based on the filter wheel closed images.
This image (the left panel of Figure~\ref{fig:xmm}) covers the northeast part of the SNR.

The PN and MOS spectra are extracted from a circular region with a radius of 5.15$'$, well confined within the central CCD of the MOS1 and MOS2.
A few bright point sources are manually removed when making the mask.
The {\sl `mos/pn-spectra"} scripts are used to create spectra, and the {\sl `mos/pn-back'} scripts to generate model quiescent particle background (QPB) spectra.
Using the same method we also produce spectra from `local background' regions at the northeast corner of the field of view, where the SNR emission seems negligible.
These regions are slightly different for the MOS1, MOS2, and PN as shown in the left panel of Figure~\ref{fig:xmm}, due to differences in the effective boundaries of different detectors.
We present the PN source spectrum and the `local background' spectrum in the middle panel of Figure~\ref{fig:xmm} with their QPB spectra subtracted, and compared with the model of the unresolved AGN emission in the cosmic X-ray background \citep{2004A&A...419..837D, 2008A&A...478..575K}.
The model is an absorbed power law component with a spectral index of 1.46 and a normalization of $\sim 11.6\, \rm photons$ $\rm cm^{-2} s^{-1} sr^{-1} keV^{-1}$ at 1 keV, and the absorption is determined by the Galactic HI column density towards this SNR of about $1.1\times10^{21}$ \cmsq~\citep{2016A&A...594A.116H}. 
In the hard X-ray band (2-7 keV), the PN source spectrum is slightly under the cosmic X-ray background curve, suggesting absence of prominent  non-thermal emission.

Nevertheless, to estimate an upper limit of the non-thermal emission, we fit the PN and MOS spectra jointly, but using the `local background' spectra as the background spectra.
A single temperature APEC model represents the SNR thermal emission, and a power law model is for the non-thermal emission.
The redshift is set to zero, and the Galactic HI column density is assigned to be the maximum column density for the foreground absorption.
The best-fit spectra are shown in the right panel of Figure~\ref{fig:xmm}.
The best-fit temperature of the thermal emission is 0.15 keV, and the metal abundances are slightly sub-solar (Table~\ref{tab:xmmpar}).
The non-thermal flux in the range of 0.5--10 keV is $6.0(\pm1.5)\times10^{-13}$ \ergcm, with a photon index of $1.4\pm0.4$.
Since that the `local background' spectra do not have a good statistic in the hard X-ray band, this component is likely from the unresolved AGN emission.
Note that the photon index of 1.4 for SNR is too hard if the hard tail X-ray component is synchrotron X-rays.
We will treat this fitted flux as an upper limit to nonthermal emission from the SNR, which in turn suggests an upper limit of the non-thermal flux of $\sim1.4\times10^{-11}$ \ergcm\ for the entire SNR region \citep{1987MNRAS.225..199K}.

\begin{deluxetable*}{l|ccccccc}[htb!]
\tablecaption{The best-fit results for the \xmm~data of G296.5+10.0}
\tablecolumns{8}
\tablehead{
 \multicolumn{8}{c}{Model: TBabs*(APEC+Powerlaw)} }
\startdata
TBabs  &   \multicolumn{2}{c}{$n_{\rm H}$} &  \multicolumn{5}{c}{$(1.0-1.1)\times10^{21}$ \cmsq} \\
\hline
APEC & $\eta_\mathrm{apec}$\tablenotemark{*} & kT [keV] &  C  & N & O & Ne, Mg, Si, S & Fe, Ni \\
 & $0.04\pm0.01$ & $0.150\pm0.001$ & $<0.6$ & $0.2\pm0.1$ & $0.3\pm0.1$ & $0.7\pm0.2$ & $1.1\pm0.3$ \\
\hline
Powerlaw &  \multicolumn{2}{c}{$flux_{(0.5-10\,\mathrm{keV})}$} & \multicolumn{5}{c}{$6.0(\pm1.5)\times10^{-13}$ \ergcm} \\
 & \multicolumn{2}{c}{Photon Index} & \multicolumn{5}{c}{$1.4\pm0.4$} \\
\hline
$\chi^2$/$d.o.f.$  &   \multicolumn{7}{c}{967/480 $\sim$ 2}
\enddata
\tablecomments{Compared to the MOS spectra, the PN spectrum may have some systematic deficit in flux. The total flux of the PN is allowed to vary during the joint fitting, and the best-fit value here is $0.84\pm0.01$.}
\tablenotetext{*}{Normalization parameter of the APEC model has the physical meaning of $\frac{10^{-14}}{4\pi[D_{\rm A}(1+z)^2]} \int n_{\rm e}n_{\rm H} dV$, where $D_{\rm A}$ is the angular diameter distance to the source (cm) and $z$ is the redshift, $n_{\rm e}$ and $n_{\rm H}$ are the electron and the hydrogen densities (cm$^{-3}$), and $V$ is the volume. }
\label{tab:xmmpar}
\end{deluxetable*}

Using 52 months of Pass 7 data recorded by {\em Fermi}-LAT, \citet{2013MNRAS.434.2202A} detected extended $\gamma$-ray emission toward the region of SNR G296.5+10.0 with a $\sim$ 5$\sigma$ confidence level. 
The $\gamma$-ray spectrum of it can be fitted by a power-law with an index of 1.85$\pm$0.13.
In the fourth Fermi-LAT source catalog \citep[4FGL;][]{2020ApJS..247...33A}, G296.5+10.0 corresponds to the extended source 4FGL J1208.5-5243e, and the spatial template of it is adopted to be an uniform disk centered at (R.A.$=182.13^\circ$, Dec.$=-52.73^\circ$) with an radius of 0.76$^\circ$.

\begin{figure}[!htb]
\centering
\includegraphics[width=3.5in]{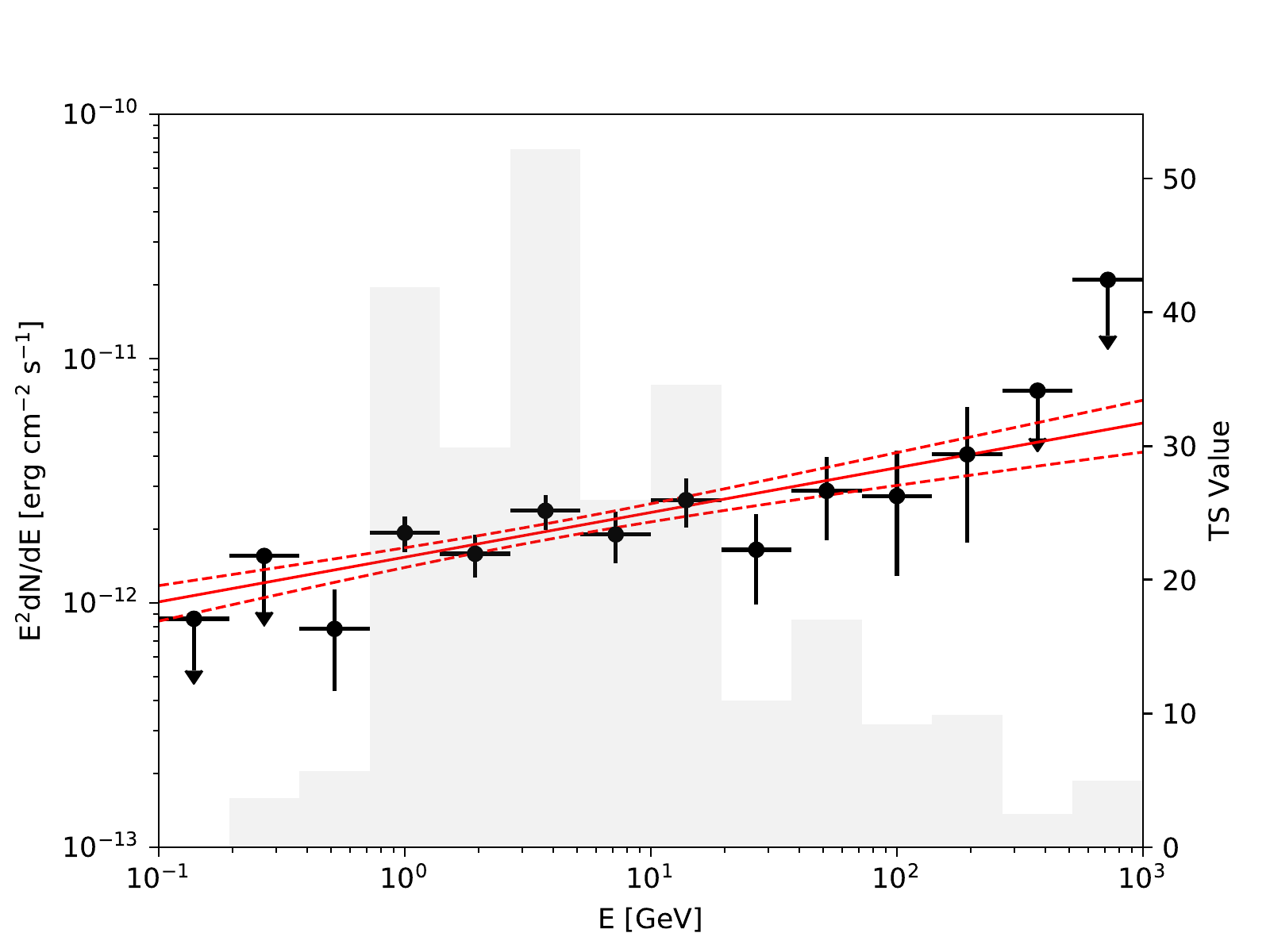}
\caption{The $\gamma$-ray SED of G296.5+10.0. The red solid line shows the best-fitting power-law spectrum in the energy range of 100 MeV - 1 TeV, and the red dashed lines show its 1$\sigma$ statistic error. The gray histogram denotes the TS value for each energy bin and the arrows indicate the upper limits with 95\% significance level.}
\label{fig:sed-g296}
\end{figure}

\begin{figure}[hbt!]
\plottwo{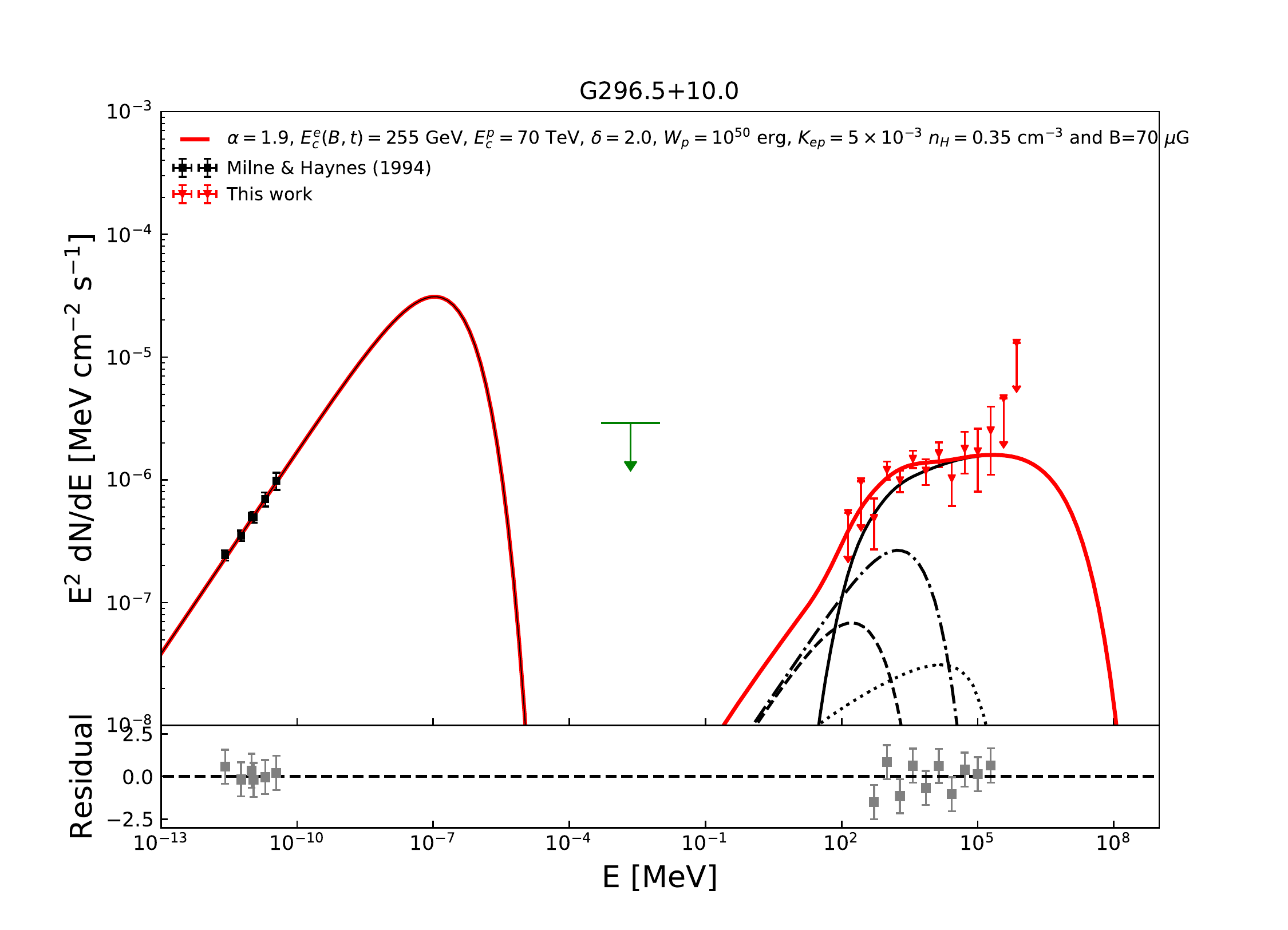}{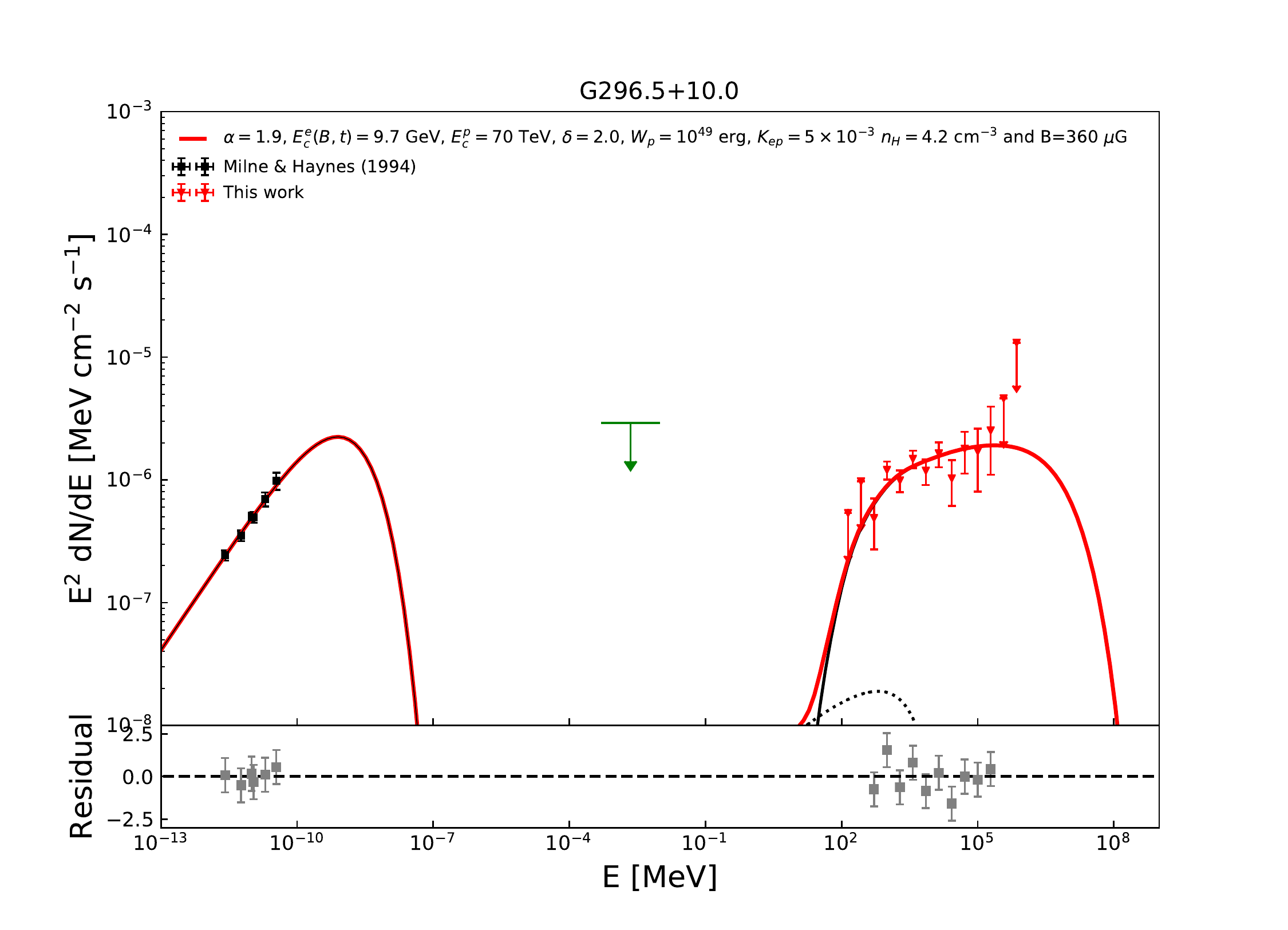}
\caption{Same as Figure \ref{fig:1912} but for G296.5+10.0. \label{fig:296}}
\end{figure}

With the latest {\em Fermi}-LAT Pass 8 data collected from 2008 August 4 to 2018 August 4, we updated the $\gamma$-ray morphology and spectrum of G296.5+10.0. Figure \ref{fig:tsmap-g296} shows the TS maps of a $3^\circ \times 3^\circ$ region around G296.5+10.0. 
In each panel, the magenta dashed circle is the size of spatial template of G296.5+10.0 in 4FGL \citep{2020ApJS..247...33A}. As can be seen, the $\gamma$-ray emission of G296.5+10.0 has a significant deviation with this spatial template.
Therefore, we use the {\em Fermi}-py tool to refit the data and get the updated centered position and radius of the uniform disk,
which is shown as the white solid circle in Figure \ref{fig:tsmap-g296}. The best-fit position and radius of the uniform disk are (R.A.$=182.11^\circ$, Dec.$=-52.38^\circ$) and 0.79$^\circ$, respectively. With this spatial template, the TS value of G296.5+10.0 is fitted to be 203.68, and the spectrum can be well fitted by a power-law with an index of 1.90$\pm$0.06. 
In the energy range from 1 GeV to 1 TeV, the integral photon flux is $(1.19\pm0.11)\times10^{-9}$ photon cm$^{-2}$ s$^{-1}$ 
with statistical error only. Moreover, to derive the $\gamma$-ray SED of G296.5+10.0, all of the data from 100 MeV to 1 TeV are divided to be 14 equal logarithmic bins. And for any energy bin with TS value of G296.5+10.0 smaller than 5.0, an upper limit at 95\% confidence level is calculated. The SED and fit results are shown in Figure \ref{fig:sed-g296}.

G296.5$+$10.0 has an age of $\sim 10$ kyrs \citep{1997ApJ...476L..43V} and a GeV $\gamma$-ray spectrum similar to that of HESS J1912+101. Adopting the hadronic model for HESS J1912+101, one can fit the radio and $\gamma$-ray spectra of G296.5$+$10.0. Due to the absence of TeV observations, the high-energy cutoff of the proton distribution is not well constrained and we adopt a value of 70 TeV derived from the spectral fit of G296.5$+$10.0. Figure \ref{fig:296} shows the spectral fit, and the model parameters are listed in Table \ref{tab:fitpatameters}, which are very similar to those for HESS J1912+101 except for a slightly stronger magnetic field. As we will see below, this is due to the higher radio to $\gamma$-ray flux ratio of G296.5$+$10.0 than that of HESS J1912+101. Similarly, we favor the model with a weaker magnetic field (left panel of Figure \ref{fig:296}).

\section{Hadronic and/or Leptonic models for hard GeV spectra of other 10 SNRs} \label{sec:sample}

\begin{figure}[ht!]
\plotone{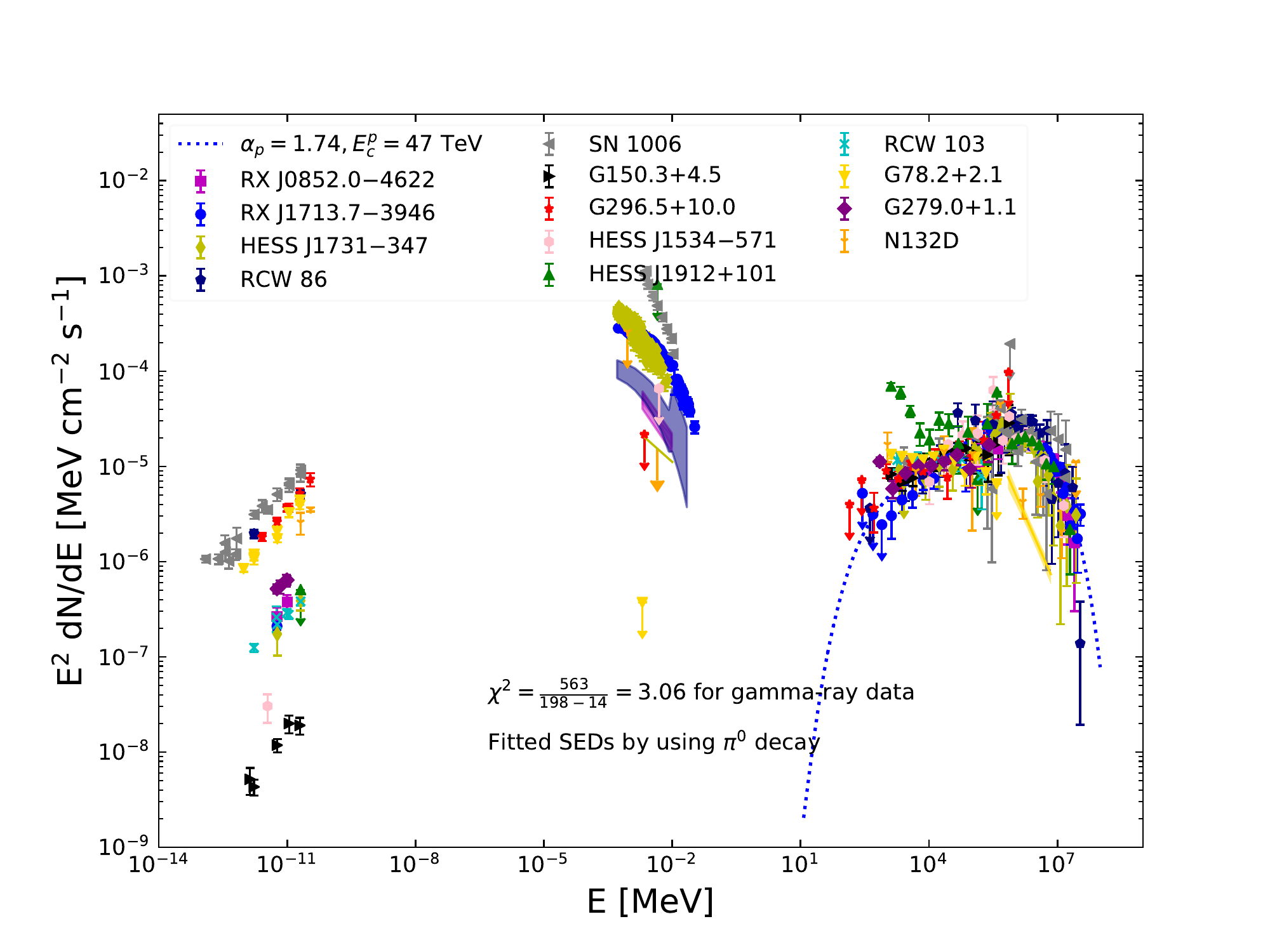}
\caption{The multi-wavelength spectral data of 13 SNRs with hard $\gamma$-ray spectra. The $\gamma$-ray spectra are fitted with a hadronic model with the normalization of individual spectrum as free parameters. The model assumes that protons have a single power-law energy distribution with an exponential high-energy cutoff. Note that the TeV spectra of G78.2$+$2.1 (HAWC) and N132D (HESS) cut off at relatively lower energies, and the soft spectral component of GeV of HESS 1912$+$101 may be from other contributors, and are not considered in SED fitting. The best-fit model parameters are indicated on the Figure.
References for the observational data are as follows. RX J0852.0$-$4622: radio \citep{2000A&A...364..732D},
GeV \citep{2011ApJ...740L..51T}, X-ray \citep{2007ApJ...661..236A}, TeV \citep{2018A&A...612A...7H}; RX J1713.7$-$3946: radio \citep{2004ApJ...602..271L}, X-ray \citep{2008ApJ...685..988T}, GeV and TeV \citep{2018A&A...612A...6H}; HESS J1731$-$347: radio \citep{2008ApJ...679L..85T}, GeV \citep{2017ApJ...851..100C,2018ApJ...853....2G}, X-ray \citep{2017A&A...608A..23D}, TeV \citep{2011A&A...531A..81H}; RCW 86: radio \citep{1975AuJPA..37....1C,2012AA...545A..28L}, X-ray \citep{2012AA...545A..28L}, GeV \citep{2016ApJ...819...98A}, TeV \citep{2018A&A...612A...4H}; SN 1006: radio \cite{2009AJ....137.2956D}, X-ray \citep{2008PASJ...60S.153B}, GeV \citep{2017ApJ...851..100C}, TeV \citep{2010AA...516A..62A}; G$150.3+4.5$: radio \citep{2014A&A...566A..76G}, X-ray and GeV \citep{2020arXiv200908397D}; G$296.5+10.0$: radio \citep{1994MNRAS.270..106M}, GeV (this work), HESS J1534$-$571: radio \citep{2018MNRAS.480..134M}, GeV \citep{2017ApJ...843...12A}, X-ary and TeV \citep{2018A&A...612A...8H}; RCW 103: radio \citep{1996AJ....111..340D}, GeV \citep{2014ApJ...781...64X}; G78.2$+$2.1: radio \citep{1991A&A...241..551W,1997A&A...324..641Z,2006A&A...457.1081K,2011A&A...529A.159G}, X-ray \citep{2013MNRAS.436..968L}, GeV \citep{2018ApJ...861..134A} and TeV \citep{2019ICRC...36..675F};
G279.0$+$1.1: radio \citep{1988MNRAS.234..971W,1995MNRAS.277..319D}, GeV \cite{2020MNRAS.492.5980A}; N132D: radio \citep{1995AJ....109..200D}, X-ray \citep{1998ApJ...505..732H,2018ApJ...854...71B}, GeV (Xin et al. 2020, in preparation) and TeV \citep{2015Sci...347..406H}.
\label{fig:sample}}
\end{figure}

We now extend the above study to all SNRs with hard $\gamma$-ray spectra. Figure \ref{fig:sample} shows the multi-wavelength non-thermal spectra of 12 SNRs with hard GeV spectra. The spectra have been normalized at $\gamma$-ray band by fitting the $\gamma$-ray spectra with a hadronic model. 
The model assumes that protons have a single power law with an exponential high energy cutoff. 
 The normalization of the SED of each source is adjusted to minimize the $\chi^2$ of the $\gamma$-ray data of all sources. We then have an index of 1.74 and $E_{\rm cut}^p=47$ TeV.
Then we may classify these sources based on their radio and/or X-ray spectra. 

Based on the radio flux, these SNRs may be divided into three categories. SN 1006, RCW 86, G296.5+10.0, G78.2+2.1 and N132D have strong radio emission, while the radio emission from G150.3+4.5 and HESS J1534-571 are very weak. Their normalized radio flux densities can differ by about two orders of magnitudes. The famous TeV bright SNRs RX J1713-3946, RX J0852.0-4622, HESS J1731-347, and HESS J1912+101, RCW 103 have normalized radio flux densities between the above two categories. In Table \ref{tab:fitpatameters}, we use double horizontal lines to separate these categories and give the estimated age, distance, radius of these SNRs and the related references. Non-thermal X-ray emission is detected from five relatively young SNRs: SN 1006, RCW 86, RX J1713-3946, RX J0852.0-4622, and HESS J1731-347. As will be shown below, a broken power-law spectrum is needed to explain their multi-wavelength non-thermal emission spectra, while for other sources, the above hadronic model for HESS J1912+101 and G296.5+10.0 with a single power-law high-energy particle spectrum is sufficient (See Table \ref{tab:fitpatameters}).

\begin{deluxetable*}{lcccccccccccccc}[htb]
\tablecaption{Physical and fitting parameters for our sample.}
\tablewidth{0pt}
\tablehead{
\colhead{Source} & \colhead{Age} & \colhead{Distance} & \colhead{Radius}&\colhead{$W_{\rm p}$} &\colhead{$n_{\rm H}$}& \colhead{$\alpha$} &\colhead{$E^{\rm p}_{\rm cut}$} & \colhead{$\Delta \alpha$}&\colhead{$\delta$} & \colhead{$E^{\rm e}_{\rm br}$}&\colhead{$E^{\rm e}_{\rm cut}$} &\colhead{$B$}&\colhead{$W_B$}&\colhead{$W_B/W_{\rm e}$}\\
\colhead{} & \colhead{kyr} & \colhead{kpc} & \colhead{pc} & \colhead{$10^{49}$ erg}&\colhead{cm$^{-3}$}& \colhead{}& \colhead{TeV} &
\colhead{} & \colhead{}& \colhead{GeV}& \colhead{TeV}& \colhead{$\mu$G}&\colhead{10$^{49}$ erg}&\colhead{10$^{2}$}
}
\startdata
\hline
G150.3$+$04.5 & 6.0 & 1.0 &24& 1.0& 4.2& 1.75 & 70& $-$ & 2.0& $-$ & 2.0$^a$      & 32 & $6.8$ & {\bf 34.6}\\
& & & & 10.0& 0.48& 1.75 & 70& $-$ & 2.0& $-$ & 0.2     & 6.0 & $0.24$&0.26\\
\hline                        
HESS J1534$-$571 & 10.0 & 3.5 &22.4& 1.0& 12& 1.45 & 20& $-$ & 2.0& $-$ & 0.14$^a$   &300 & 496 & $7.2 \times 10^4$\\
G323.7$-$01.0    &      &      &   & 10.0& 1.2& 1.45& 20 & $-$ & 2.0& $-$ & 0.59$^a$   & 46  &11.6& ${\bf 16}$\\
\hline    
\hline
HESS J1912$+$101 & 170 & 4.1 & 15 & 1.0& 27& 1.9 & 70& $-$ & 2.0& $-$ & 0.0051$^a$   & 120  & 23.9& $6.0 \times 10^2$\\
G044.5$-$00.2      &        &   &    & 10.0& 2.7& 1.9 & 70& $-$ & 2.0& $-$ & 0.1225$^a$    & 24.5  & 1.0&{\bf 0.63}\\
\hline
G279.0$+$1.1 & $100$ & 3.0 &41.5& 5.0& 5.2& 1.9 & 40& $-$ & 2.0& $-$ & 0.006$^a$  &145& $7.3\times10^2$& $3.0 \times 10^3$\\
              &      &     &    & 10.0& 2.6& 1.9 & 40& $-$ & 2.0& $-$ & 0.0186$^a$   & 82 &$2.3\times10^2$ & ${\bf 256}$\\
\hline
RCW 103 & 4.4 & 3.3 & 4.8 &1.0& 65& 2.05 & 70& $-$ & 2.0& $-$ & 0.56$^a$     & 226 & 2.8 & {12}\\
G332.4$-$00.4 &    & &    &10.0& 6.5& 2.05 & 70& $-$ & 2.0& $-$ & 1.14$^a$     & 50  & 0.14&{\bf 0.034}\\
\hline
RX J1713.7$-$3946 & 1.6 & 1.0 & 8.7  & 0.6& 11.7& 1.6 & 40& 1.0 & 2.0& 25.8$^a$  &13.5   & 550 & 97.4& $3.9 \times 10^3$\\
&    &            && 5.0& 1.2 & 1.55 & 40& 1.75 & 2.0 & 390$^a$ & 40    & 142 & 6.5 & ${14}$\\
G347.3$-$00.5    &    &            && 10.0& 0.6 & 1.6 & 40& 1.0 & 2.0 & 82 & 38    & 62 & 1.24 & ${\bf 1.73}$\\
(Leptonic)   &    &       && 7.7& 0.1 & 1.89 & 70& 1.0 & 2.0 & 1800 & 70    & 22 & 0.16& {\bf 0.055} \\
\hline                  
RX J0852.0$-$4622 & 2.7 & 0.75 &13& 1.0& 5.5& 1.75 & 70& 1.0 & 2.0& 28  & 20    & 165  & 29.9& $3.9 \times 10^2$\\
&  &  &  & 1.0& 5.5& 1.75 & 70& 1.4 & 2.0& 170$^a$  & 20    & 165  & 29.9& $2.4 \times 10^2$\\
 &     &        && 10.0& 0.45& 1.5  & 35& 2.32 & 2.0& 460$^a$  & 30    & 100 & 11 & ${ 14}$\\
 G266.2$-$01.2              &     &        && 10.0& 0.55& 1.75  & 70& 1.0 & 2.0& 55  & 50    & 31 & 1.06 & ${\bf 1.08}$\\
 (Leptonic)     &     &        && 350& 0.001& 1.5 & 70& $1.0$ & 2.0& $20$ & 70    & 8 & 0.7&0.012\\
(Leptonic)     &     &        && 6.0& 0.2& 2.2 & 70& $-$ & 1.0& $-$ & 22    & 11 & 0.13&{\bf 0.037}\\
\hline                  
HESS J1731$-$347 & 2.5 & 3.2 &14&  1.0& 7.5& 1.5 & 30& 1.0 & 2.0& 38  & 6.5    & 550  &403& $1.1 \times 10^4$\\
 &  & && 
1.0& 7.5& 1.5 & 30& 0.85 & 2.0& 16.5$^a$  & 6.4    & 550  &403& $1.5 \times 10^4$\\
    &    &      &  & 10.0& 0.6& 1.5 & 30& 1.82 & 2.0& 692$^a$  & 25   & 85  &9.6& ${ 8.3}$\\
G353.6$-$00.7 &    &      &  & 10.0& 0.75& 1.5 & 30& 1.0 & 2.0& 100 & 17   & 85  &9.6& ${\bf 16.2}$\\
(Leptonic)    &    &      &  & 55& 0.05& 1.5 & 70& 1.0 & 2.0& 170  & 25   & 32  &1.05&  0.29\\
(Leptonic)    &    &      &  & 5.5& 0.05& 1.95 & 70& $-$ & 1.0& $-$  & 13   & 23  &0.71& {\bf 0.30}\\
\hline
\hline
G296.5$+$10.0 & 10.0 & 2.1 &24.1& 1.0& 4.2& 1.9 & 70& $-$ & 2.0& $-$ & 0.0097$^a$  &360& $8.9\times10^2$& $1.5 \times 10^4$\\
              &      &     &    & 10.0& 0.35& 1.9 & 70& $-$ & 2.0& $-$ & 0.26$^a$   & 70 & 33.7& ${\bf 18}$\\
\hline              
SN 1006 & 1.0 & 2.2 & 9.6 & 1.0& 1.0& 1.90 & 70& 1.0 & 1.0& 386$^a$ & 8.1    & 180 &14& $52$\\
G327.6$+$14.6(Lep)   &     &  &&1.5& 0.1& 2.1 & 70& $-$ & 1.0& $-$ & 7.3    & 65 &1.8& ${\bf 2.4}$ \\
\hline        
RCW 86 & 1.8 & 2.5 &15.3& 1.0& 6.5& 1.45& 20 & 0.85& 1.0 & 0.048$^a$ & 0.85    & $1.2\times10^4$ &$2.5 \times 10^5$& $9.0 \times 10^{8}$\\
G315.4$-$02.3(Lep)  &     & &&  10& 0.05& 2.25 & 70& 1.0 & 1.0& 7200$^a$ & 30  & 31 &1.68&{\bf 0.27} \\
\hline
Gamma Cygni & 8.25 & 2.0 &17& 1.0& 29& 2.00& 10 & $-$& 2.0 & $-$& 0.0036$^a$  & 650 &$1.0\times 10^3$& $1.6 \times 10^{4}$\\
G78.2$+$01.2  &     & &&  10& 2.9& 2.00 & 10& $-$ & 2.0&$-$& 0.077$^a$  & 140 &47.2&${\bf 18}$ \\ 
\hline 
N132D & 2.5 & 50 &11.4& 10.0& 32& 2.10& 70 &$-$ & 2.0 &$-$ &0.028$^a$   & 423 &$1.3 \times 10^2$&  {\bf 56}\\
N132D & 2.5 & 50 &11.4& 50.0& 6.4& 2.10& 70 &$-$ & 2.0 &$-$ &0.23$^a$   & 148 &15.9& {\bf 0.89}\\
\hline
\enddata
\tablecomments{References of physical parameters $-$ HESS J1912$+$101 \citep{Su_2017,2020ApJ...889...12Z}; 
RCW 103 \citep{2004PASA...21...82R,2019MNRAS.489.4444B}; 
RX J1713.7$-$3946 \citep{2003PASJ...55L..61F,2016PASJ...68..108T}; RX J0852.0$-$4622
\citep{2008ApJ...678L..35K}; HESS J1731$-$347 \citep{2008ApJ...679L..85T,2011AA...531A..81H};
G150.3$+$4.5 \citep{2016PhDT.......190C,2020arXiv200908397D}; HESS J1534$-$571 \citep{2018MNRAS.480..134M}; 
G279.0$+$1.1 \citep{1995MNRAS.277..319D,2020MNRAS.492.5980A};
G296.5$+$10 \citep{1997ApJ...476L..43V,2000AJ....119..281G}; 
SNR 1006 \citep{2003ApJ...585..324W,2009ApJ...692L.105K}; RCW 86 \citep{2000AA...360..671B,2013MNRAS.435..910H}; $\gamma$ Cygni \citep{1977AJ.....82..329H,2013MNRAS.436..968L}; N132D \citep{1995AJ....109..200D,2011ApSS.331..521V}.}
$a:$ Determined by requiring the synchrotron energy loss time being equal to the SNR age;
\label{tab:fitpatameters}
\end{deluxetable*}

\begin{figure}[ht!]
\plottwo{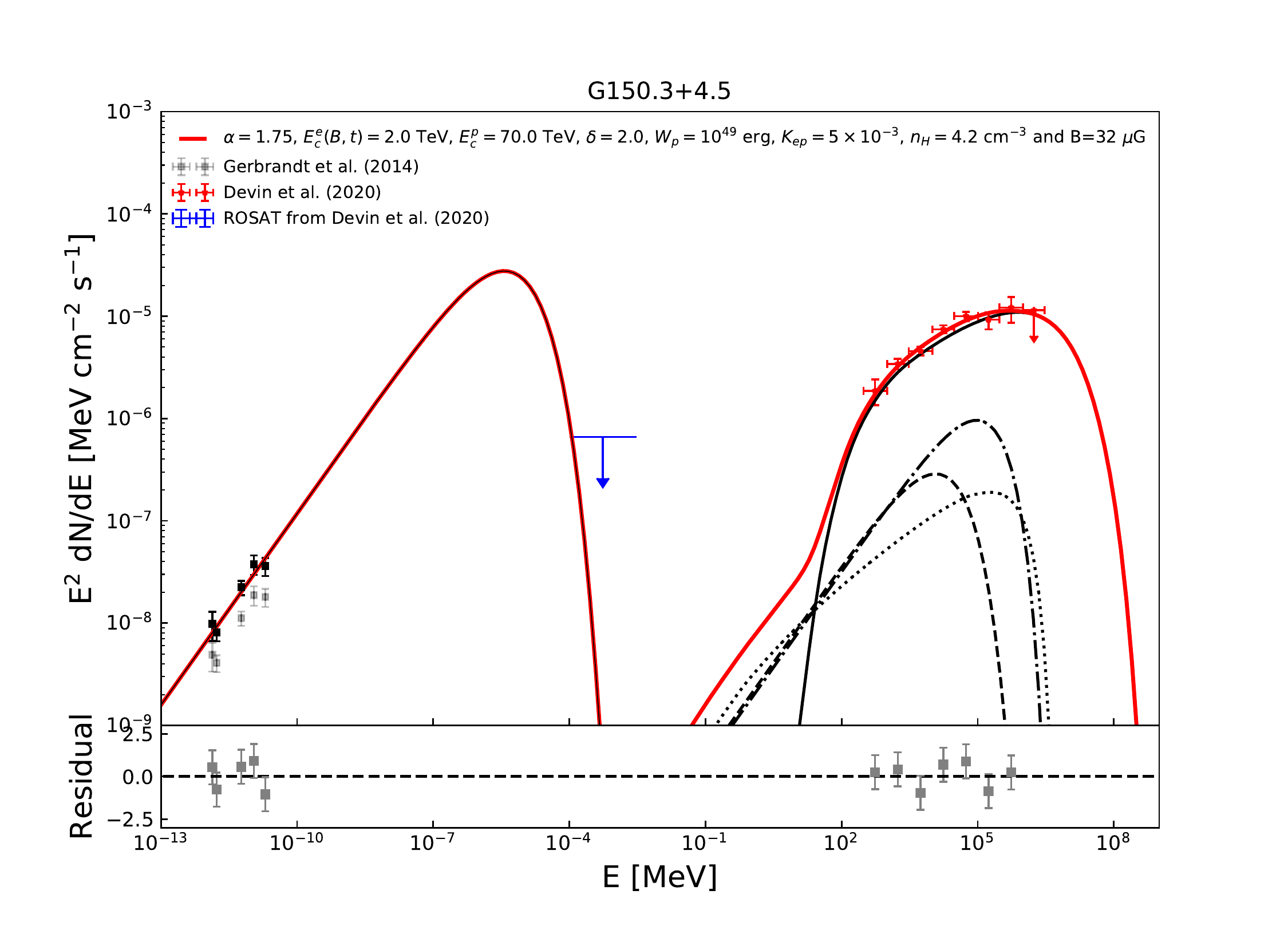}{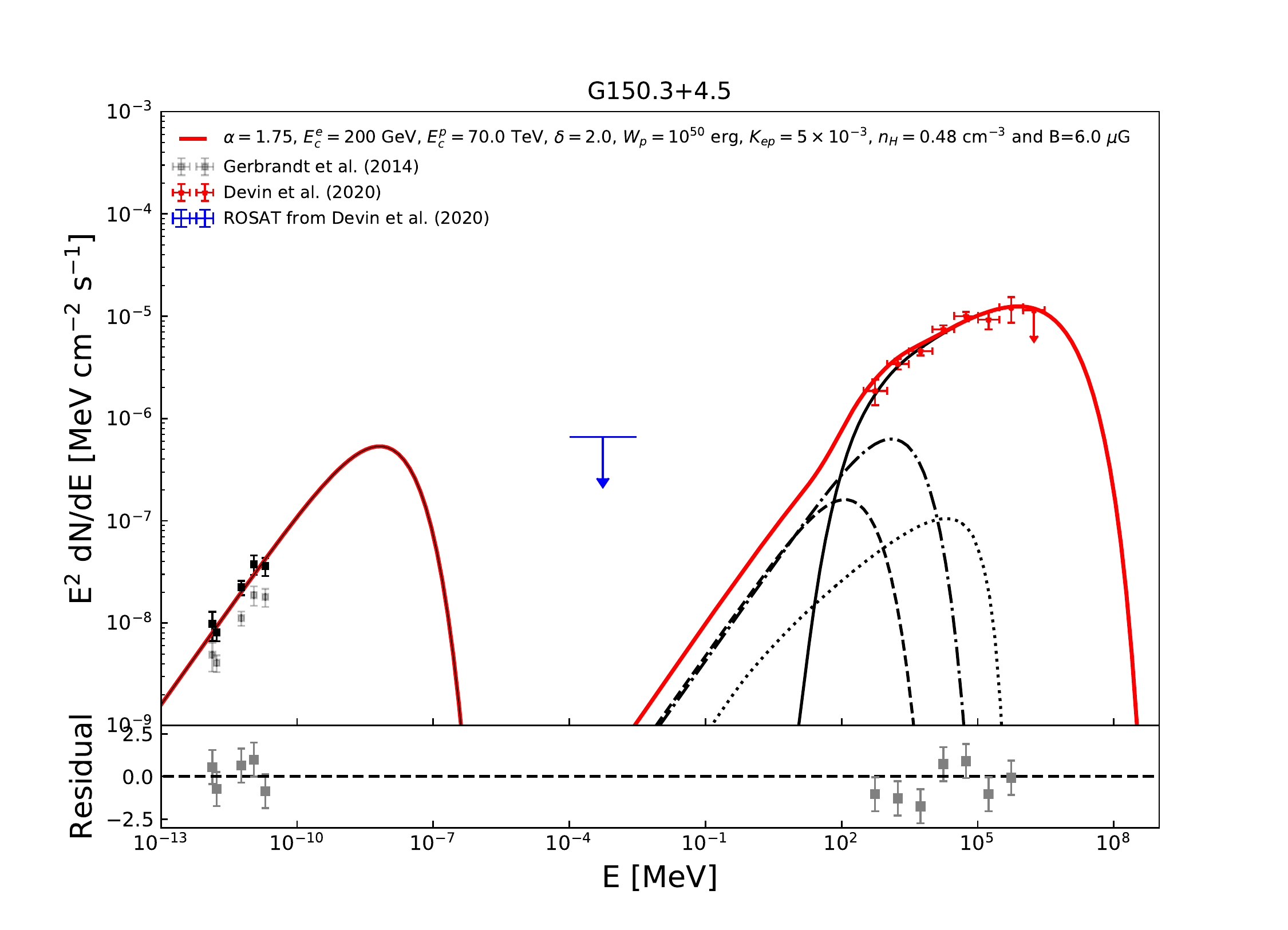}
\plottwo{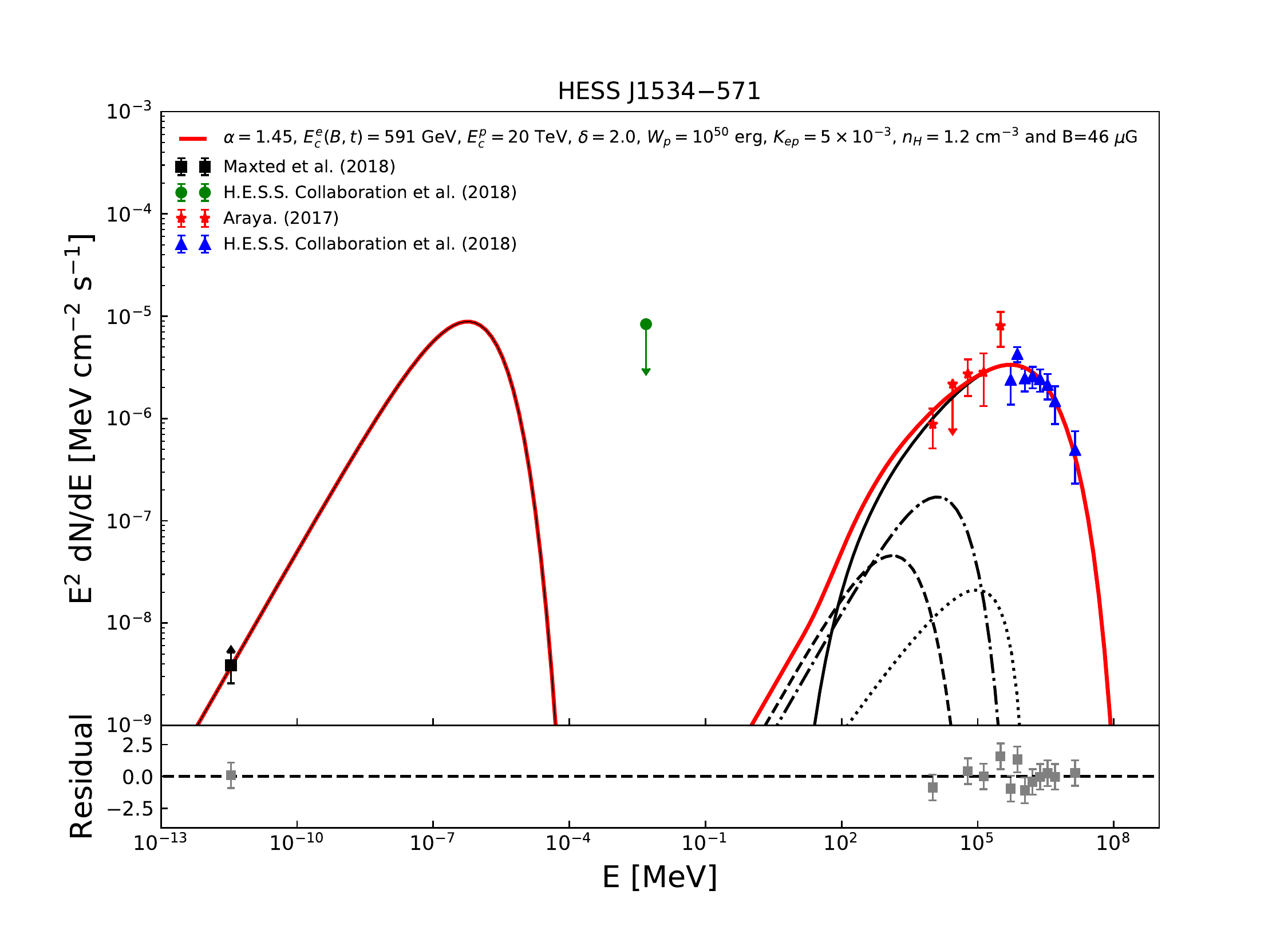}{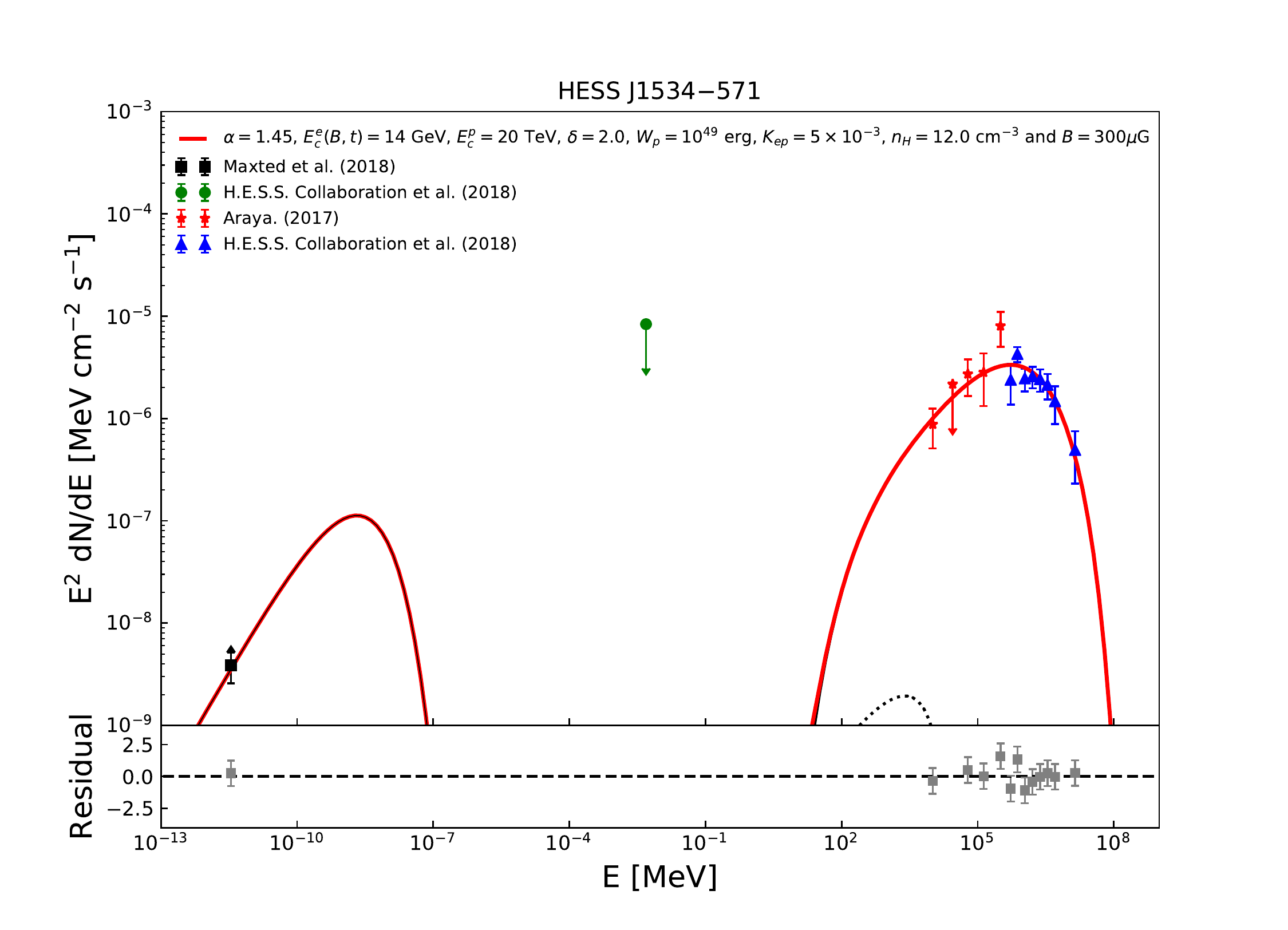}
\caption{Same as Figure \ref{fig:1912} but for G150.3$+$4.5 (upper) and HESS J1534$-$571 (lower) with very weak radio emission.  \label{fig:weak}}
\end{figure}

For the two SNRs with very weak radio emission, G150.3$+$4.5 is similar to G296.5+10.0 in the sense that there is no TeV data. We therefore set the high-energy cutoff of the proton distribution at 70 TeV. The spectral fits are shown in the upper panels of Figure \ref{fig:weak}. We favor the model with a low value of $10^{49}$ erg for the total proton energy $W_{\rm p}$ (left panel) with a magnetic field of $32\ \mu$G. An even lower value of $W_{\rm p}$ is disfavored since it implies an even stronger magnetic field and therefore an energy ratio of the magnetic field to the high-energy electrons greater than 3500. For higher values of the total proton energy, the electron cutoff energy needs to be lower than that determined by requiring the synchrotron energy loss time being equal to the age of the SNR. Otherwise the IC emission will dominate the $\gamma$-ray emission with a harder spectrum than the observed one. The right panel shows such a case with the electron cutoff energy of 200 GeV to suppress contributions to the $\gamma$-ray via the IC processes. Equating the synchrotron energy loss time to the SNR age will lead to a cutoff energy of electron $E^{\rm e}_{\rm cut}$ greater than 50 TeV for the relatively weaker magnetic field of $6\, \mu$G.

We notice that there are some degeneracies among $W_{\rm p}$, $n_{\rm H}$, $B$ and $K_{\rm ep}$. The product of $W_{\rm p}$ and $n_{\rm H}$ is determined by the $\gamma$-ray spectrum. The total proton energy discussed above should be re-scaled by the actual value of the mean number density of the background $n_{\rm H}$. We fix $K_{\rm ep}$ to $5\times 10^{-3}$ in the spectral fit above. For a given radio flux and $W_{\rm p}$, an increase of $K_{\rm ep}$ will lead to more energetic electrons and a weaker magnetic field.

The lower panels of Figure \ref{fig:weak} show the spectral fits for HESS J1534$-$571. The cutoff energy of the proton distribution for HESS J1534$-$571 is well constrained by TeV observations. However, Table \ref{tab:fitpatameters} shows that the cutoff energy of $20$ TeV is much lower than the value of 70 TeV for the three sources studied above. Similar to G296.5+10.0 and HESS J1912+101, we favor the model with a higher value of $W_{\rm p}$ (left panel) for the weaker magnetic field and higher synchrotron cutoff energy. The ratio of $W_{B}/W_{\rm e}=1600$ is also more reasonable. However, the magnetic field is not as well constrained as for G150.3$+$4.5. We also notice that with a spectral index of 1.45, HESS J1534$-$571 has a much harder energetic particle spectrum than other sources. Such a hard spectrum is needed to fit the $\gamma$-ray spectrum. More radio flux density measurements are needed to test this model.

\begin{figure*}
\plottwo{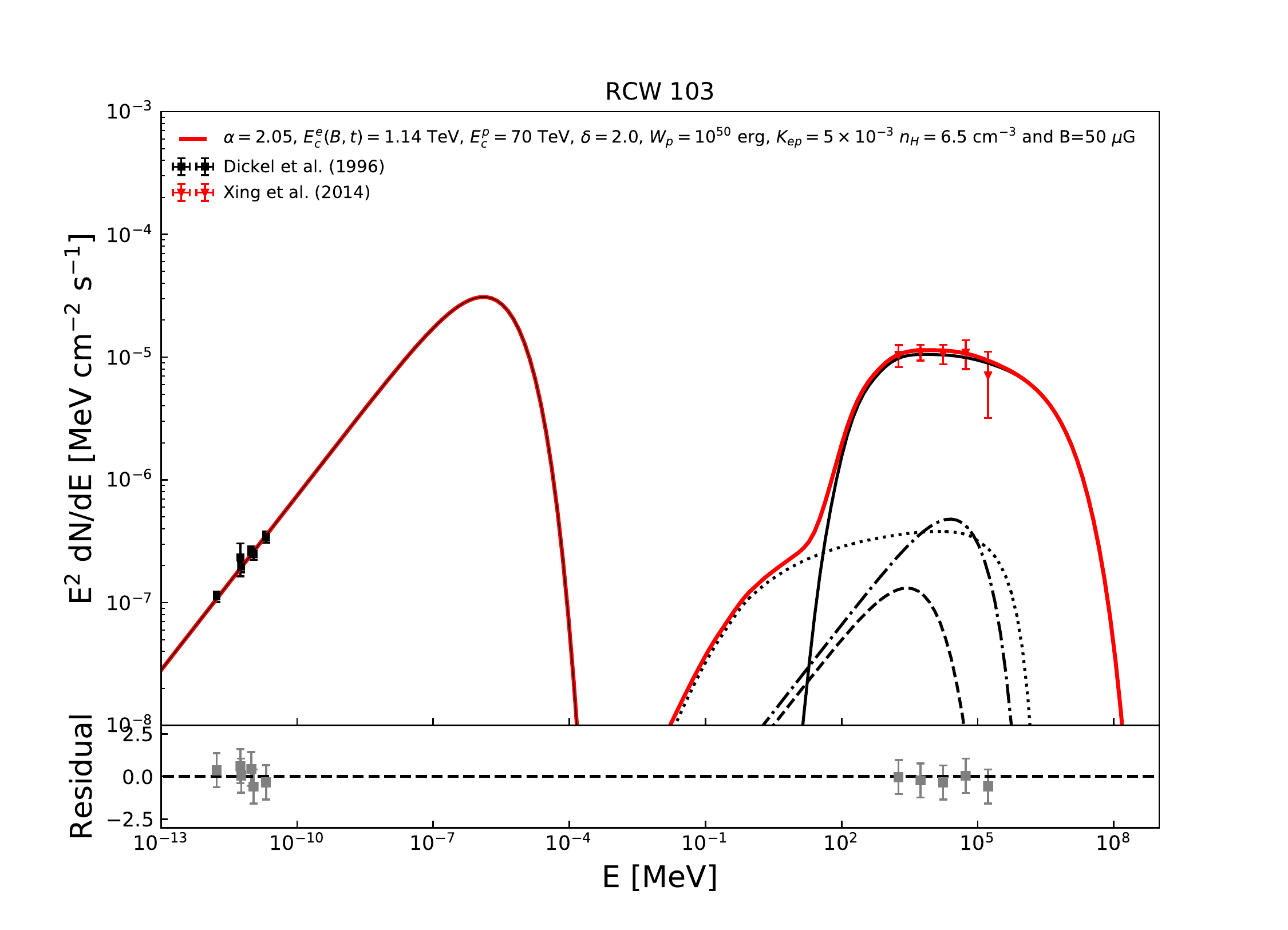}{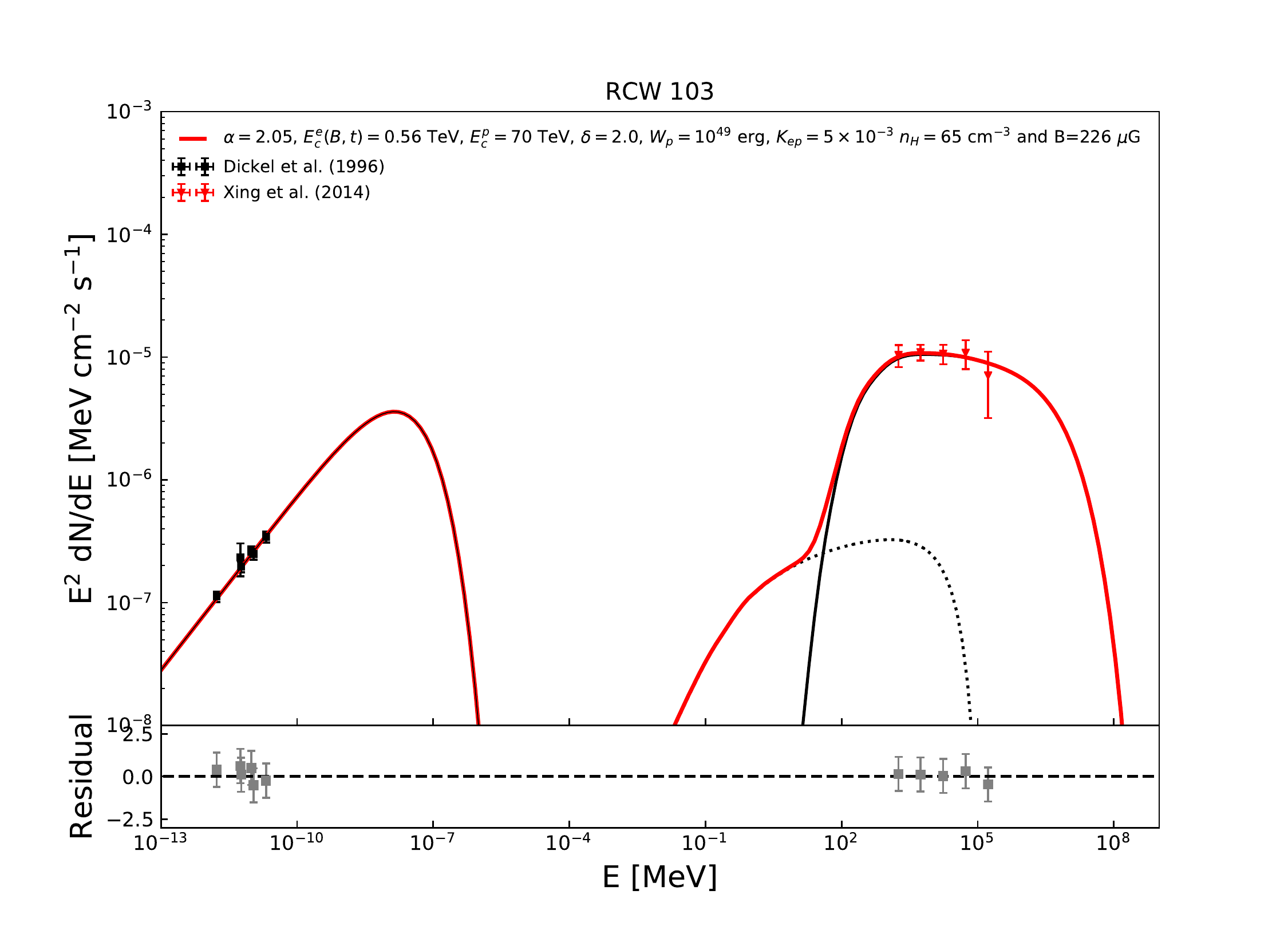}
\caption{Same as Figure \ref{fig:1912} but for RCW 103.
\label{fig:intermediate0}}
\end{figure*}

Among the 5 SNRs with intermediate radio emission, RCW 103 does not have non-thermal X-ray emission, similar to HESS J1912$+$101. There is also no TeV data, we then fix the proton cutoff energy at $70$ TeV. Although RCW 103 is relatively young with an age of 4.4 kyrs, radio and GeV observations lead to a high-energy particle spectrum slightly softer than that for HESS J1912$+$101.
The GeV flux of RCW 103 is actually about 3 times higher than that of HESS J1912$+$101. Since the distance to the two source is comparable, the $\gamma$-ray luminosity of RCW 103 of about 2 times higher. For the same proton energy $W_{\rm p}$ and comparable radio flux densities,  RCW 103 therefore has a higher background density $n_{\rm H}$ and stronger magnetic field. However, the radius of RCW 103 is more than 3 times smaller than HESS J1912$+$101, leading to a factor of 30 difference in the volume. We therefore have a much lower value of $W_{B}/W_{\rm e}$. Even the case with a $W_{\rm p}$ of $10^{50}$ erg (left panel of Figure \ref{fig:intermediate0}) has a very low value of 0.034 for $W_{B}/W_{\rm e}$, we still favor this model for its relatively higher cutoff energy of the synchrotron spectrum.
To increase the value of $W_{B}/W_{\rm e}$, one needs to consider lower values for $W_{\rm p}$ as discussed above (See right panel of Figure \ref{fig:intermediate0}).

\begin{figure*}
\centering
\plottwo{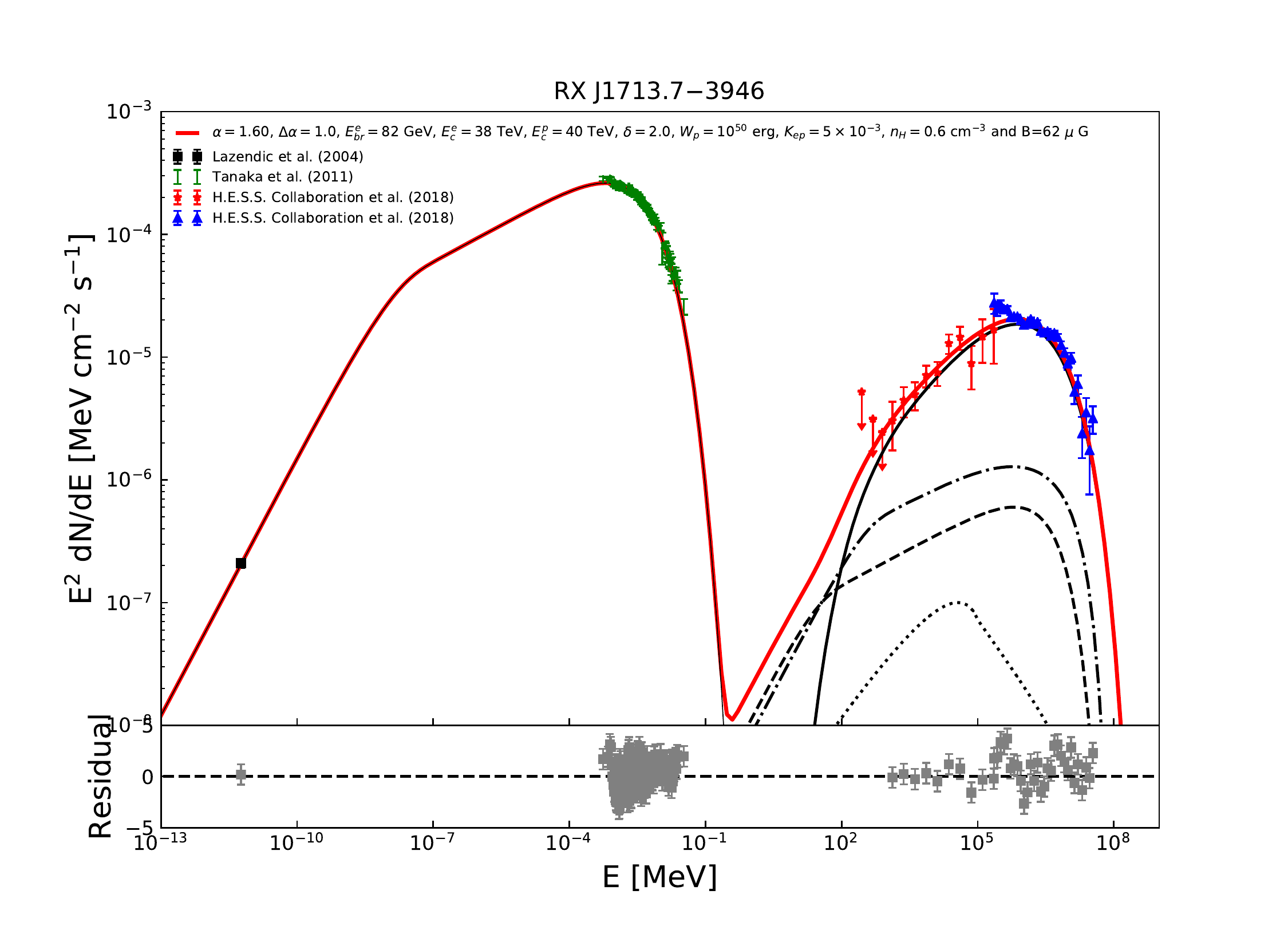}{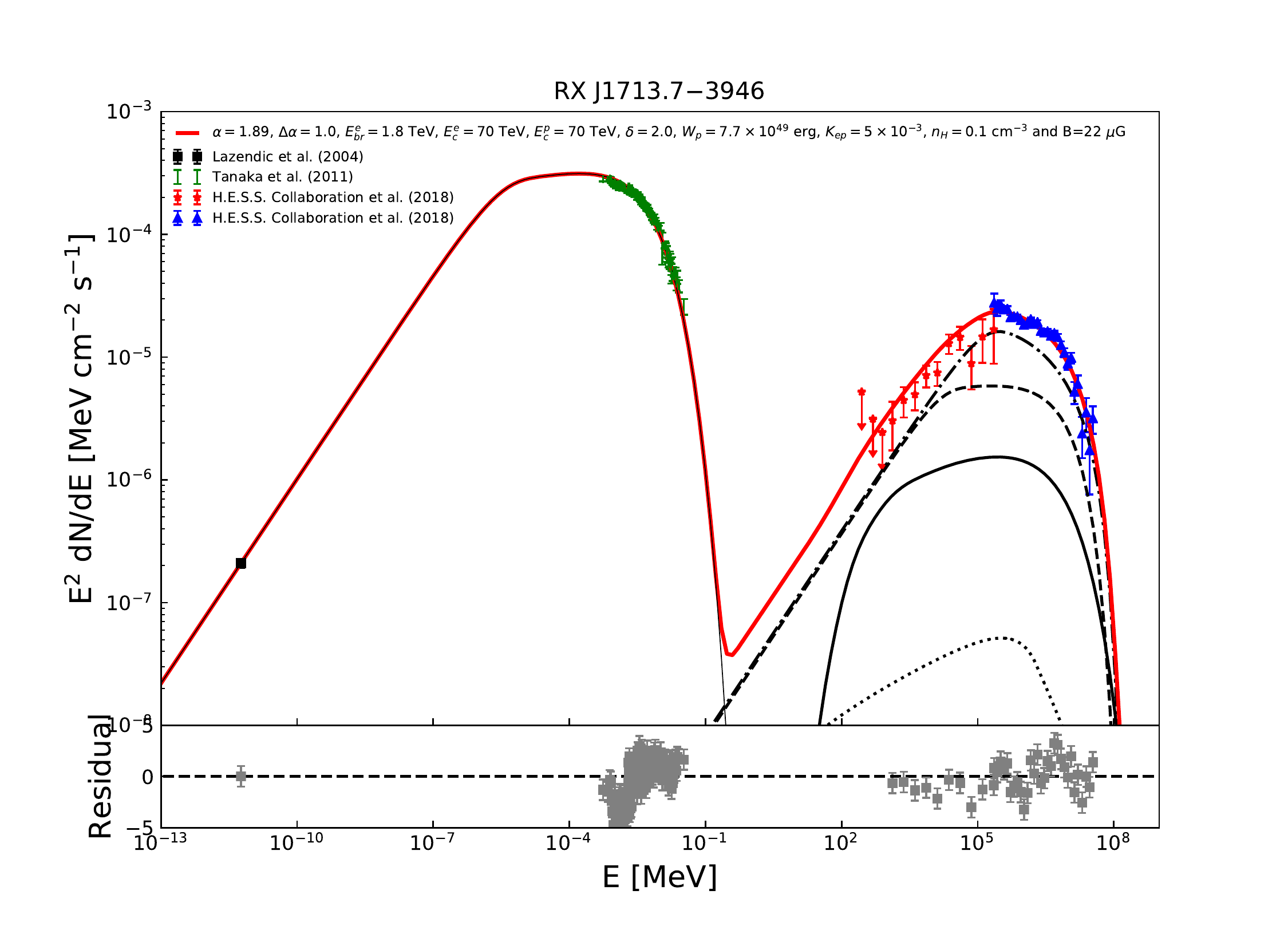}
\plottwo{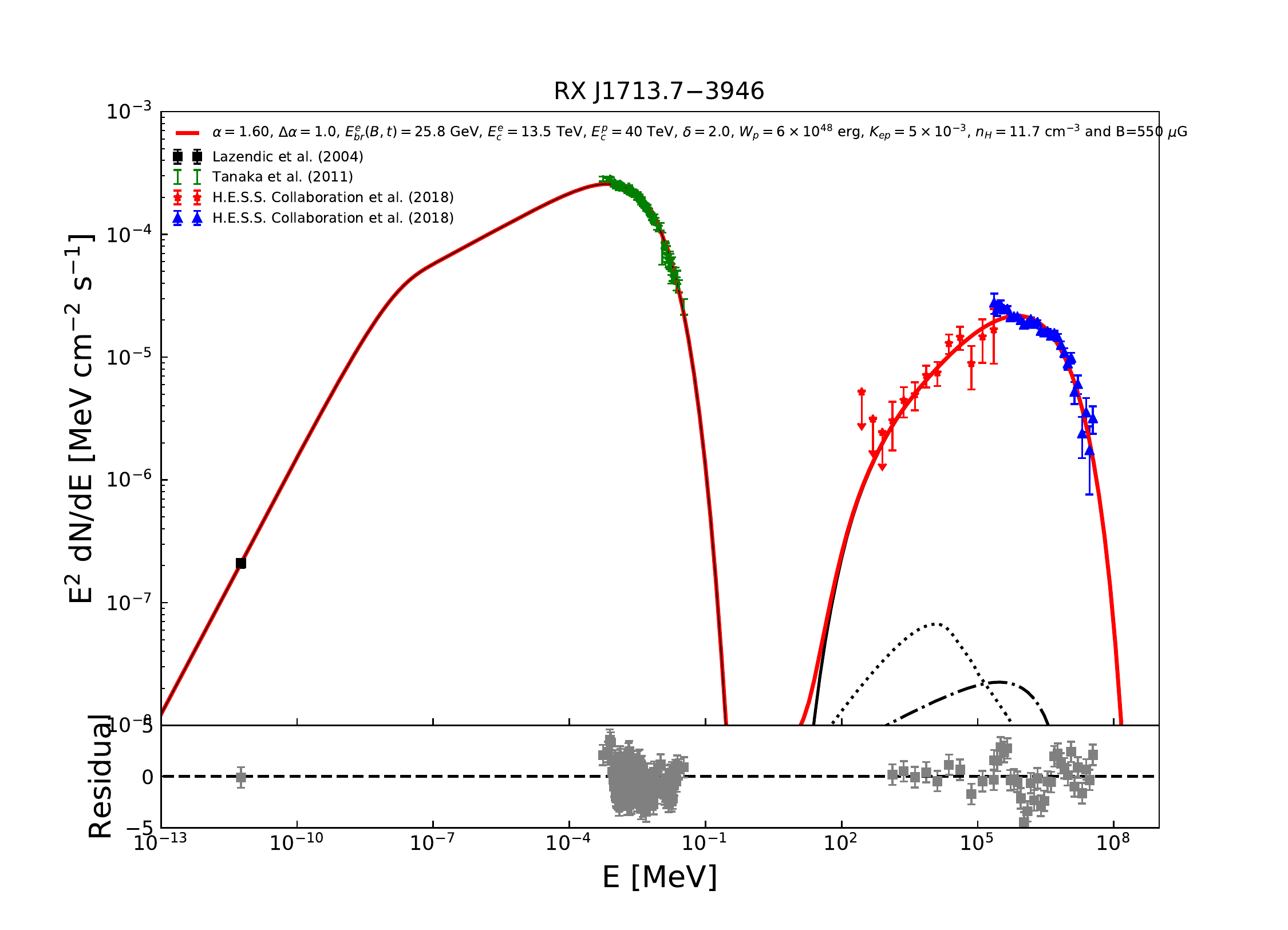}{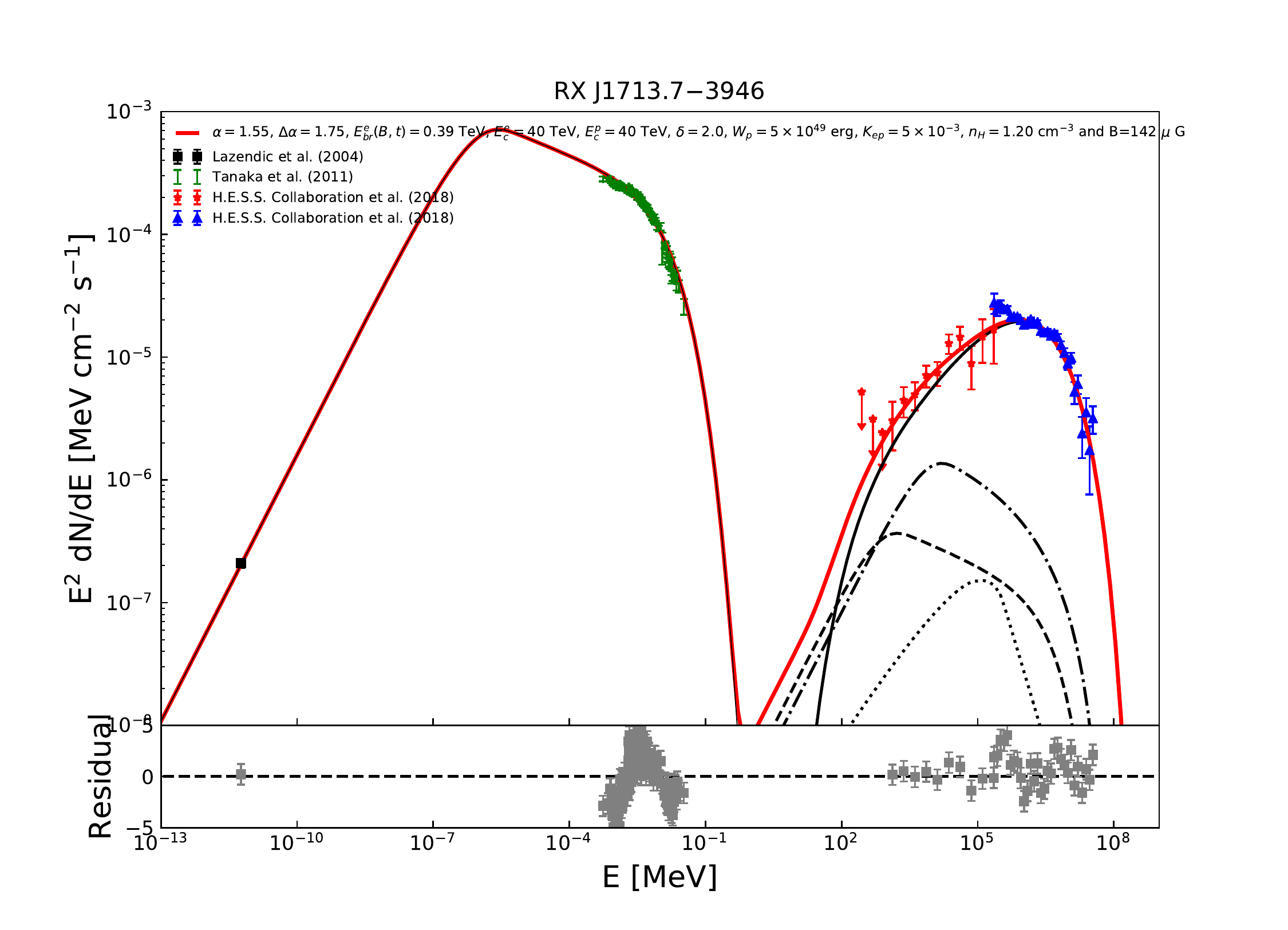}
\caption{Several spectral fits to the SED of RX J17137.7$-$3946. The upper left is our favored model. The upper right corresponds to the leptonic scenario for the $\gamma$-ray emission. Although the model shown in the lower left has fewer parameters, the magnetic field appears to be too strong. The model in the lower right panel has a relatively weaker magnetic field but gives a poorer fit to the SED.
\label{fig:intermediate1}}
\end{figure*}

The other 3 SNRs with intermediate radio emission RX J1713.7$-$3946, RX J0852$-$4622, and HESS J1731$-$347 have been studied extensively for their prominent non-thermal X-rays and TeV emission. There are still debates on the nature of the $\gamma$-ray emission. Detailed TeV observations of RX J17137.7$-$3946 have shown that a broken power-law distribution is needed to fit the $\gamma$-ray spectrum in both the leptonic and hadronic scenarios for the $\gamma$-ray emission \citep{2018A&A...612A...6H}. In general, for sources with strong non-thermal X-ray emission, a single power law particle distribution will lead to a poor fit to the multi-wavelength SED. But for the sake of simplicity and considering the overall quality of $\gamma$-ray data, we will still adopt a single power-law distribution with an exponential high-energy cutoff for the rigidity of ions. The electron distribution, however, can be a broken power law with an exponential ($\delta =1$) or super exponential ($\delta=2.0$) cutoff. Since the acceleration of low energy particles are not affected by radiative energy loss, we assume that the electrons and ions have the same spectral index at low energies.

The upper-right panel of Figure \ref{fig:intermediate1} shows a leptonic scenario for the $\gamma$-ray emission. The corresponding model parameters are given in the fourth row for RX J1713.7$-$3946 in Table \ref{tab:fitpatameters}. Since we adopt a much higher energy density of 1 eV cm$^{-3}$ for the infrared background photons and ion processes also have significantly contribution to the $\gamma$-ray emission (solid line), a high value of 22 $\mu$G is inferred for the magnetic field. However, the total energy of the magnetic field is on the order of $10^{48}$ ergs, which is still much lower than that for ions. Since the ion spectral cutoff is not well-constrained by the data, we set it at $70$ TeV as we did above for other sources. An increase of the magnetic field will lead to a shift to the dominance of $\gamma$-ray fluxes by the hadronic processes. The upper left panel of Figure \ref{fig:intermediate1} shows our favored hadronic model for the $\gamma$-ray emission. The corresponding model parameters are given in the third row for RX J1713.7$-$3946 in Table \ref{tab:fitpatameters}.
The cutoff energy of protons is 40 TeV, which is lower than that inferred from $\gamma$-ray observations \citep{2018A&A...612A...6H} for the adoption of a single power-law ion distribution here. The spectral index of 1.6 is comparable to the low energy spectral index inferred from $\gamma$-ray observations and the product of $W_{\rm p}$ and $n_{\rm H}$ is also compatible with these observations. For such a hard spectrum, a broken power law electron distribution is needed to fit the radio to X-ray spectrum via the synchrotron process. For the magnetic field of 62 $\mu$G, we find a break energy of $82$ GeV and a cutoff energy of $38$ TeV. Although the electron cutoff energy is very close to the proton cutoff energy, their distributions are quite different at high energies for the differences in the spectral index and shape of the cutoff $\delta$. The total energy of the magnetic field is comparable to that of ions and is more than 2 orders of magnitude higher than that of electrons. Note that although $K_{\rm ep}=0.005$, the total energy of protons is more than 3 orders of magnitude higher than that of electron for the hard spectrum and much lower break energy of the electron distribution than the cutoff energy of protons \citep{2008MNRAS.386L..20B}. Further increase of the magnetic field will lead to a higher value of $W_B/W_{\rm e}$ and a slight decrease in $E_{\rm br}^{\rm e}$, giving rise to a slightly higher ratio of $W_{\rm p}/W_{\rm e}$.

\begin{figure*}[htb]
\plottwo{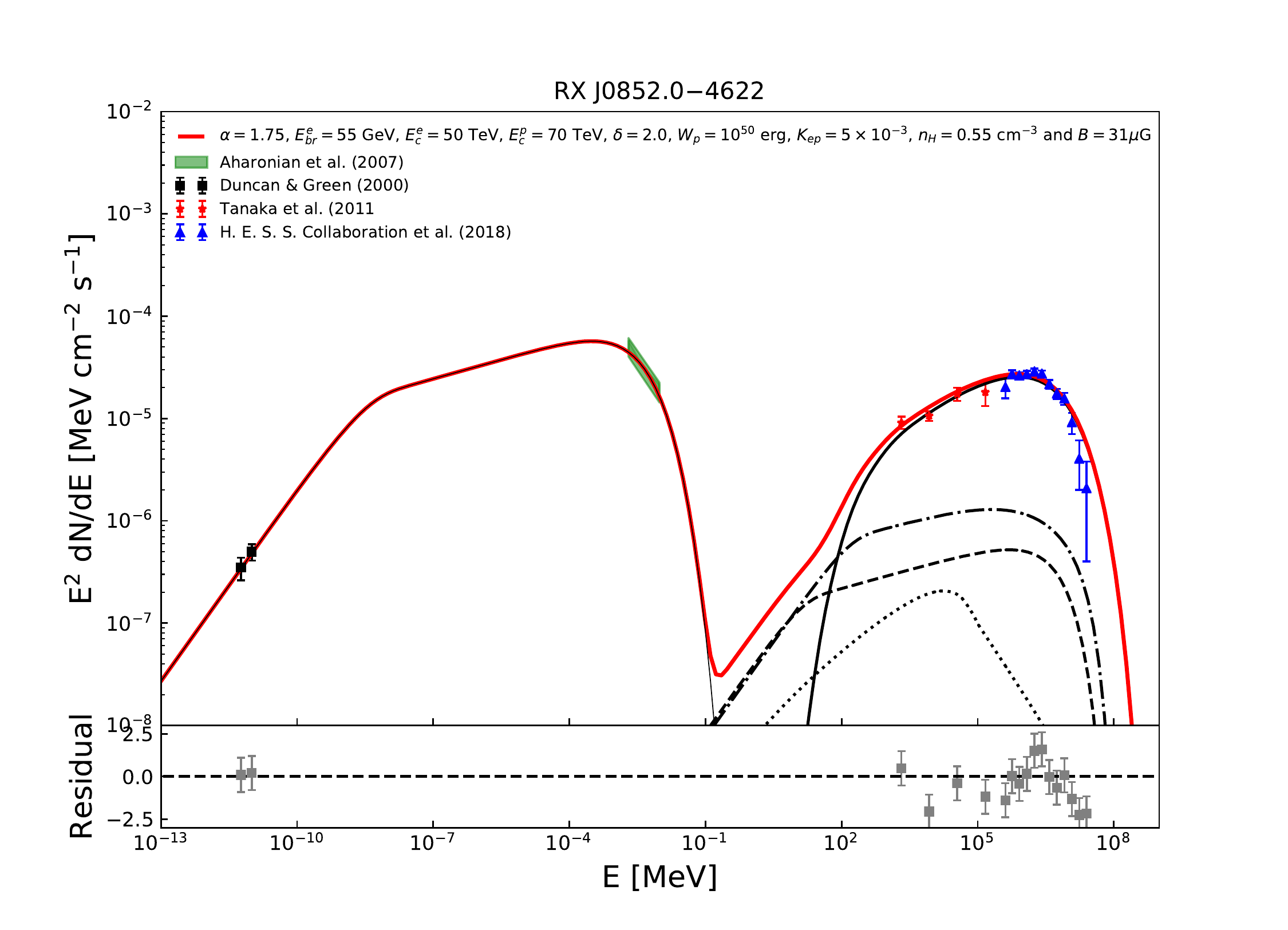}{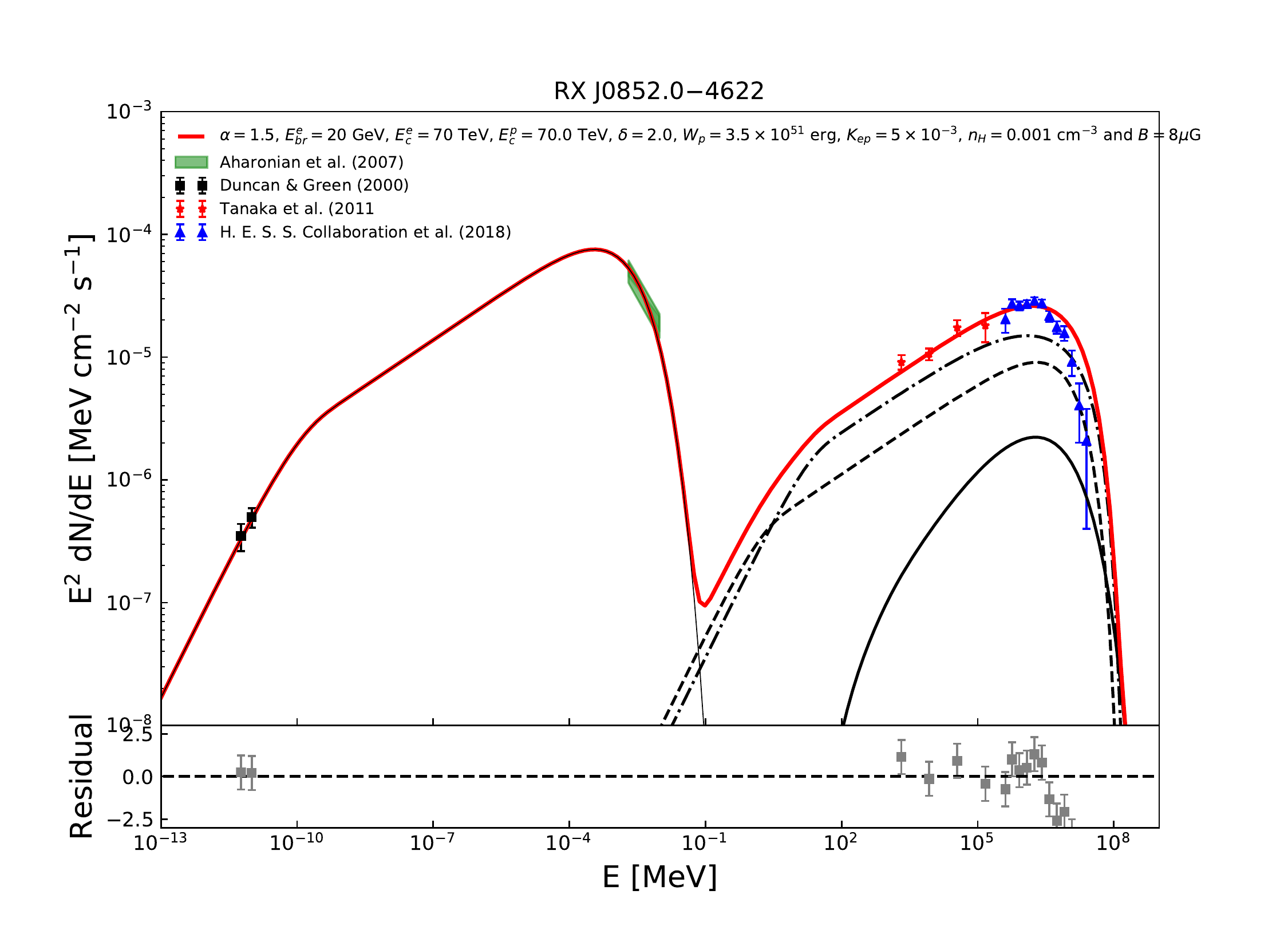}
\plottwo{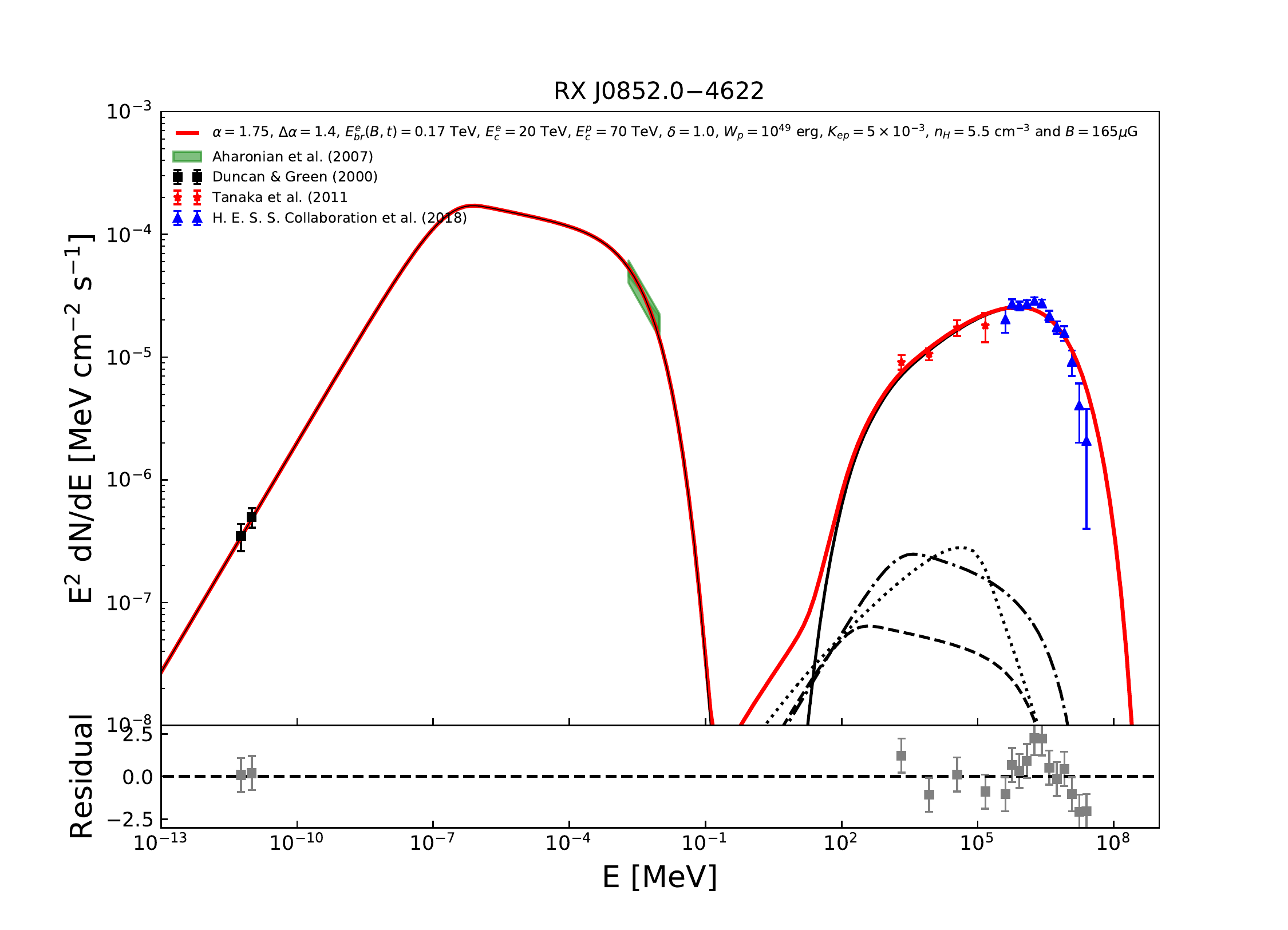}{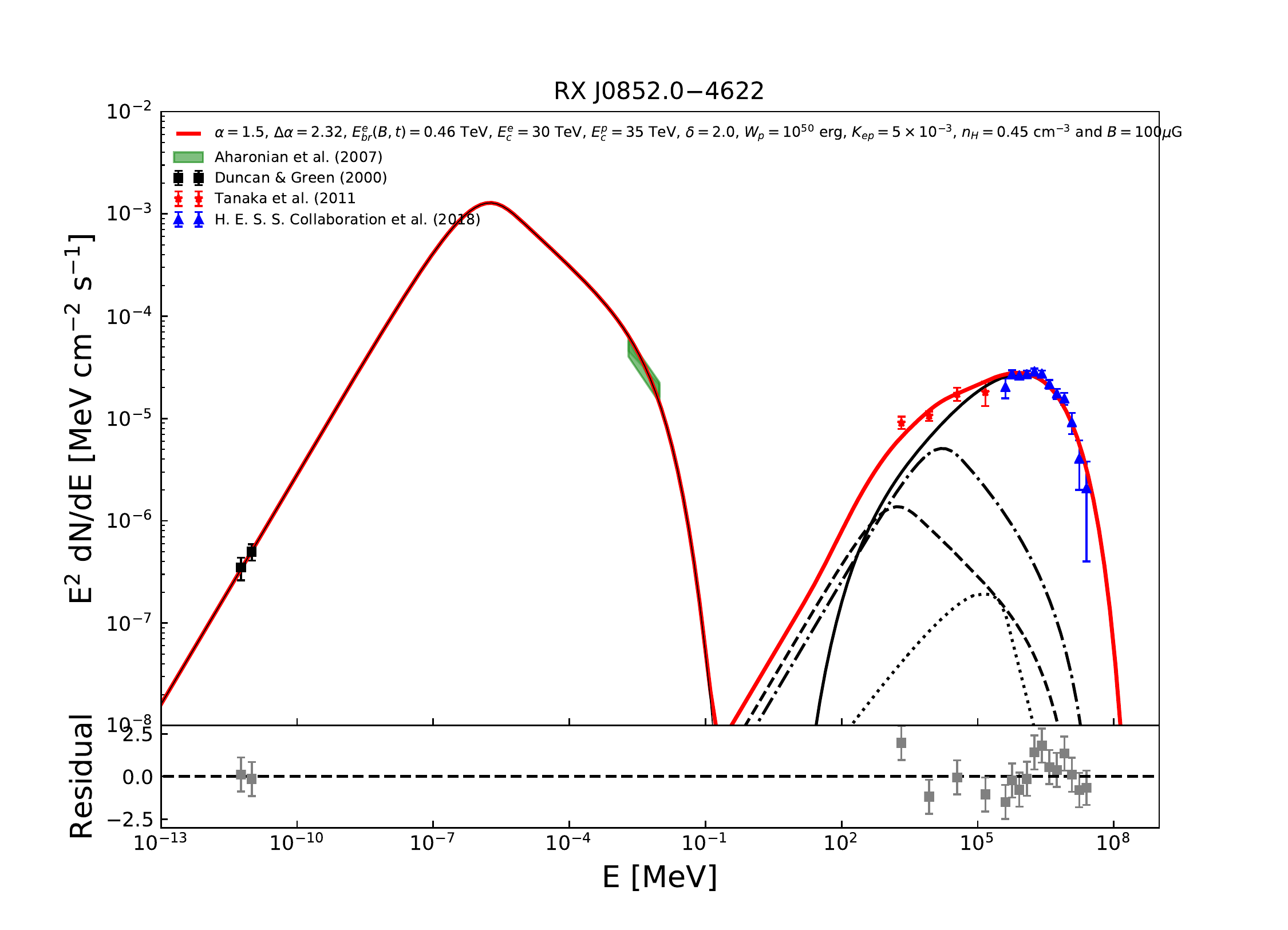}
\plottwo{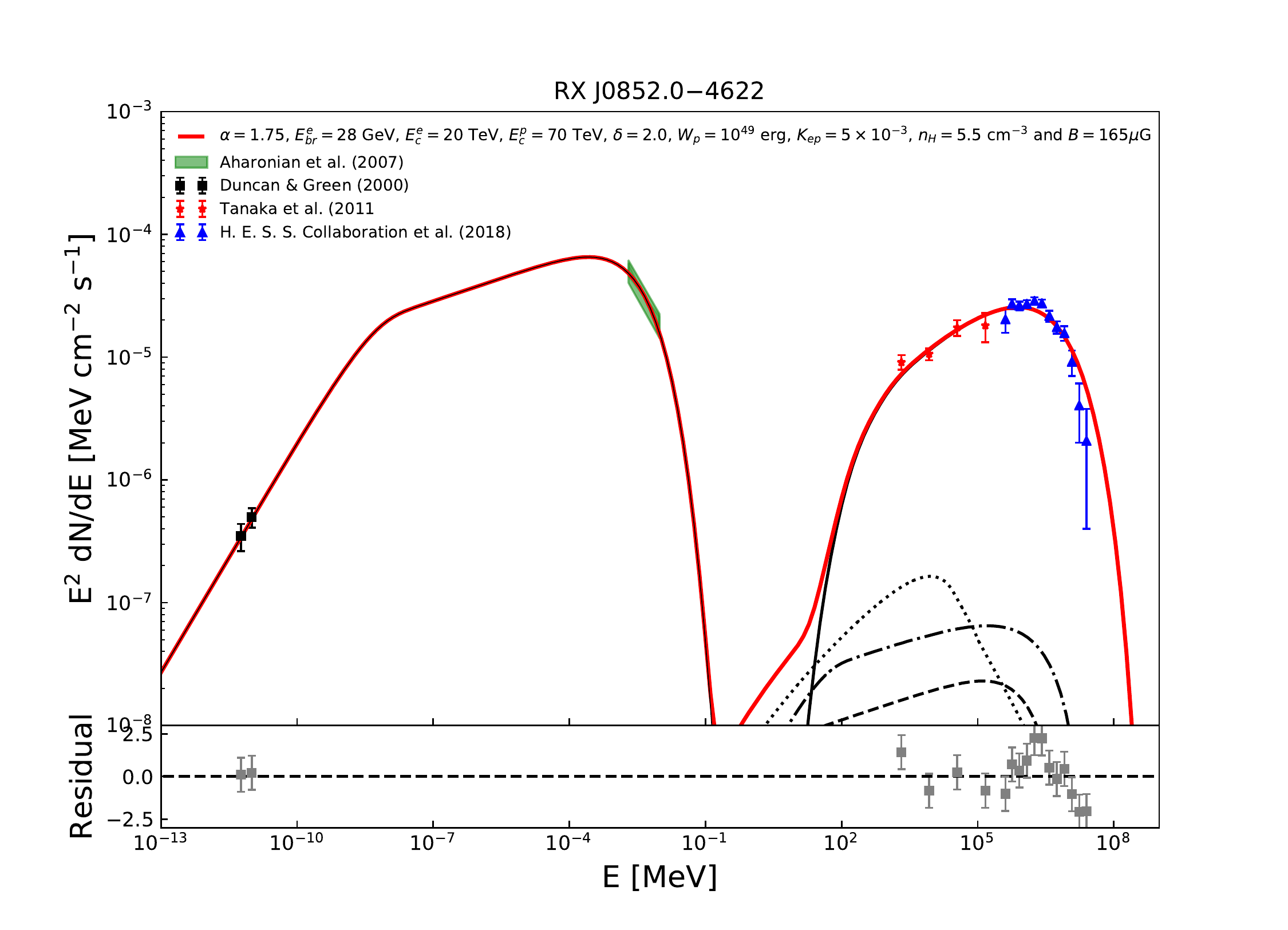}{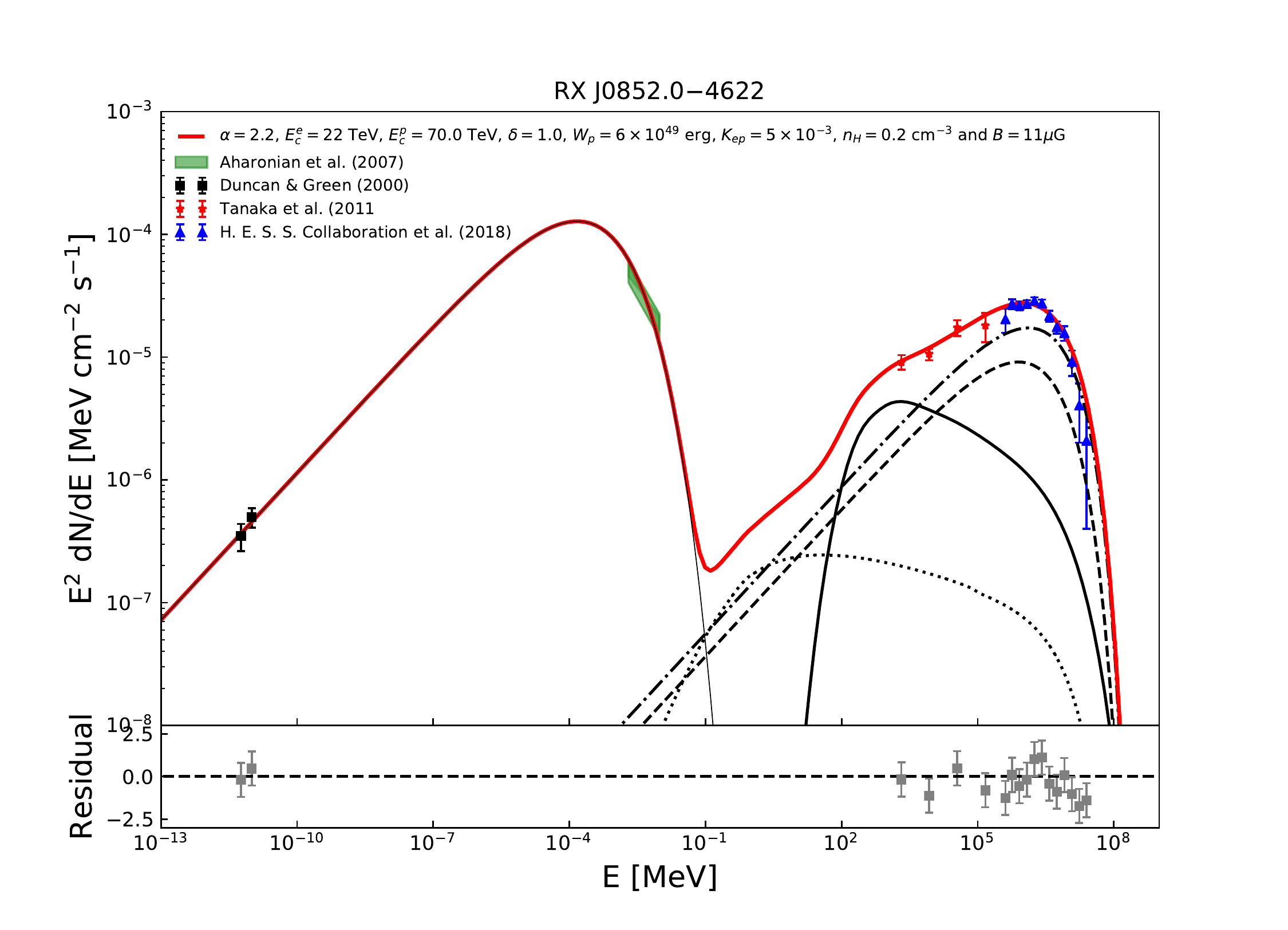}          
\caption{The upper two rows are the same as Figure \ref{fig:intermediate1} but for RX J0852$-$4622. The bottom left panel is similar to the top left panel except for a lower $W_{\rm p}$ and a stronger magnetic field. The bottom right panel is for a relatively simple single power-law leptonic model. However, the cutoff of the electron distribution is exponential in this case.
\label{fig:intermediate2}}
\end{figure*}

To reduce the number of model parameters, one may increase the magnetic field and set the electron radiative energy loss timescale at the break energy to be equal to the age of the SNR. This leads to the fit in the lower left panel of Figure \ref{fig:intermediate1}. The corresponding model parameters are given in the first row for RX J1713.7$-$3946 in Table \ref{tab:fitpatameters}. The $\gamma$-ray emission is completely dominated by the hadronic processes. However, the total energy of the magnetic field $W_B$ is more than 5 orders of magnitude higher than that of electrons and more than 2 orders of magnitude higher than the energy of ions $W_{\rm p}$. One may adjust the change of the spectral index $\Delta \alpha$ from low to high energies to reduce the magnetic field. This leads to the fit in the lower right panel of Figure \ref{fig:intermediate1}, which is not as good as the others, especially in the X-ray band. The corresponding model parameters are given in the second row for RX J1713.7$-$3946 in Table \ref{tab:fitpatameters}. The slight decrease in $\alpha$ is due to contribution to GeV $\gamma$-ray via the leptonic processes.

\begin{figure*}
\plottwo{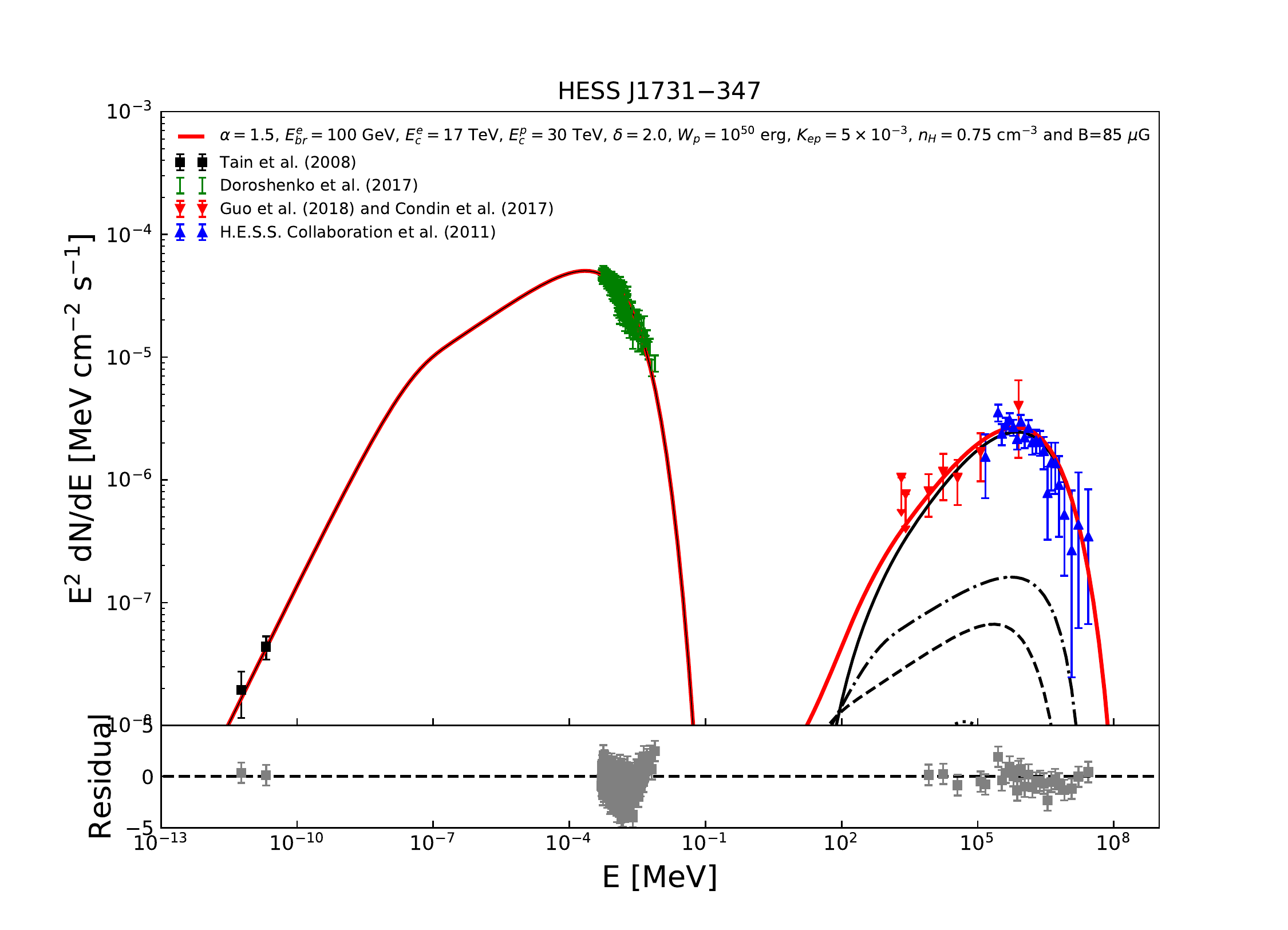}{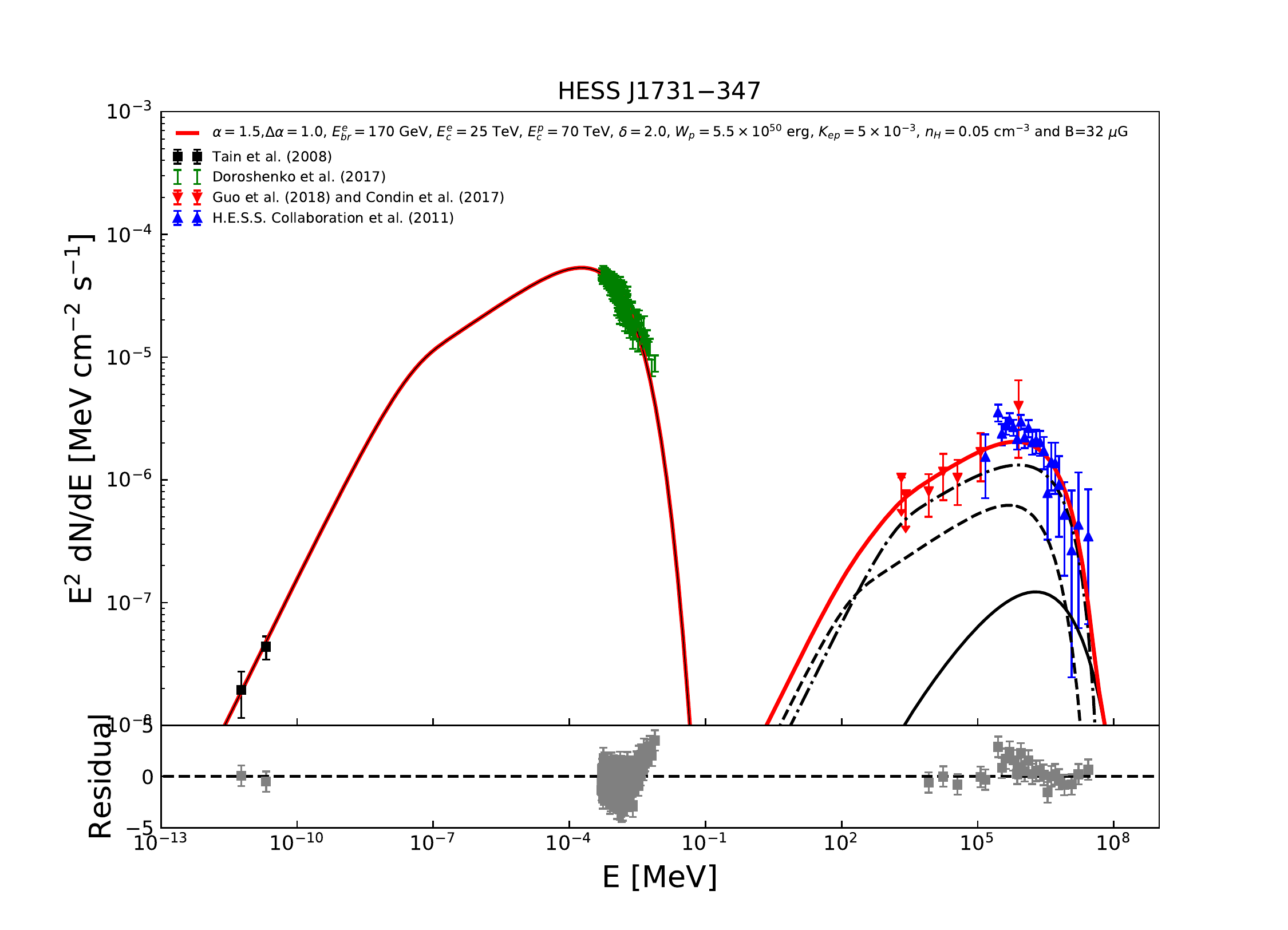}
\plottwo{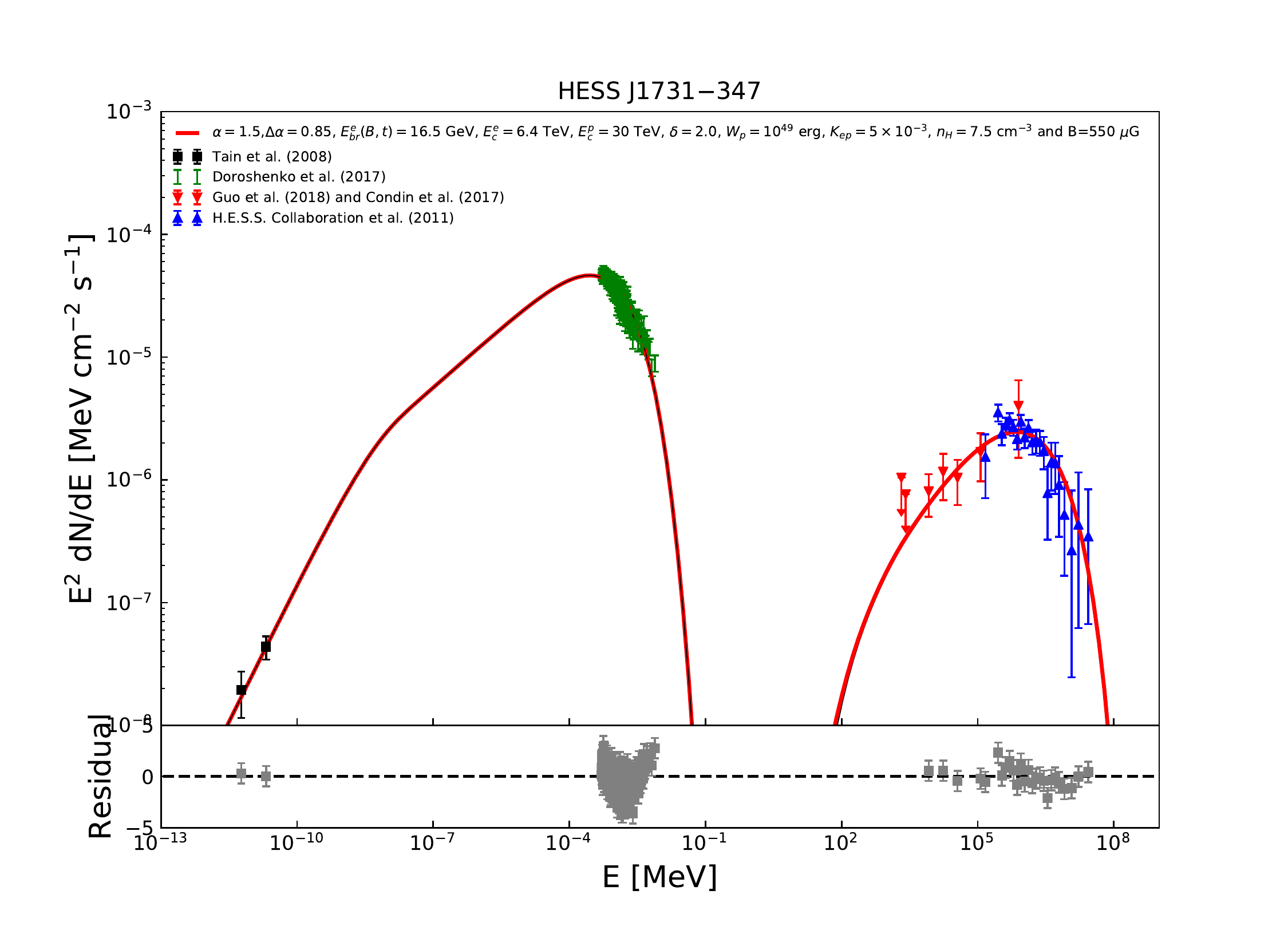}{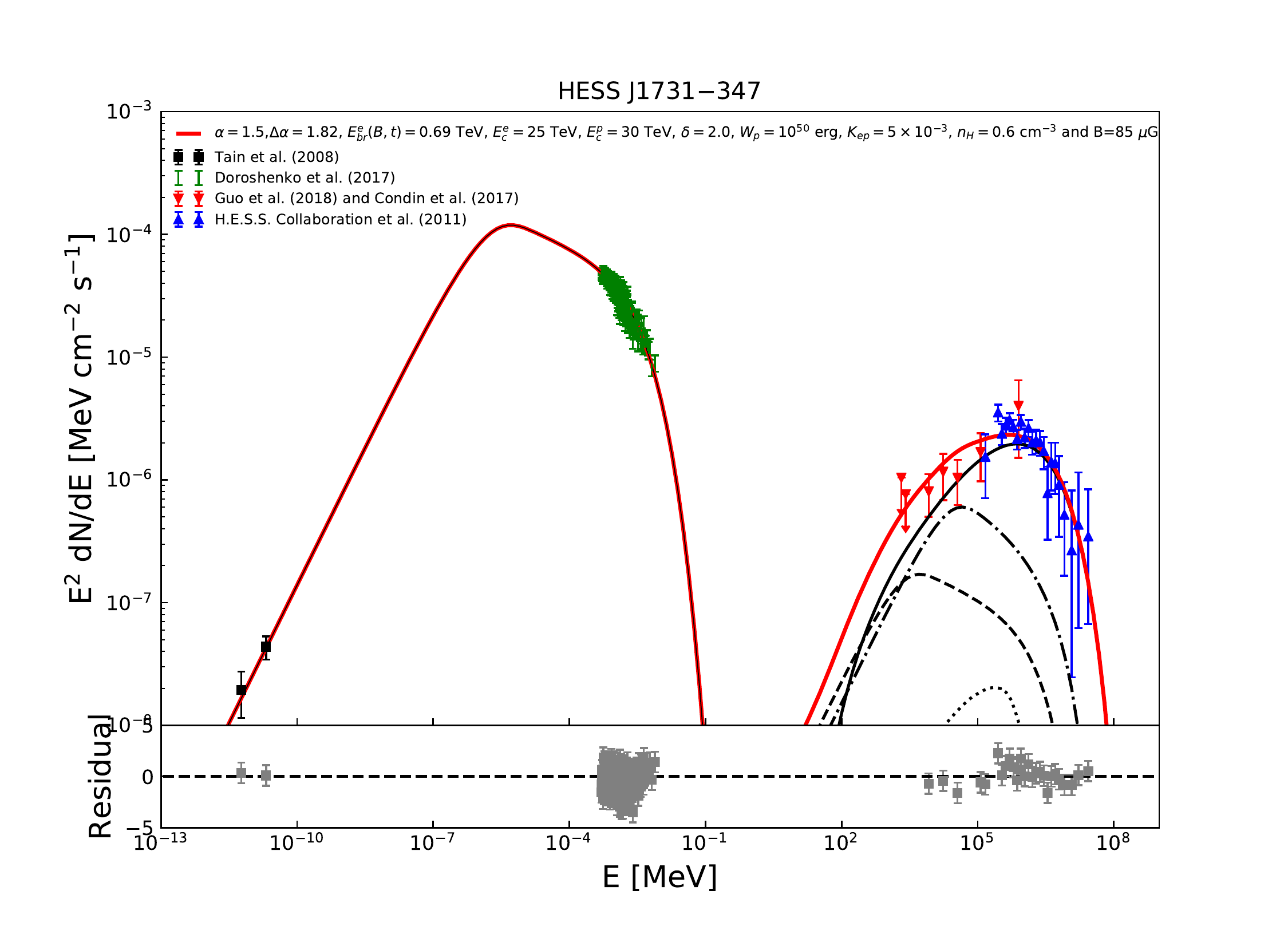} 
\plottwo{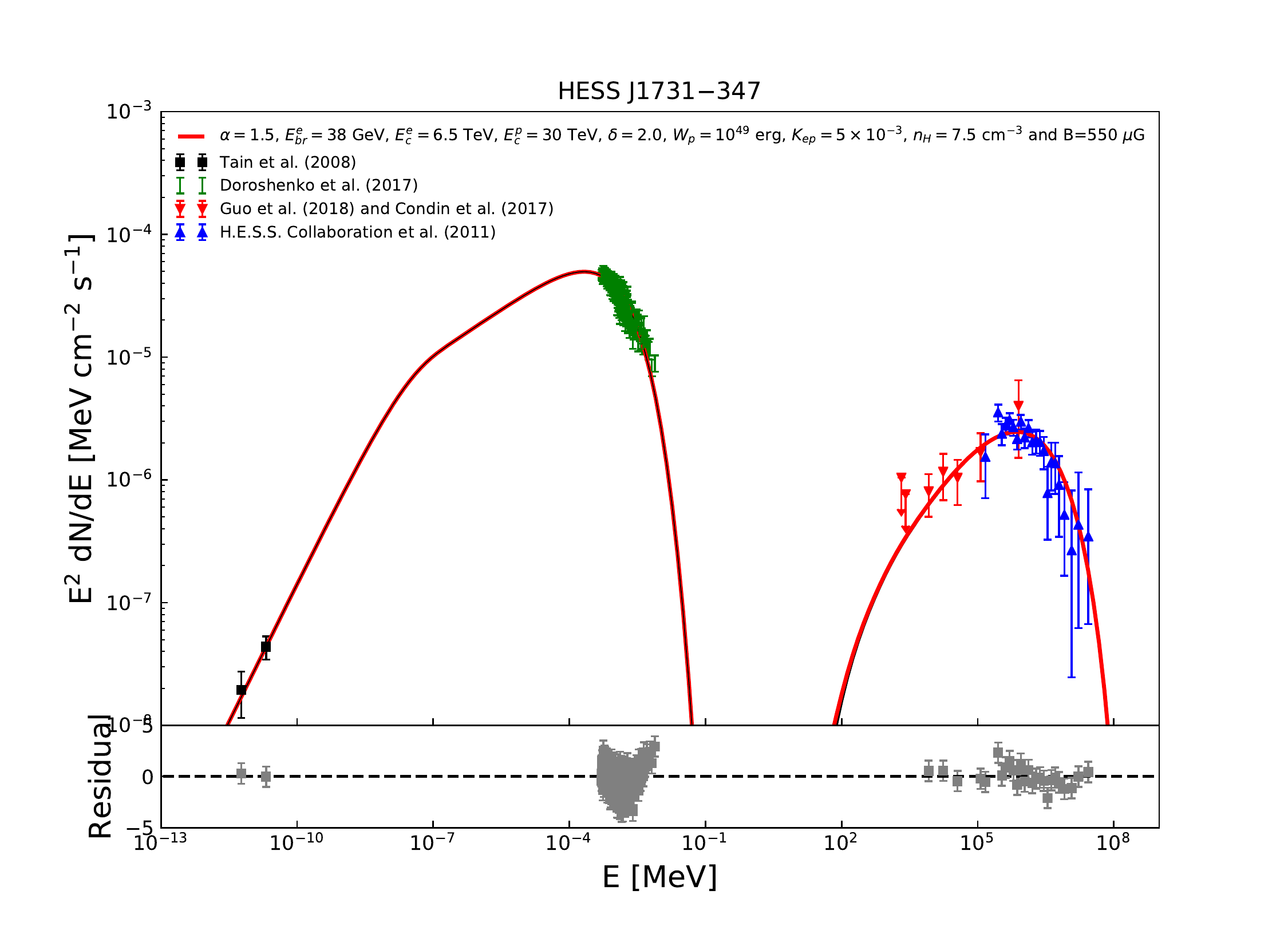}{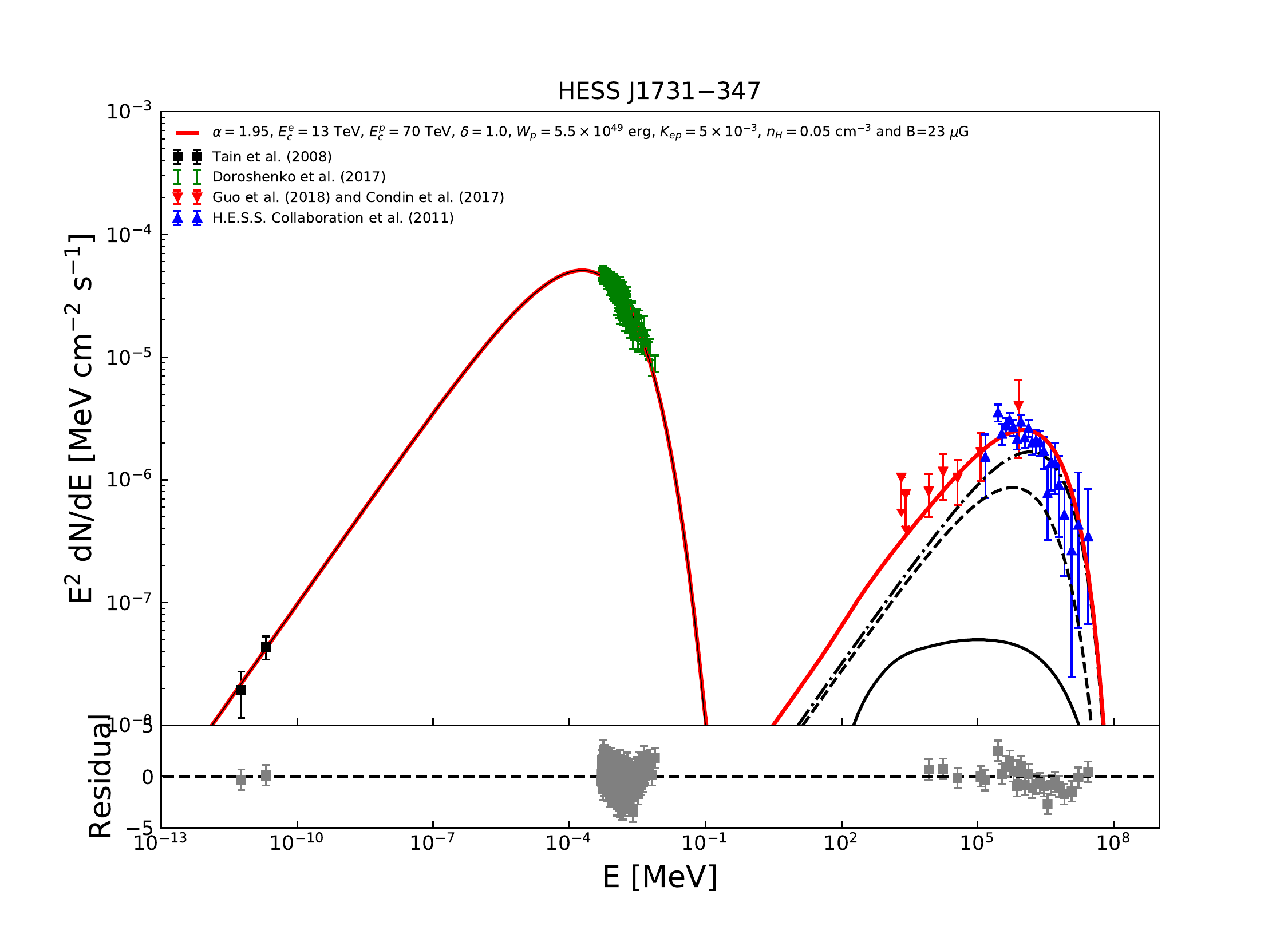}
\caption{Same as Figure \ref{fig:intermediate2} but for HESS J1731$-$347.
\label{fig:intermediate3}}
\end{figure*}

The spectral fits to the SED of RX J0852$-$4622 shown in Figure \ref{fig:intermediate2} are very similar to those for RX J1713.7$-$3946. 
The upper left panel shows our favored model with the model parameters given in the fourth row for RX J0852$-$4622 in Table \ref{tab:fitpatameters}. The spectral index of 1.75 for the ions is slightly larger than 1.6 for the favored model of RX J1713.7$-$3946, and the cutoff energy of the proton distribution is also slightly higher. With a magnetic field of $31\ \mu$G, the magnetic field energy is about one tenth of the proton energy and about 100 times higher than the energy of electrons. The upper right panel shows a leptonic model with the model parameters given in the fifth row for RX J0852$-$4622 in Table \ref{tab:fitpatameters}. The model slight overproduces $\gamma$-rays at tens of TeV. Compared with the leptonic model for RX J1713.7$-$3946, its break energy is about 100 times smaller, leading to a 20 times higher total energy of electrons. The total energy of protons is about $3.5\times 10^{51}$ erg for the very hard spectrum and very high cutoff energy, which is a bit too high to be reasonable. The total energy of the magnetic field however is about $10^{49}$ erg and comparable to that of the electrons. 

The model of the middle right panel of Figure \ref{fig:intermediate2} is similar to that of the lower right panel of Figure \ref{fig:intermediate1} for RX J1713.7$-$3946. The model parameters are given in the third row for RX J0852$-$4622 in Table \ref{tab:fitpatameters}. The model of the middle left panel of Figure \ref{fig:intermediate2} is similar to that of the lower right panel of Figure \ref{fig:intermediate1} for RX J1713.7$-$3946 with the model parameters given in the second row for RX J0852$-$4622 in Table \ref{tab:fitpatameters}. To reduce the ratio of $W_B/W_{\rm e}$, $\Delta \alpha = 1.4$ instead of 1 is adopted for RX J1713.7$-$3946. 

The model of the lower left panel of Figure \ref{fig:intermediate2} is similar to our favored model in the upper left with the model parameter given in the first row for RX J0852$-$4622 in Table \ref{tab:fitpatameters}. Here the total energy of proton $W_{\rm p}$ is 10 times smaller, leading to a very strong magnetic field of 165 $\mu$G and a very high value of $3.9\times 10^4$ for $W_B/W_{\rm e}$. The model of the lower right panel of Figure \ref{fig:intermediate2} has a single power-law electron distribution with the model parameter given in the sixth row for RX J0852$-$4622 in Table \ref{tab:fitpatameters}. However to fit the SED, the shape of the cutoff needs to be exponential instead of super exponential for other models. The energy of the magnetic field is comparable to that of electrons, reminiscence of the leptonic model with a broken power law electron distribution (upper right panel of Figure \ref{fig:intermediate2}).

\begin{figure}[ht!]
\plottwo{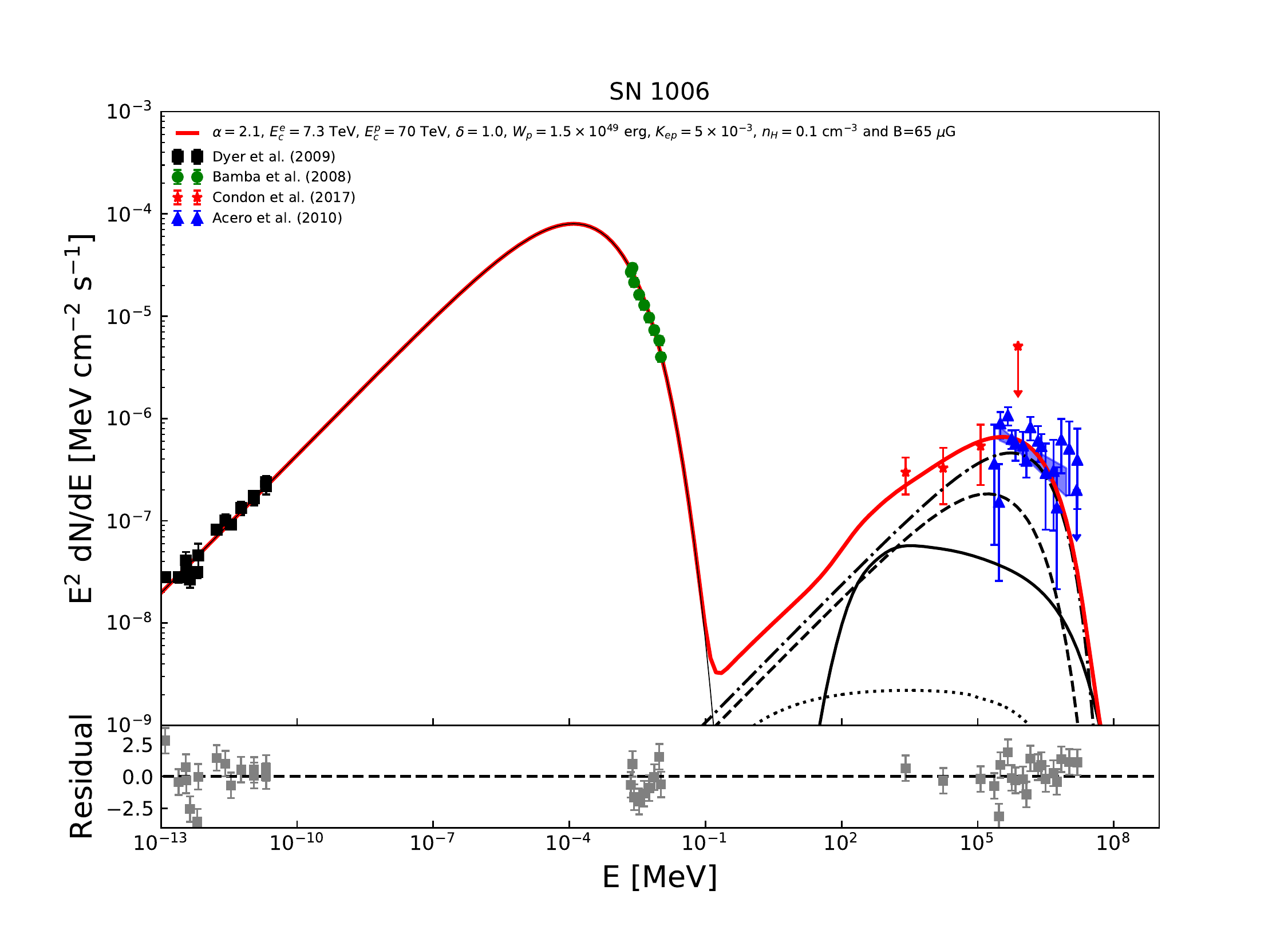}{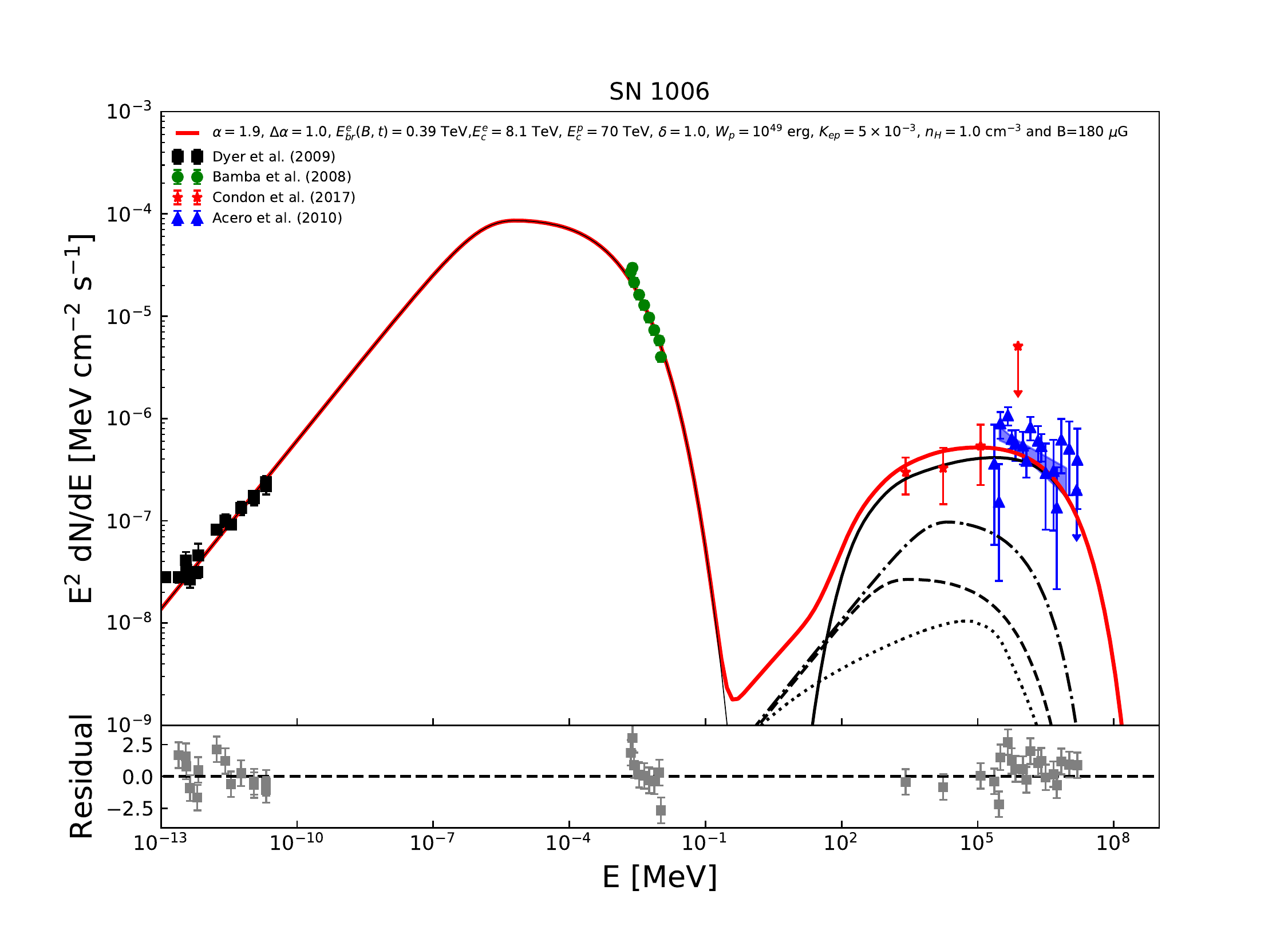}
\plottwo{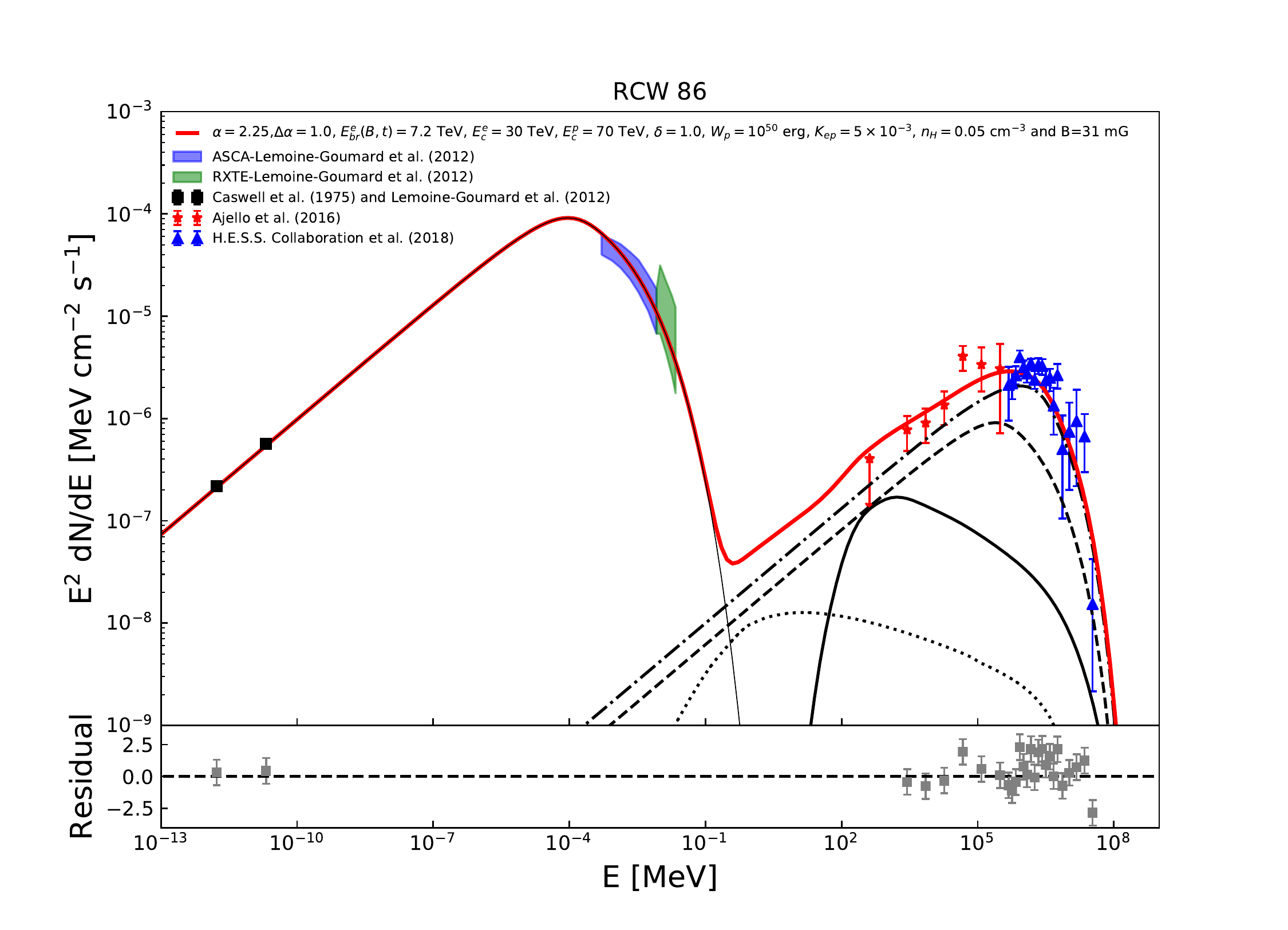}{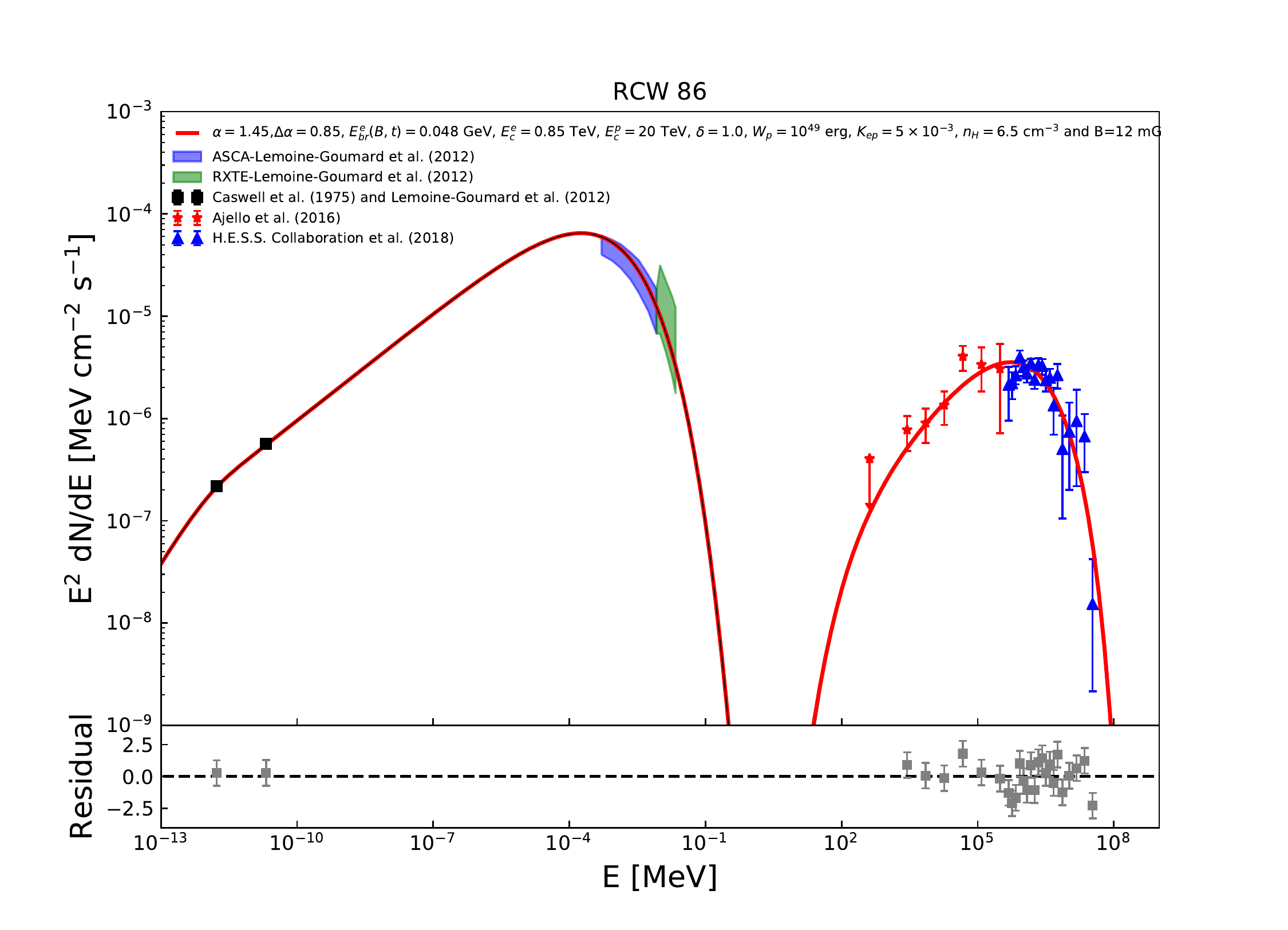}
\caption{The spectral fits for SN 1006 and RCW 86. The left panels correspond to the favored leptonic scenarios for the $\gamma$-ray emission. \label{fig:strong}}
\end{figure}

The six spectral fits for HESS J1731$-$347 shown in Figure \ref{fig:intermediate3} are very similar to those in Figure \ref{fig:intermediate2}. The model parameters are given in Table \ref{tab:fitpatameters}. Due to its relatively higher X-ray to $\gamma$-ray flux ratio, the magnetic fields in the leptonic scenarios are higher than those for RX J0852$-$4622. For the leptonic models with a broken power-law electron distribution, the break energy for HESS J1731$-$347 is between those for the other two sources, so is the total energy of protons. For the hadronic models, we first notice that the cutoff energy of protons is the lowest among these three sources and its particle distributions are also harder than the other two sources. The $\gamma$-ray luminosity of these three sources are comparable as can be seen from the product of $W_{\rm p}$ and $n_{\rm H}$.

Compared with G296.5$+$10.0, SN 1006 and RCW 86 have prominent non-thermal X-ray emission and are relatively younger. They all have relatively strong radio emission. Figure \ref{fig:strong} shows the spectral fits to these two SNRs with the model parameters given in Table \ref{tab:fitpatameters}.
Although both the leptonic (left panels) and hadronic (right panels) models give reasonable fits to the SEDs, the hadronic models are disfavored for their relatively stronger magnetic fields, especially for RCW 86, whose hadronic model requires a magnetic field of more than 10 mG. This is because the hard GeV spectrum is not compatible to the soft radio spectrum if we assume electrons and ions have the same spectral index. To fit the spectrum, the synchrotron spectrum needs to have a spectrum break below the radio band, implying very strong magnetic field if this break is associated with radiative energy loss processes.
However, we notice that multi-wavelength images of RCW 86 reveal complicated structure \citep{2016ApJ...819...98A}. Multi-zone hadronic models may still work. More detailed studies are warranted.
For SN 1006,  $W_{p}$ is on the order of $10^{49}$ erg. A much higher value can be ruled out without reducing $K_{\rm ep}$ since electrons already have significant contributions to the $\gamma$-rays via the IC emission process in both scenarios. 
We also notice that both leptonic models for these two SNRs require an exponential cutoff instead of the super exponential one and SNR 1006 has a single power law electron distribution while RCW 86 has a broken one. The leptonic models for RWC 86 explored by \citet{2016ApJ...819...98A} only consider the CMB for the IC process and have a single power law electron distribution, which is different from our favored model.

\begin{figure*}
\plottwo{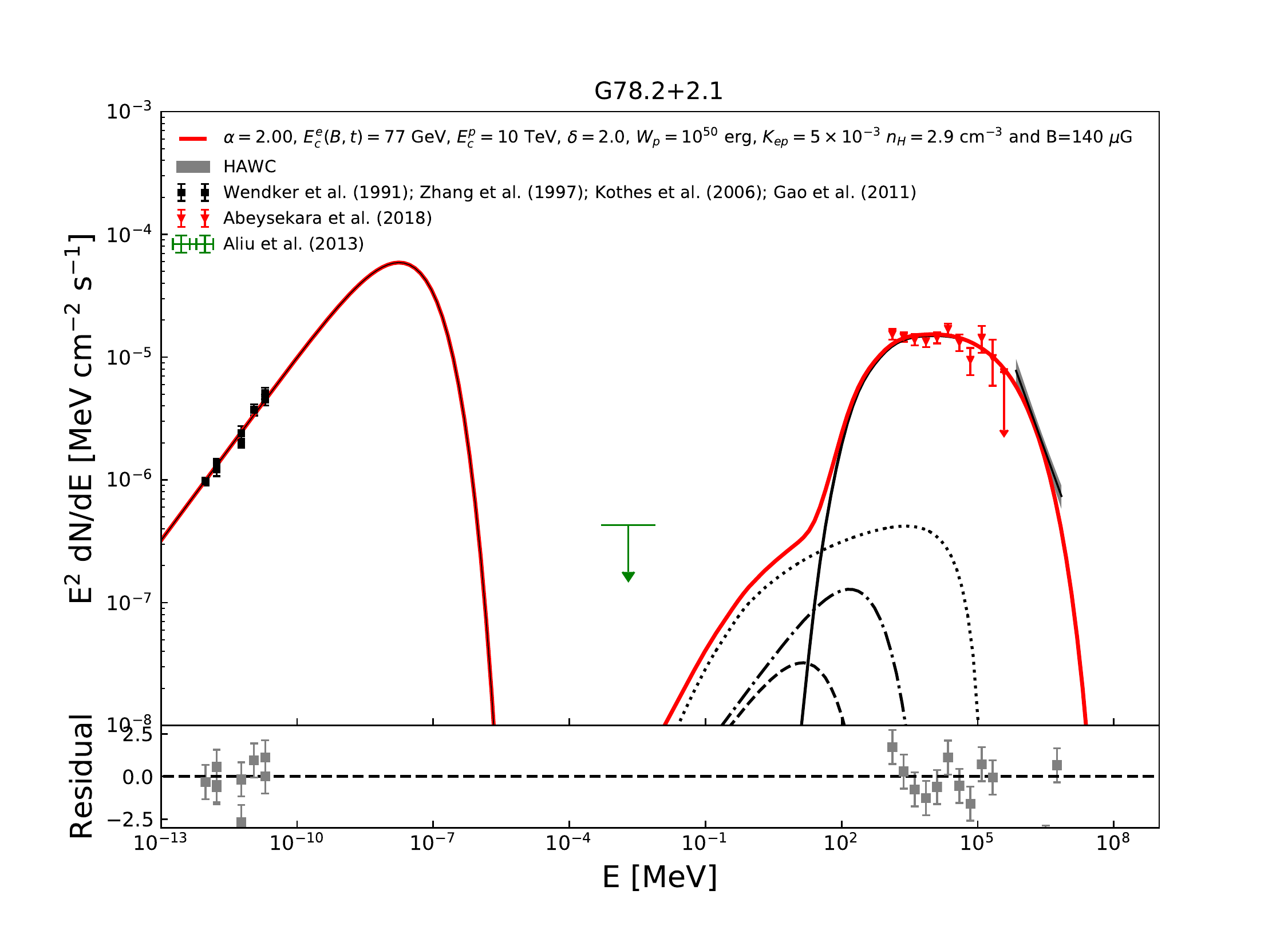}{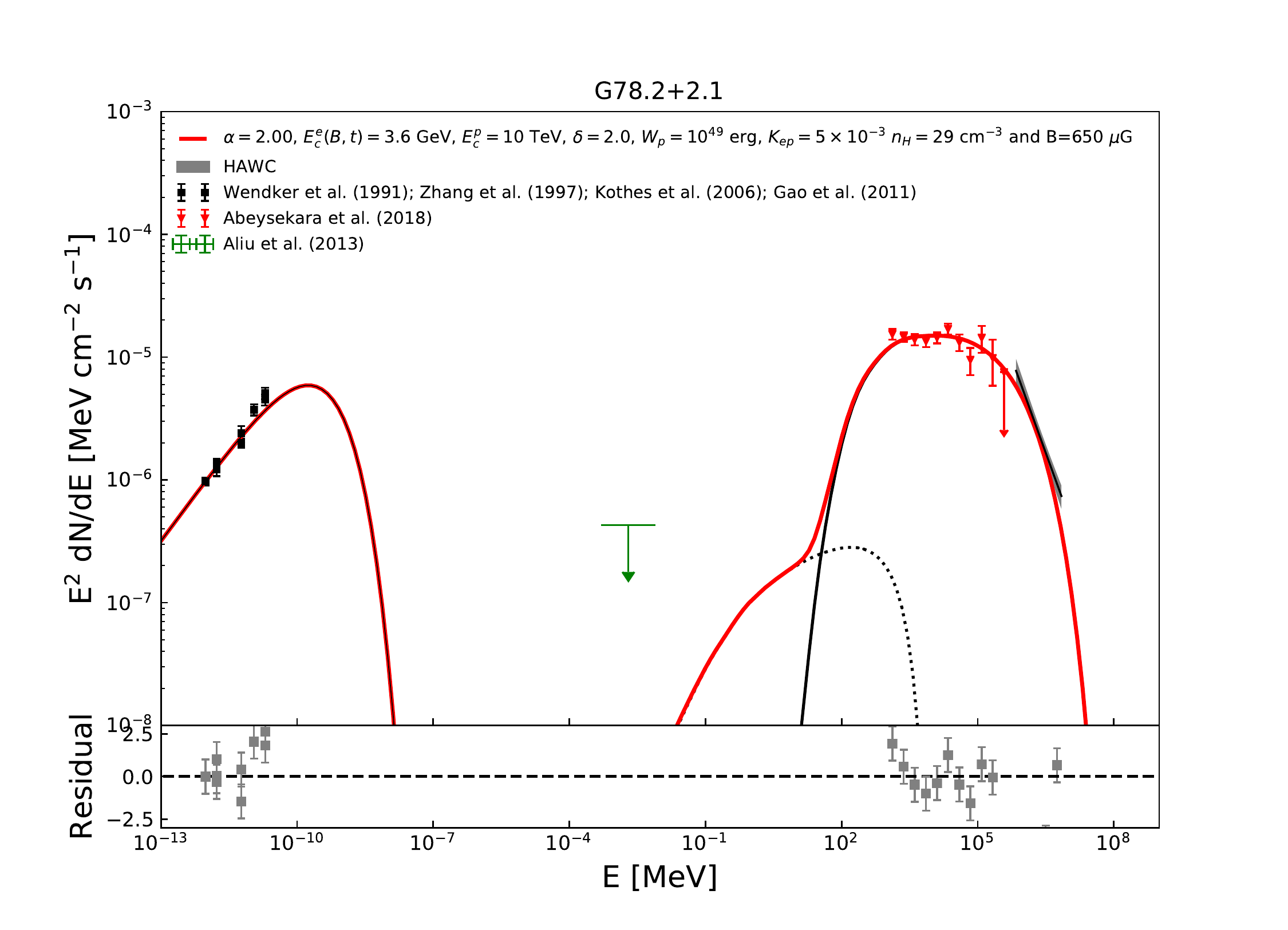}
\plottwo{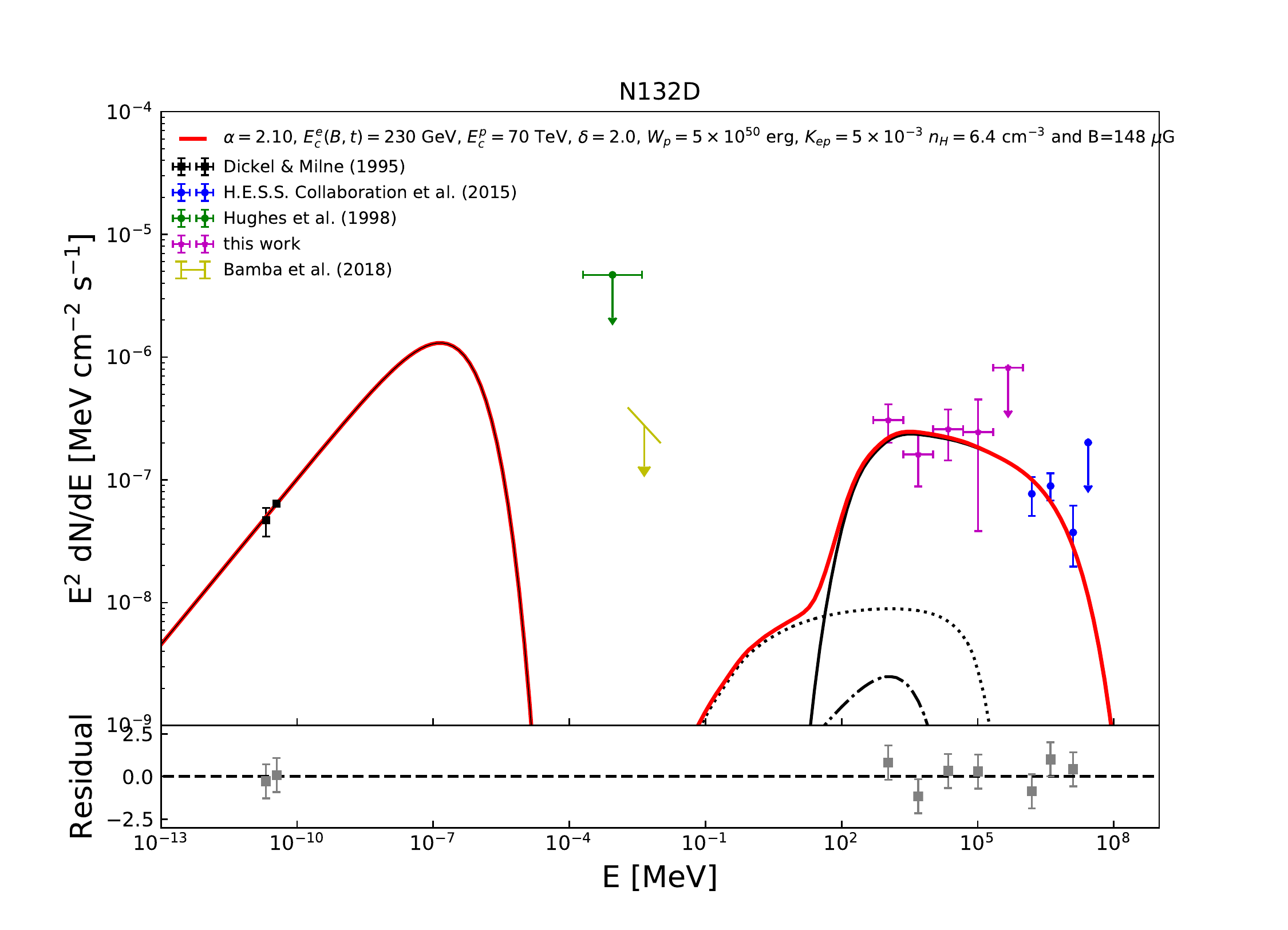}{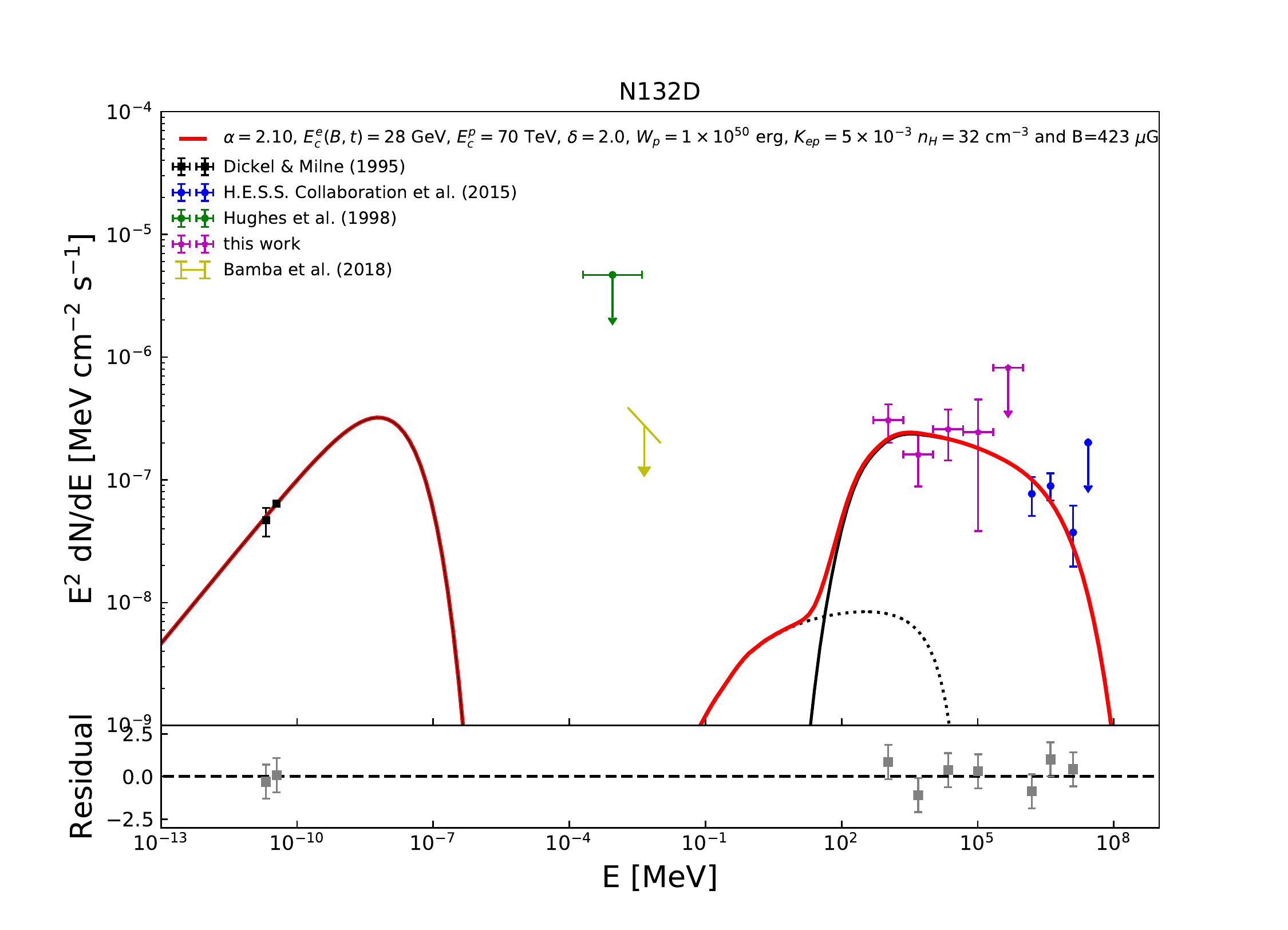}
\caption{Same as Figure \ref{fig:1912} but for G78.2+01.2 (upper) and N132D (lower).
\label{fig:strong2}}
\end{figure*}

Besides the three SNRs studied above, there are two more radio bright SNRs: G78.2+01.2 and N132D. Both of them have very soft spectra in the TeV band, indicating a spectral cutoff. The latter is an SNR in the Large Magellanic Cloud and is the most powerful SNR in our sample \citep{2018ApJ...854...71B}. Although thermal X-ray emission has been detected from both sources, there is no evidence for non-thermal X-rays. A simple single power-law model can readily fit their SEDs. The model parameters are given in Table \ref{tab:fitpatameters}. For G78.2+01.2, we favor the model with $W_{\rm p} = 10^{50}$ erg for the more reasonable value of magnetic field and $W_B/W_{\rm e}$. Stronger magnetic fields are needed for lower values of $W_{\rm p}$. Considering the fact that N132D is the most powerful SNR, a value of $5\times 10^{50}$ erg appears to be reasonable for $W_{\rm p}$. To reduce $W_{\rm p}$, the magnetic field needs to be increased to reproduce the observed radio flux. The strong magnetic field of $423 \mu$ G appears to be reasonable to this SNR. So both models for N132D shown in Figure \ref{fig:strong2} are favored.
N132D has a relatively soft $\gamma$-ray spectrum. The proton distribution cuts off at 70 TeV. The cutoff energy of the proton distribution for G78.2+01.2, however, is only about 10 TeV, the lowest in our sample. It is likely that protons above 10 TeV have already escaped from the SNR and may illuminate surrounding molecular clouds. More observations in the TeV band are warranted.

\section{Discussions}
\label{dis}

In general, all the models presented above give good fits to the SEDs. We picked up the favored ones mostly based on the model parameters. Firstly, we favor models with a total energy of protons below $10^{50}$ erg and a magnetic field between 10 to 100 $\mu$G or as close as to this range as possible.
Since these SNRs are expected to dominate the flux of TeV cosmic rays, on average each SNR should inject less than $10^{50}$ erg of energy to the cosmic rays \citep{2017ApJ...844L...3Z}.
Secondly, the total energy of the magnetic field should be comparable to that of the electrons or protons. In Table \ref{tab:fitpatameters}, we highlighted these models with a bold face for the ratio of $W_B/W_{\rm e}$. It can be seen that most of the favored models satisfies the first criteria except for SNRs G78.2+01.2 and N132D, where the magnetic field is above 100 $\mu$G. A magnetic field below $100\ \mu$G will require a total energy of protons exceeding $10^{50}$ ergs. Most of the favored models have a $W_{\rm p}$ of $10^{50}$ erg. Only for G150.3+04.5 and SN 1006, $W_{\rm p}$ is one order of magnitude lower. This is reasonable since these SNRs are likely resulting from Type Ia SNs for their lack of compact neutron stars within the remnants. N132D is the most powerful SNR, a value of $W_{\rm p}$ exceeding $10^{50}$ erg is acceptable.

\begin{figure}[ht!]
\plotone{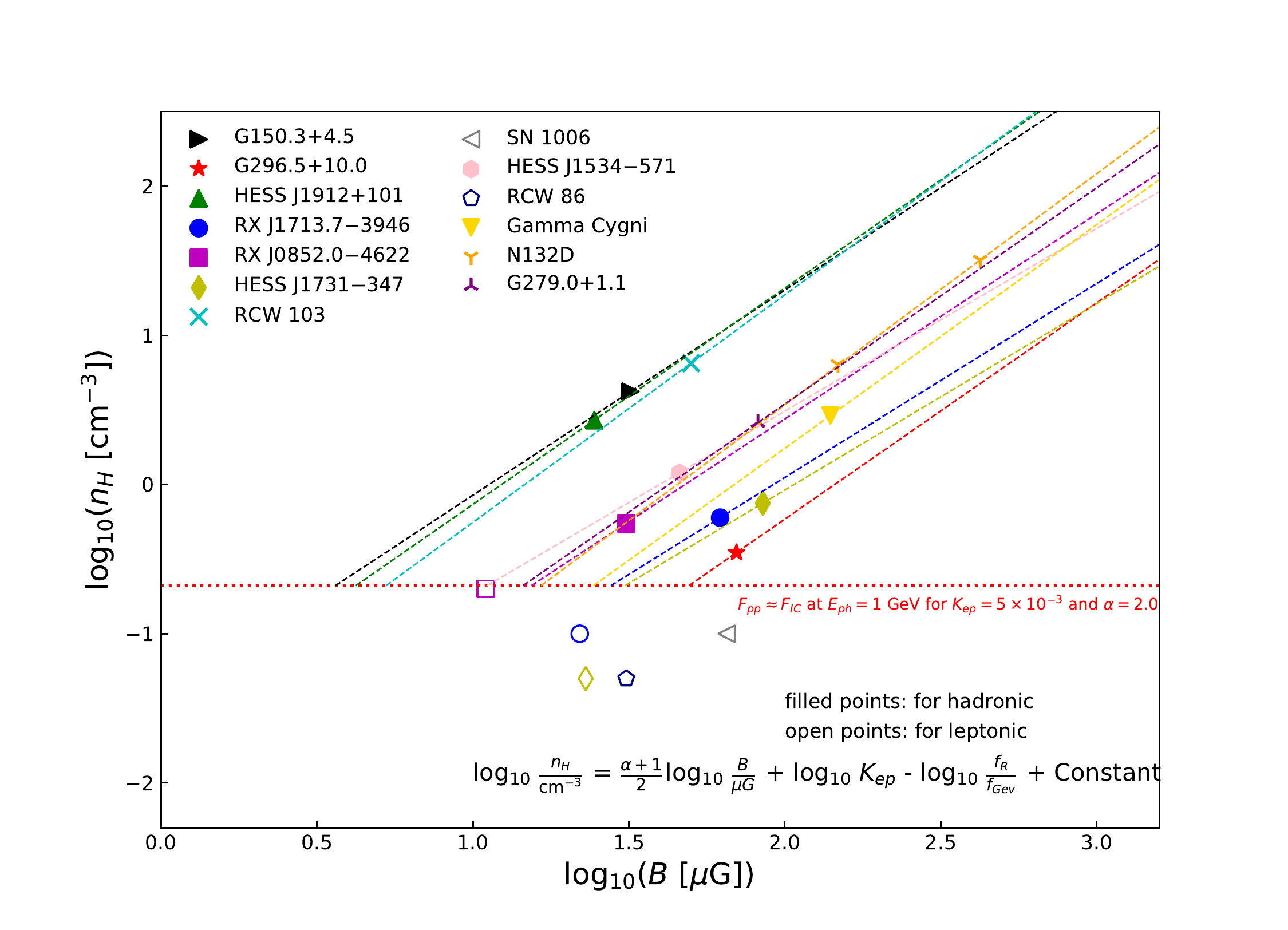}
\caption{Gas density $n_{\rm H}$ vs magnetic field $B$ for the favored models for the 13 SNRs studied in this paper. The open signs below $n_{\rm H}=0.21 $cm$^{-3}$ are for the leptonic models.
The dashed lines satisfy the inserted equation, which assumes synchrotron process for the radio flux density $f_{\rm R}$ and hadronic processes for the $\gamma$-ray flux density $f_{\rm GeV}$ and
gives the model-predicted relationship between $n_{\rm H}$ and $B$ in the hadronic scenario for different SNRs.
The dotted line indicates the gas density that can give a $\gamma$-ray flux density at 1 GeV via hadronic processes $F_{\rm pp}$ equal to the 1 GeV flux density produced via IC scattering of CMB $F_{\rm IC}$ by energetic electrons with $K_{\rm ep}=0.005$ and $\alpha=2.0$.    
\label{fig:had1}}
\end{figure}

Figure \ref{fig:had1} shows the scatter plot between $B$ vs $n_{\rm H}$ for the favored models. The mean density of the emission region is always less than $10$ cm$^{-3}$ except for N132D, which is consistent with the X-ray observations. It appears that $n_{\rm H} =0.21$ cm$^{-3}$ gives the dividing line between the leptonic and hadronic models with the former having a lower density. Assuming a single power-law distribution with $\alpha=2$ and $K_{\rm ep}=5\times10^{-3}$, the dotted line shows the density when the $\gamma$-ray flux at 1 GeV produced via the hadronic processes equals to that produced via IC of the CMB. With the decrease of $\alpha$, this line will shift toward high densities since the GeV emission efficiency via the IC process increases.
Along the dotted line, the radio to $\gamma$-ray flux density ratio increases with the increase of $B$, which is consistent with Figure \ref{fig:sample} with radio brighter sources having stronger magnetic fields. The radio flux density also depends on the spectral index with weaker radio emission for sources with harder spectra. This explains the relative strong magnetic field for the two weak radio sources HESS J1534-571 and G150.3+4.5.
The exact strength of the magnetic field also depends on contributions to the $\gamma$-ray via the leptonic processes in the hadronic scenario.
Below the dotted line, the leptonic process dominates the $\gamma$-ray emission. 
The dashed lines indicate the correlation between $n_{\rm H}$ and $B$ if the $\gamma$-ray emission is solely produced via the hadronic scenario for 13 SNRs studied in the paper, where $f_{\rm R}$ and $f_{\rm GeV}$ represent the flux density at 1 GHz and 1 GeV, respectively. Equally good fits to the SEDs can be obtained along these lines.
 Note that via the CO and/or HI observations, much higher densities are obtained for some sources in our sample, e.g., $\sim 130$ cm$^{-3}$ for RX J1713.7-3946 \citep{2012ApJ...746...82F}, $\sim 60$ cm$^{-3}$ for HESS J1731-347 \citep{2014ApJ...788...94F}, $\sim 100$ cm$^{-3}$ for RX J0852-4622 \citep{2017ApJ...850...71F}, $\sim 75$ cm$^{-3}$ for RCW 86 \citep{2019ApJ...876...37S}, $\sim 30-80$ cm$^{-3}$ for N132D \citep{2018ApJ...854...71B,2020ApJ...902...53S} and  $\sim 45$ cm$^{-3}$ for Gamma Cygni \citep{2020arXiv201015854M}. 
 A higher average density in the $\gamma$-ray emission region will result in a larger magnetic field and a lower total energy of relativistic protons. We obtained a lower limit to the total energy of cosmic rays of ~$10^{48}-10^{49}$ ergs by adopting these densities. The number density of some individual cloud can be as high as $10^4-10^5$ cm$^{-3}$ in  RX J1713.7-3946 \citep{2010ApJ...724...59S, 2012MNRAS.422.2230M}. Most of these high density regions are associated with local clouds with a small volume filling factor, implying that the shock acceleration mainly operates in a low density inter-cloud medium as we assumed (see Table \ref{tab:fitpatameters}). Under the strong magnetic field of these clouds, the broad-band SED can be well reproduced without considering the higher value of $n_{\rm H}$ due to the energy depended-penetration of cosmic-rays into the dense clouds. Considering that the distribution of high energy particles should be more or less uniform due to diffusion, a good spatial correspondence between TeV gamma-rays and interstellar neutral gases is expected in the hadronic scenario for the $\gamma$-ray emission.

To explain the origin of hard $\gamma$-ray spectra from supernova remnants in the hadronic scenario, \citet{2014MNRAS.445L..70G} first showed that the energy dependent penetration of cosmic rays into dense emission regions can lead to a very hard $\gamma$-ray spectrum at low energies. Detailed studies of shock interaction with molecular clouds show that the magnetic field strength will be enhanced not only on the surface of targeted clouds, but also inside these clouds \citep{2010ApJ...708..965Z,2012ApJ...744...71I, 2019ApJ...872...46I}.
The model was further developed by \citet{2019MNRAS.487.3199C}. Detailed modeling of RX J1713.7-3946 also favors the hadronic scenario \citep{2016ApJ...821...43Z}. Our results suggest that the hard spectra may be due to very efficient particle acceleration in a low density environment \citep{2017ApJ...844L...3Z, 2019MNRAS.482.5268Z}. A softer spectrum can be produced when shocks slow down dramatically due to interaction with molecular clouds \citep{2013MNRAS.431..415B, 2014ApJ...784L..35T}, which may explain the low energy spectral component seen in HESS J1912+101 and G279.0+1.1.

\begin{figure}[ht!]
\plotone{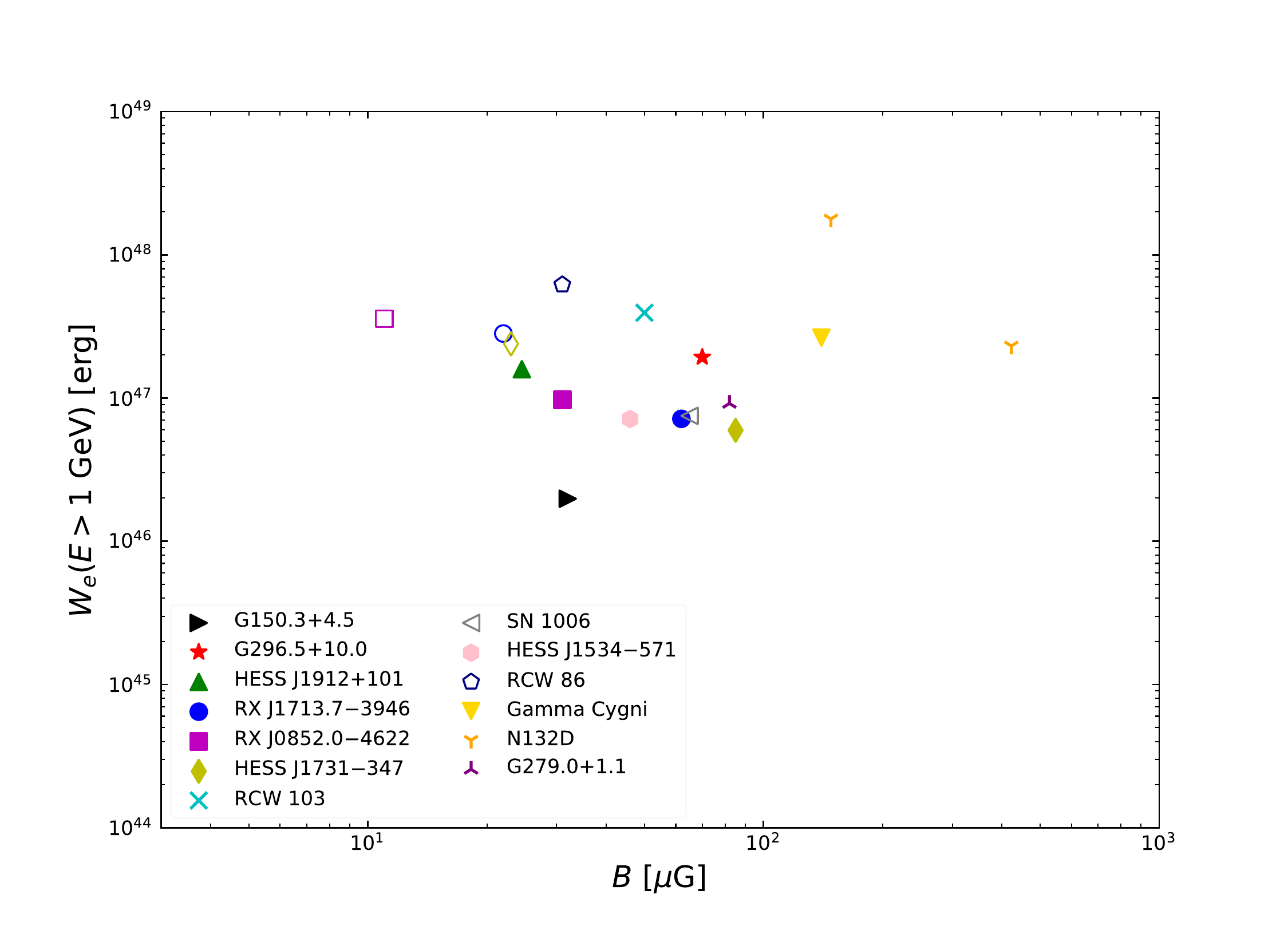}
\caption{The total energy content of electron above 1 GeV $W_e$ vs magnetic field $B$ for the 13 SNRs studied in this paper.
\label{fig:B_We}}
\end{figure}

Figure \ref{fig:B_We} shows the scatter plot between $B$ and $W_{\rm e}$. The total energy of electrons $W_{\rm e}$ is about $10^{47}$ erg, with N132D having the highest value of $1.8\times 10^{48}$ erg and G150.3+04.5 having the lowest value of $2.0\times 10^{46}$ erg. These results look reasonable. Table \ref{tab:fitpatameters} shows that the cutoff energies of these electrons are always around 1 TeV for sources without nonthermal X-ray emission, which may explain the spectral break near 1 TeV in the cosmic ray electron spectrum \citep{2017Natur.552...63D}. For the three sources RX J1713.7$-$3946, RX J0852$-$4622, and HESS J1731$-$347, both the leptonic and hadronic scenarios can lead to reasonable model parameters. The leptonic models always have weaker magnetic fields and more energy in electrons than the corresponding hadronic models. However, Table \ref{tab:fitpatameters} shows that leptonic models always have less energetic protons even we fix $K_{\rm ep}$ for the difference in their spectral shape. The leptonic model always has a softer spectrum than the hadronic one. Better spectral measurements in the radio band may distinguish these models.

\begin{figure}[ht!]
\plotone{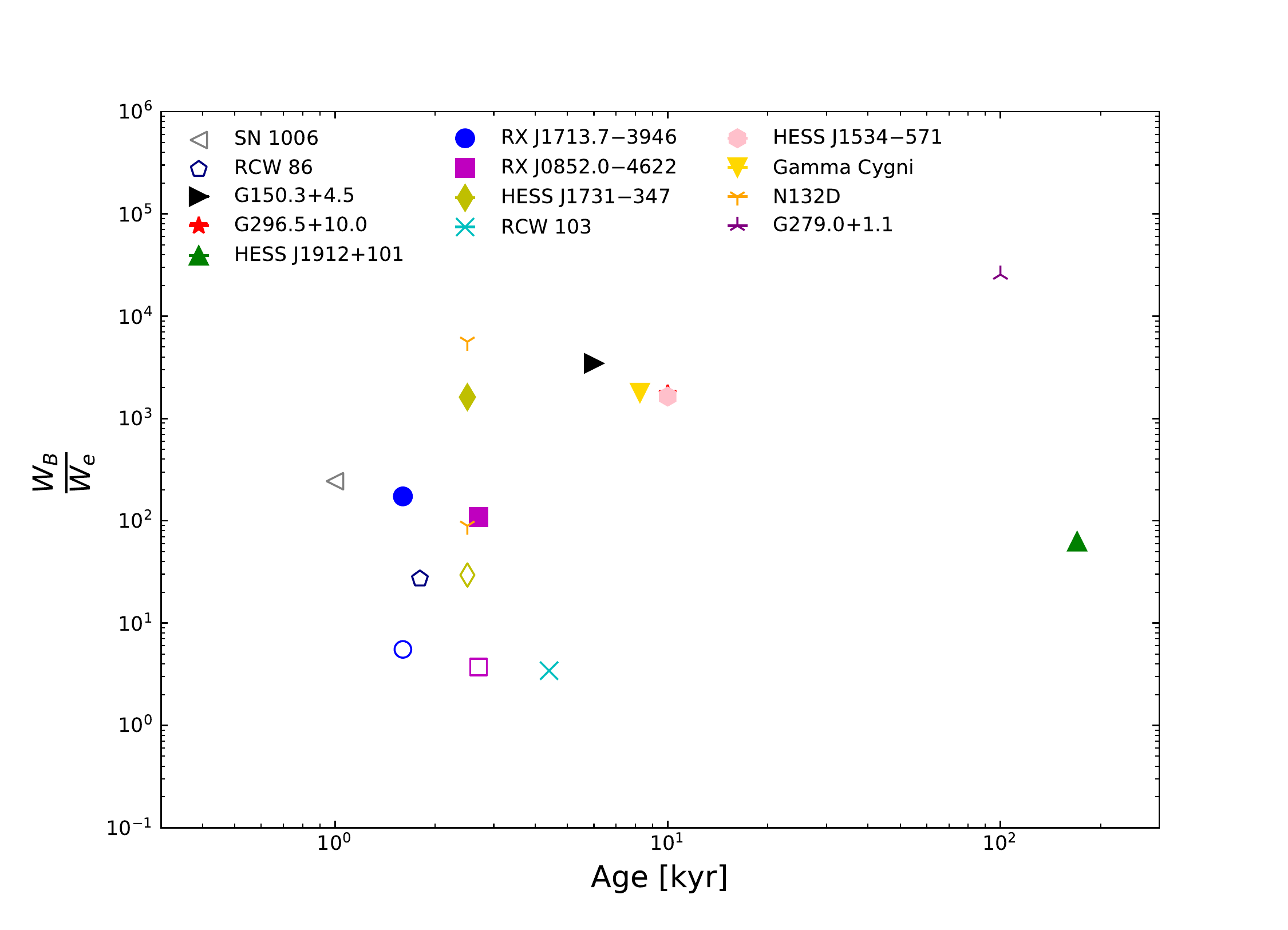}
\caption{The ratio of the magnetic energy $W_B$ to the total energy content of electron above 1 GeV $W_{\rm e}$ for the 13 SNRs studied in this paper.
\label{fig:B_We1}}
\end{figure}

Figure \ref{fig:B_We1} shows the dependence of $W_{B}/W_{\rm e}$ on the age of the SNR. In the leptonic models, the magnetic field energy is comparable to the total energy of electrons. In the hadronic scenario, the magnetic energy is comparable to that of protons and  $W_{B}/W_{\rm e}$ appears to increase with the age of the SNR.

\section{Conclusions}\label{sec:conclusion}

In this paper, we carried out detailed spectral modeling of 13 SNRs with hard GeV spectra. We re-analyzed the Fermi data of HESS J1912+101, and found its TeV emission can not be attributed to leptonic process for the old age of the SNR inferred from molecular cloud observations and the spin-down age of the associated pulsar. The same is for SNR G279.0+1.1. A detailed analysis of XMM-Newton observations of G296.5+10.0 failed to uncover nonthermal emission, which in combination with Fermi data analyses also favors the hadronic scenario for the $\gamma$-ray.  

Of the 13 sources studied here, only SN 1006 and RCW 86 favor the leptonic scenario for their $\gamma$-ray emission. RX J1713.7$-$3946, RX J0852$-$4622, and HESS J1731$-$347 can be explained with both leptonic and hadronic models. In the leptonic models, the total energy of the magnetic field is comparable to that of the electrons. In the hadronic models, the magnetic fields and protons are close to energy equipartition. All these sources have prominent nonthermal X-ray emission. For the other 8 sources without evident nonthermal X-ray emission, the hadronic models with a single power-law particle distribution are favored.  And in the hadronic scenario, the magnetic field of older remnants tends to contain more energy than relativistic particles, which may be attributed to escape of high energy particles from SNRs \citep{2018PhRvD..97l3008P, 2020arXiv201015854M}.

Although our results do not completely address the origin of hard $\gamma$-ray spectra from SNRs, young remnants with prominent nonthermal X-ray emission favors the leptonic scenario, while absence of nonthermal X-ray emission strongly favors the hadronic scenario. For RX J1713.7$-$3946, RX J0852$-$4622, and HESS J1731$-$347, both scenarios can fit the SEDs with reasonable parameters. The leptonic model always predicts a softer radio spectrum than the corresponding hadronic one, which may be tested with future observations.

The proton spectrum always cuts off below 70 TeV, implying that SNRs may not be able to accelerate cosmic ray to PeV energies. Alternative PeV sources are needed to explain CR observations. SNRs of Type Ia SNs produce much fewer TeV ions than core collapse SNs. TeV cosmic ray fluxes are therefore likely dominated by more powerful SNRs with compact neutron stars in the middle. The total energy of relativistic protons is on the order of $10^{50}$ erg for each core collapse SNR, which indicates very efficient ion acceleration and is compatible to the relatively hard spectra inferred from observations \citep{2019MNRAS.482.5268Z}. 
G78.2+01.2 has the lowest cutoff energy indicating escape of particles beyond 10 TeV from the SNR. Nearby molecular clouds may be illuminated by these escaping particles and produce $\gamma$-ray in the TeV band. HESS 1912+101 has the oldest age in our sample, yet the cutoff energy of protons is relatively high, indicating structure of the magnetic field may play an important role on the escape process. The strong linear polarization of the radio emission indeed indicates presence of large scale regular magnetic field, which may trap high-energy particles in this source effectively.
For sources with prominent nonthermal X-ray emission, the electron distribution always cuts off above 7 TeV, while those without nonthermal X-ray emission, their electron distribution always cuts off near 1 TeV, which may explain the cosmic ray electron spectrum in the TeV range \citep{2017Natur.552...63D}. Further exploration of this issue is warranted.

\begin{figure}[ht!]
\plotone{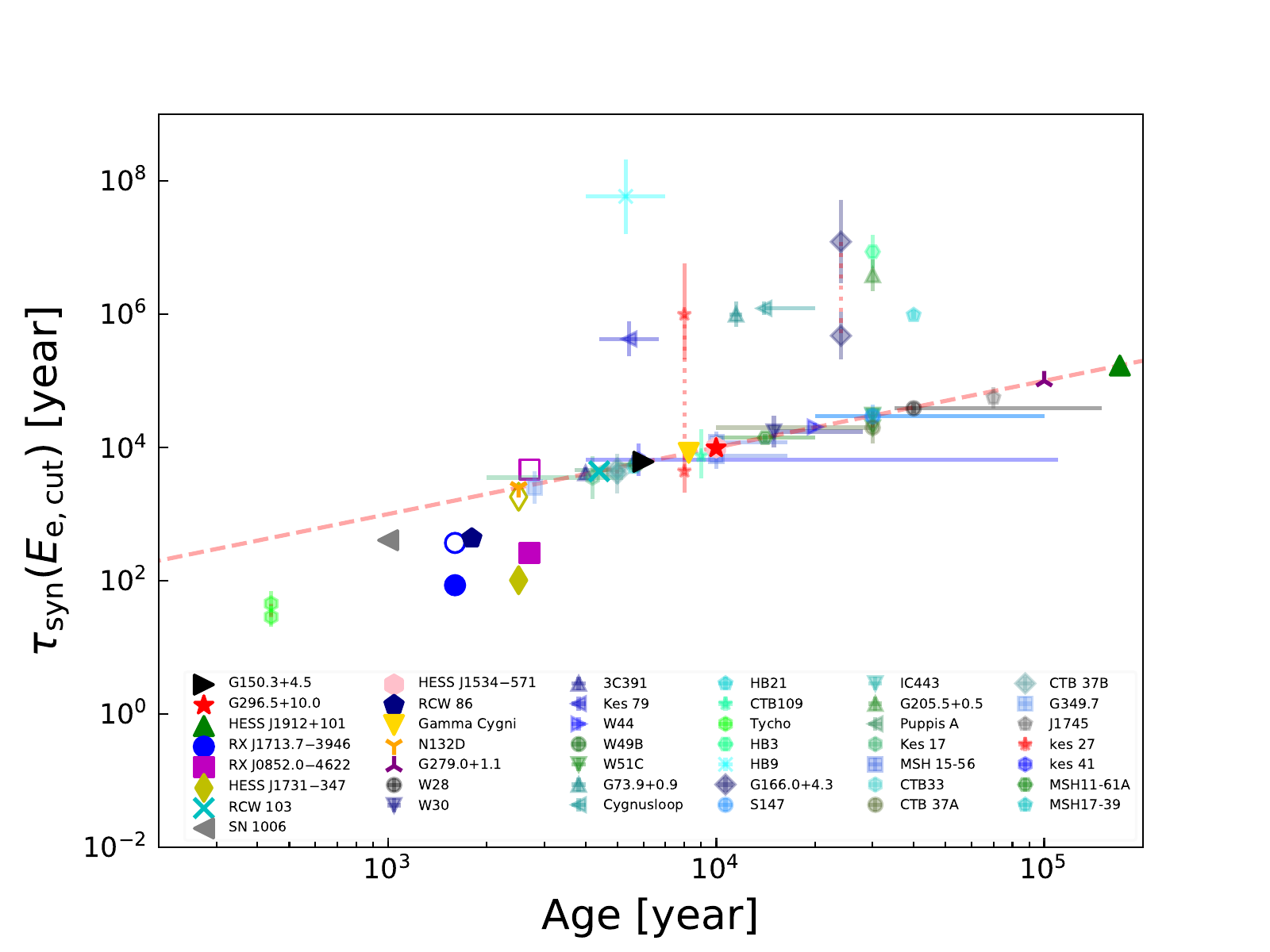}
\caption{The correlation between the electron synchrotron cooling time at the cutoff energy and the age of SNRs.
\label{fig:Syntau_age}}
\end{figure}

To study the acceleration of high-energy particles in SNRs, \citet{2019ApJ...874...50Z} plotted the synchrotron energy loss time at the cutoff energy of electron distribution vs the ages of a sample of SNRs. This figure is updated here as shown in Figure \ref{fig:Syntau_age}, confirming the early discovery that high-energy electrons are mostly accelerated in young SNRs and the radiative energy loss and escape processes dominate in old SNRs.

\acknowledgments

We thank the anonymous referee for very helpful suggestions that help to improve the manuscript significantly. This work is partially supported by National Key R\&D Program
of China: 2018YFA0404203, NSFC grants: U1738122, U1931204,
11761131007, 11573070, the Natural Science Foundation for Young Scholars of Jiangsu Province, China (No. BK20191109), and by the International Partnership Program of
Chinese Academy of Sciences, grant No. 114332KYSB20170008.

%






\bibliography{ms}{}

\begin{thebibliography}{}
\expandafter\ifx\csname natexlab\endcsname\relax\def\natexlab#1{#1}\fi
\providecommand{\url}[1]{\href{#1}{#1}}
\providecommand{\dodoi}[1]{doi:~\href{http://doi.org/#1}{\nolinkurl{#1}}}
\providecommand{\doeprint}[1]{\href{http://ascl.net/#1}{\nolinkurl{http://ascl.net/#1}}}
\providecommand{\doarXiv}[1]{\href{https://arxiv.org/abs/#1}{\nolinkurl{https://arxiv.org/abs/#1}}}

\bibitem[{{Abdo} {et~al.}(2010){Abdo}, {Ackermann}, {Ajello}, {Baldini},
  {Ballet}, {Barbiellini}, {Baring}, {Bastieri}, {Baughman}, {Bechtol},
  {Bellazzini}, {Berenji}, {Blandford}, {Bloom}, {Bonamente}, {Borgland},
  {Bregeon}, {Brez}, {Brigida}, {Bruel}, {Burnett}, {Buson}, {Caliandro},
  {Cameron}, {Caraveo}, {Casandjian}, {Cecchi}, {{\c{C}}elik}, {Chekhtman},
  {Cheung}, {Chiang}, {Ciprini}, {Claus}, {Cognard}, {Cohen-Tanugi},
  {Cominsky}, {Conrad}, {Cutini}, {Dermer}, {de Angelis}, {de Palma}, {Digel},
  {do Couto e Silva}, {Drell}, {Dubois}, {Dumora}, {Espinoza}, {Farnier},
  {Favuzzi}, {Fegan}, {Focke}, {Fortin}, {Frailis}, {Fukazawa}, {Funk},
  {Fusco}, {Gargano}, {Gasparrini}, {Gehrels}, {Germani}, {Giavitto},
  {Giebels}, {Giglietto}, {Giordano}, {Glanzman}, {Godfrey}, {Grenier},
  {Grondin}, {Grove}, {Guillemot}, {Guiriec}, {Hanabata}, {Harding},
  {Hayashida}, {Hays}, {Hughes}, {Jackson}, {J{\'o}hannesson}, {Johnson},
  {Johnson}, {Johnson}, {Kamae}, {Katagiri}, {Kataoka}, {Katsuta}, {Kawai},
  {Kerr}, {Kn{\"o}dlseder}, {Kocian}, {Kramer}, {Kuss}, {Lande}, {Latronico},
  {Lemoine-Goumard}, {Longo}, {Loparco}, {Lott}, {Lovellette}, {Lubrano},
  {Lyne}, {Madejski}, {Makeev}, {Mazziotta}, {McEnery}, {Meurer}, {Michelson},
  {Mitthumsiri}, {Mizuno}, {Monte}, {Monzani}, {Morselli}, {Moskalenko},
  {Murgia}, {Nakamori}, {Nolan}, {Norris}, {Noutsos}, {Nuss}, {Ohsugi},
  {Omodei}, {Orlando}, {Ormes}, {Paneque}, {Parent}, {Pelassa}, {Pepe},
  {Pesce-Rollins}, {Piron}, {Porter}, {Rain{\`o}}, {Rando}, {Razzano},
  {Reimer}, {Reimer}, {Reposeur}, {Rochester}, {Rodriguez}, {Romani}, {Roth},
  {Ryde}, {Sadrozinski}, {Sanchez}, {Sander}, {Parkinson}, {Scargle},
  {Sgr{\`o}}, {Siskind}, {Smith}, {Smith}, {Spand re}, {Spinelli}, {Stappers},
  {Stecker}, {Strickman}, {Suson}, {Tajima}, {Takahashi}, {Takahashi},
  {Tanaka}, {Thayer}, {Thayer}, {Theureau}, {Thompson}, {Tibaldo}, {Tibolla},
  {Torres}, {Tosti}, {Tramacere}, {Uchiyama}, {Usher}, {Vasileiou}, {Venter},
  {Vilchez}, {Vitale}, {Waite}, {Wang}, {Winer}, {Wood}, {Yamazaki}, {Ylinen},
  \& {Ziegler}}]{2010Sci...327.1103A}
{Abdo}, A.~A., {Ackermann}, M., {Ajello}, M., {et~al.} 2010, Science, 327,
  1103, \dodoi{10.1126/science.1182787}

\bibitem[{{Abdollahi} {et~al.}(2020){Abdollahi}, {Acero}, {Ackermann},
  {Ajello}, {Atwood}, {Axelsson}, {Baldini}, {Ballet}, {Barbiellini},
  {Bastieri}, {Becerra Gonzalez}, {Bellazzini}, {Berretta}, {Bissaldi}, {Bland
  ford}, {Bloom}, {Bonino}, {Bottacini}, {Brandt}, {Bregeon}, {Bruel},
  {Buehler}, {Burnett}, {Buson}, {Cameron}, {Caputo}, {Caraveo}, {Casandjian},
  {Castro}, {Cavazzuti}, {Charles}, {Chaty}, {Chen}, {Cheung}, {Chiaro},
  {Ciprini}, {Cohen-Tanugi}, {Cominsky}, {Coronado-Bl{\'a}zquez}, {Costantin},
  {Cuoco}, {Cutini}, {D'Ammando}, {DeKlotz}, {Torre Luque}, {de Palma},
  {Desai}, {Digel}, {Lalla}, {Mauro}, {Venere}, {Dom{\'\i}nguez}, {Dumora},
  {Dirirsa}, {Fegan}, {Ferrara}, {Franckowiak}, {Fukazawa}, {Funk}, {Fusco},
  {Gargano}, {Gasparrini}, {Giglietto}, {Giommi}, {Giordano}, {Giroletti},
  {Glanzman}, {Green}, {Grenier}, {Griffin}, {Grondin}, {Grove}, {Guiriec},
  {Harding}, {Hayashi}, {Hays}, {Hewitt}, {Horan}, {J{\'o}hannesson},
  {Johnson}, {Kamae}, {Kerr}, {Kocevski}, {Kovac'evic'}, {Kuss}, {Landriu},
  {Larsson}, {Latronico}, {Lemoine-Goumard}, {Li}, {Liodakis}, {Longo},
  {Loparco}, {Lott}, {Lovellette}, {Lubrano}, {Madejski}, {Maldera},
  {Malyshev}, {Manfreda}, {Marchesini}, {Marcotulli}, {Mart{\'\i}-Devesa},
  {Martin}, {Massaro}, {Mazziotta}, {McEnery}, {Mereu}, {Meyer}, {Michelson},
  {Mirabal}, {Mizuno}, {Monzani}, {Morselli}, {Moskalenko}, {Negro}, {Nuss},
  {Ojha}, {Omodei}, {Orienti}, {Orlando}, {Ormes}, {Palatiello}, {Paliya},
  {Paneque}, {Pei}, {Pe{\~n}a-Herazo}, {Perkins}, {Persic}, {Pesce-Rollins},
  {Petrosian}, {Petrov}, {Piron}, {Poon}, {Porter}, {Principe}, {Rain{\`o}},
  {Rando}, {Razzano}, {Razzaque}, {Reimer}, {Reimer}, {Remy}, {Reposeur},
  {Romani}, {Parkinson}, {Schinzel}, {Serini}, {Sgr{\`o}}, {Siskind}, {Smith},
  {Spandre}, {Spinelli}, {Strong}, {Suson}, {Tajima}, {Takahashi}, {Tak},
  {Thayer}, {Thompson}, {Tibaldo}, {Torres}, {Torresi}, {Valverde}, {Klaveren},
  {Zyl}, {Wood}, {Yassine}, \& {Zaharijas}}]{2020ApJS..247...33A}
{Abdollahi}, S., {Acero}, F., {Ackermann}, M., {et~al.} 2020, \apjs, 247, 33,
  \dodoi{10.3847/1538-4365/ab6bcb}

\bibitem[{{Abeysekara} {et~al.}(2018){Abeysekara}, {Archer}, {Aune}, {Benbow},
  {Bird}, {Brose}, {Buchovecky}, {Bugaev}, {Cui}, {Daniel}, {Falcone}, {Feng},
  {Finley}, {Fleischhack}, {Flinders}, {Fortson}, {Furniss}, {Gotthelf},
  {Grube}, {Hanna}, {Hervet}, {Holder}, {Huang}, {Hughes}, {Humensky},
  {H{\"u}tten}, {Johnson}, {Kaaret}, {Kar}, {Kelley-Hoskins}, {Kertzman},
  {Kieda}, {Krause}, {Kumar}, {Lang}, {Lin}, {Maier}, {McArthur}, {Moriarty},
  {Mukherjee}, {O'Brien}, {Ong}, {Otte}, {Pandel}, {Park}, {Petrashyk}, {Pohl},
  {Popkow}, {Pueschel}, {Quinn}, {Ragan}, {Reynolds}, {Richards}, {Roache},
  {Rousselle}, {Rulten}, {Sadeh}, {Santander}, {Sembroski}, {Shahinyan},
  {Tyler}, {Vassiliev}, {Wakely}, {Ward}, {Weinstein}, {Wells}, {Wilcox},
  {Wilhelm}, {Williams}, \& {Zitzer}}]{2018ApJ...861..134A}
{Abeysekara}, A.~U., {Archer}, A., {Aune}, T., {et~al.} 2018, \apj, 861, 134,
  \dodoi{10.3847/1538-4357/aac4a2}

\bibitem[{{Abeysekara} {et~al.}(2020){Abeysekara}, {Archer}, {Benbow}, {Bird},
  {Brose}, {Buchovecky}, {Buckley}, {Chromey}, {Cui}, {Daniel}, {Das},
  {Dwarkadas}, {Falcone}, {Feng}, {Finley}, {Fortson}, {Gent}, {Gillanders},
  {Giuri}, {Gueta}, {Hanna}, {Hassan}, {Hervet}, {Holder}, {Hughes},
  {Humensky}, {Kaaret}, {Kar}, {Kelley-Hoskins}, {Kertzman}, {Kieda}, {Krause},
  {Krennrich}, {Kumar}, {Lang}, {Maier}, {Moriarty}, {Mukherjee},
  {Nievas-Rosillo}, {O'Brien}, {Ong}, {Park}, {Petrashyk}, {Pfrang}, {Pohl},
  {Pueschel}, {Quinn}, {Ragan}, {Reynolds}, {Richards}, {Roache}, {Sadeh},
  {Santander}, {Sembroski}, {Shahinyan}, {Sushch}, {Weinstein}, {Wilcox},
  {Wilhelm}, {Williams}, {Williamson}, {Zitzer}, \&
  {Ghiotto}}]{2020ApJ...894...51A}
{Abeysekara}, A.~U., {Archer}, A., {Benbow}, W., {et~al.} 2020, \apj, 894, 51,
  \dodoi{10.3847/1538-4357/ab8310}

\bibitem[{{Acero} {et~al.}(2010){Acero}, {Aharonian}, {Akhperjanian}, {Anton},
  {Barres de Almeida}, {Bazer-Bachi}, {Becherini}, {Behera}, {Beilicke},
  {Bernl{\"o}hr}, {Bochow}, {Boisson}, {Bolmont}, {Borrel}, {Brucker}, {Brun},
  {Brun}, {B{\"u}hler}, {Bulik}, {B{\"u}sching}, {Boutelier}, {Chadwick},
  {Charbonnier}, {Chaves}, {Cheesebrough}, {Conrad}, {Chounet}, {Clapson},
  {Coignet}, {Dalton}, {Daniel}, {Davids}, {Degrange}, {Deil}, {Dickinson},
  {Djannati-Ata{\"i}}, {Domainko}, {O'C.~Drury}, {Dubois}, {Dubus}, {Dyks},
  {Dyrda}, {Egberts}, {Eger}, {Espigat}, {Fallon}, {Farnier}, {Fegan},
  {Feinstein}, {Fiasson}, {F{\"o}rster}, {Fontaine}, {F{\"u}{\ss}ling},
  {Gabici}, {Gallant}, {G{\'e}rard}, {Gerbig}, {Giebels}, {Glicenstein},
  {Gl{\"u}ck}, {Goret}, {G{\"o}ring}, {Hauser}, {Hauser}, {Heinz},
  {Heinzelmann}, {Henri}, {Hermann}, {Hinton}, {Hoffmann}, {Hofmann},
  {Hofverberg}, {Holleran}, {Hoppe}, {Horns}, {Jacholkowska}, {de Jager},
  {Jahn}, {Jung}, {Katarzy{\'n}ski}, {Katz}, {Kaufmann}, {Kerschhaggl},
  {Khangulyan}, {Kh{\'e}lifi}, {Keogh}, {Klochkov}, {Klu{\'z}niak}, {Kneiske},
  {Komin}, {Kosack}, {Kossakowski}, {Lamanna}, {Lemoine-Goumard}, {Lenain},
  {Lohse}, {Marandon}, {Marcowith}, {Masbou}, {Maurin}, {McComb}, {Medina},
  {M{\'e}hault}, {Moderski}, {Moulin}, {Naumann-Godo}, {de Naurois}, {Nedbal},
  {Nekrassov}, {Nicholas}, {Niemiec}, {Nolan}, {Ohm}, {Olive}, {de O{\~n}a
  Wilhelmi}, {Orford}, {Ostrowski}, {Panter}, {Paz Arribas}, {Pedaletti},
  {Pelletier}, {Petrucci}, {Pita}, {P{\"u}hlhofer}, {Punch}, {Quirrenbach},
  {Raubenheimer}, {Raue}, {Rayner}, {Reimer}, {Renaud}, {de Los Reyes},
  {Rieger}, {Ripken}, {Rob}, {Rosier-Lees}, {Rowell}, {Rudak}, {Rulten},
  {Ruppel}, {Ryde}, {Sahakian}, {Santangelo}, {Schlickeiser}, {Sch{\"o}ck},
  {Sch{\"o}nwald}, {Schwanke}, {Schwarzburg}, {Schwemmer}, {Shalchi}, {Sushch},
  {Sikora}, {Skilton}, {Sol}, {Stawarz}, {Steenkamp}, {Stegmann}, {Stinzing},
  {Superina}, {Szostek}, {Tam}, {Tavernet}, {Terrier}, {Tibolla}, {Tluczykont},
  {van Eldik}, {Vasileiadis}, {Venter}, {Venter}, {Vialle}, {Vincent}, {Vink},
  {Vivier}, {V{\"o}lk}, {Volpe}, {Vorobiov}, {Wagner}, {Ward}, {Zdziarski},
  {Zech}, \& {H.E.S.S.~Collaboration}}]{2010AA...516A..62A}
{Acero}, F., {Aharonian}, F., {Akhperjanian}, A.~G., {et~al.} 2010, \aap, 516,
  A62, \dodoi{10.1051/0004-6361/200913916}

\bibitem[{Aguilar {et~al.}(2015{\natexlab{a}})Aguilar, Aisa, Alpat, Alvino,
  Ambrosi, Andeen, Arruda, Attig, Azzarello, Bachlechner, Barao, Barrau,
  Barrin, Bartoloni, Basara, Battarbee, Battiston, Bazo, Becker, Behlmann,
  Beischer, Berdugo, Bertucci, Bigongiari, Bindi, Bizzaglia, Bizzarri, Boella,
  de~Boer, Bollweg, Bonnivard, Borgia, Borsini, Boschini, Bourquin, Burger,
  Cadoux, Cai, Capell, Caroff, Casaus, Cascioli, Castellini, Cernuda, Cerreta,
  Cervelli, Chae, Chang, Chen, Chen, Cheng, Chen, Cheng, Chou, Choumilov,
  Choutko, Chung, Clark, Clavero, Coignet, Consolandi, Contin, Corti, Gil,
  Coste, Creus, Crispoltoni, Cui, Dai, Delgado, Della~Torre, Demirk\"oz,
  Derome, Di~Falco, Di~Masso, Dimiccoli, D\'{\i}az, von Doetinchem, Donnini,
  Du, Duranti, D'Urso, Eline, Eppling, Eronen, Fan, Farnesini, Feng, Fiandrini,
  Fiasson, Finch, Fisher, Galaktionov, Gallucci, Garc\'{\i}a,
  Garc\'{\i}a-L\'opez, Gargiulo, Gast, Gebauer, Gervasi, Ghelfi, Gillard,
  Giovacchini, Goglov, Gong, Goy, Grabski, Grandi, Graziani, Guandalini,
  Guerri, Guo, Haas, Habiby, Haino, Han, He, Heil, Hoffman, Hsieh, Huang, Huh,
  Incagli, Ionica, Jang, Jinchi, Kanishev, Kim, Kim, Kirn, Kossakowski,
  Kounina, Kounine, Koutsenko, Krafczyk, La~Vacca, Laudi, Laurenti, Lazzizzera,
  Lebedev, Lee, Lee, Leluc, Levi, Li, Li, Li, Li, Li, Li, Li, Li, Li, Lim, Lin,
  Lipari, Lippert, Liu, Liu, Lolli, Lomtadze, Lu, Lu, Lu, Luebelsmeyer, Luo,
  Lv, Majka, Ma\~n\'a, Mar\'{\i}n, Martin, Mart\'{\i}nez, Masi, Maurin,
  Menchaca-Rocha, Meng, Mo, Morescalchi, Mott, M\"uller, Ni, Nikonov, Nozzoli,
  Nunes, Obermeier, Oliva, Orcinha, Palmonari, Palomares, Paniccia, Papi,
  Pauluzzi, Pedreschi, Pensotti, Pereira, Picot-Clemente, Pilo, Piluso,
  Pizzolotto, Plyaskin, Pohl, Poireau, Postaci, Putze, Quadrani, Qi, Qin, Qu,
  R\"aih\"a, Rancoita, Rapin, Ricol, Rodr\'{\i}guez, Rosier-Lees, Rozhkov,
  Rozza, Sagdeev, Sandweiss, Saouter, Sbarra, Schael, Schmidt, von Dratzig,
  Schwering, Scolieri, Seo, Shan, Shan, Shi, Shi, Shi, Siedenburg, Son, Spada,
  Spinella, Sun, Sun, Tacconi, Tang, Tang, Tang, Tao, Tescaro, Ting, Ting,
  Tomassetti, Torsti, T\"urko\ifmmode~\breve{g}\else \u{g}\fi{}lu, Urban,
  Vagelli, Valente, Vannini, Valtonen, Vaurynovich, Vecchi, Velasco, Vialle,
  Vitale, Vitillo, Wang, Wang, Wang, Wang, Wang, Wang, Weng, Whitman,
  Wienkenh\"over, Wu, Wu, Xia, Xie, Xie, Xiong, Xin, Xu, Xu, Yan, Yang, Yang,
  Ye, Yi, Yu, Yu, Zeissler, Zhang, Zhang, Zhang, Zhang, Zheng, Zhuang, Zhukov,
  Zichichi, Zimmermann, Zuccon, \& Zurbach}]{PhysRevLett.114.171103}
Aguilar, M., Aisa, D., Alpat, B., {et~al.} 2015{\natexlab{a}}, Phys. Rev.
  Lett., 114, 171103, \dodoi{10.1103/PhysRevLett.114.171103}

\bibitem[{Aguilar {et~al.}(2015{\natexlab{b}})Aguilar, Aisa, Alpat, Alvino,
  Ambrosi, Andeen, Arruda, Attig, Azzarello, Bachlechner, Barao, Barrau,
  Barrin, Bartoloni, Basara, Battarbee, Battiston, Bazo, Becker, Behlmann,
  Beischer, Berdugo, Bertucci, Bindi, Bizzaglia, Bizzarri, Boella, de~Boer,
  Bollweg, Bonnivard, Borgia, Borsini, Boschini, Bourquin, Burger, Cadoux, Cai,
  Capell, Caroff, Casaus, Castellini, Cernuda, Cerreta, Cervelli, Chae, Chang,
  Chen, Chen, Chen, Chen, Cheng, Chou, Choumilov, Choutko, Chung, Clark,
  Clavero, Coignet, Consolandi, Contin, Corti, Gil, Coste, Creus, Crispoltoni,
  Cui, Dai, Delgado, Della~Torre, Demirk\"oz, Derome, Di~Falco, Di~Masso,
  Dimiccoli, D\'{\i}az, von Doetinchem, Donnini, Duranti, D'Urso, Egorov,
  Eline, Eppling, Eronen, Fan, Farnesini, Feng, Fiandrini, Fiasson, Finch,
  Fisher, Formato, Galaktionov, Gallucci, Garc\'{\i}a, Garc\'{\i}a-L\'opez,
  Gargiulo, Gast, Gebauer, Gervasi, Ghelfi, Giovacchini, Goglov, Gong, Goy,
  Grabski, Grandi, Graziani, Guandalini, Guerri, Guo, Haas, Habiby, Haino, Han,
  He, Heil, Hoffman, Hsieh, Huang, Huh, Incagli, Ionica, Jang, Jinchi,
  Kanishev, Kim, Kim, Kirn, Korkmaz, Kossakowski, Kounina, Kounine, Koutsenko,
  Krafczyk, La~Vacca, Laudi, Laurenti, Lazzizzera, Lebedev, Lee, Lee, Leluc,
  Li, Li, Li, Li, Li, Li, Li, Li, Li, Li, Lim, Lin, Lipari, Lippert, Liu, Liu,
  Liu, Lolli, Lomtadze, Lu, Lu, Lu, Luebelsmeyer, Luo, Luo, Lv, Majka,
  Ma\~n\'a, Mar\'{\i}n, Martin, Mart\'{\i}nez, Masi, Maurin, Menchaca-Rocha,
  Meng, Mo, Morescalchi, Mott, M\"uller, Nelson, Ni, Nikonov, Nozzoli, Nunes,
  Obermeier, Oliva, Orcinha, Palmonari, Palomares, Paniccia, Papi, Pauluzzi,
  Pedreschi, Pensotti, Pereira, Picot-Clemente, Pilo, Piluso, Pizzolotto,
  Plyaskin, Pohl, Poireau, Putze, Quadrani, Qi, Qin, Qu, R\"aih\"a, Rancoita,
  Rapin, Ricol, Rodr\'{\i}guez, Rosier-Lees, Rozhkov, Rozza, Sagdeev,
  Sandweiss, Saouter, Schael, Schmidt, von Dratzig, Schwering, Scolieri, Seo,
  Shan, Shan, Shi, Shi, Shi, Siedenburg, Son, Song, Spada, Spinella, Sun, Sun,
  Tacconi, Tang, Tang, Tang, Tao, Tescaro, Ting, Ting, Tomassetti, Torsti,
  T\"urko\ifmmode~\breve{g}\else \u{g}\fi{}lu, Urban, Vagelli, Valente,
  Vannini, Valtonen, Vaurynovich, Vecchi, Velasco, Vialle, Vitale, Vitillo,
  Wang, Wang, Wang, Wang, Wang, Wang, Weng, Whitman, Wienkenh\"over,
  Willenbrock, Wu, Wu, Xia, Xie, Xie, Xiong, Xu, Xu, Yan, Yang, Yang, Yang, Ye,
  Yi, Yu, Yu, Zeissler, Zhang, Zhang, Zhang, Zhang, Zhang, Zhang, Zhang, Zheng,
  Zhuang, Zhukov, Zichichi, Zimmermann, \& Zuccon}]{PhysRevLett.115.211101}
---. 2015{\natexlab{b}}, Phys. Rev. Lett., 115, 211101,
  \dodoi{10.1103/PhysRevLett.115.211101}

\bibitem[{{Aharonian} {et~al.}(2007){Aharonian}, {Akhperjanian}, {Bazer-Bachi},
  {Beilicke}, {Benbow}, {Berge}, {Bernl{\"o}hr}, {Boisson}, {Bolz}, {Borrel},
  {Braun}, {Brown}, {B{\"u}hler}, {B{\"u}sching}, {Carrigan}, {Chadwick},
  {Chounet}, {Coignet}, {Cornils}, {Costamante}, {Degrange}, {Dickinson},
  {Djannati-Ata{\"\i}}, {Drury}, {Dubus}, {Egberts}, {Emmanoulopoulos},
  {Espigat}, {Feinstein}, {Ferrero}, {Fiasson}, {Filipovic}, {Fontaine},
  {Fukui}, {Funk}, {Funk}, {F{\"u}{\ss}ling}, {Gallant}, {Giebels},
  {Glicenstein}, {Goret}, {Hadjichristidis}, {Hauser}, {Hauser}, {Heinzelmann},
  {Henri}, {Hermann}, {Hinton}, {Hiraga}, {Hoffmann}, {Hofmann}, {Holleran},
  {Hoppe}, {Horns}, {Ishisaki}, {Jacholkowska}, {de Jager}, {Kendziorra},
  {Kerschhaggl}, {Kh{\'e}lifi}, {Komin}, {Konopelko}, {Kosack}, {Lamanna},
  {Latham}, {Le Gallou}, {Lemi{\`e}re}, {Lemoine-Goumard}, {Lohse}, {Martin},
  {Martineau-Huynh}, {Marcowith}, {Masterson}, {Maurin}, {McComb}, {Moulin},
  {Moriguchi}, {de Naurois}, {Nedbal}, {Nolan}, {Noutsos}, {Orford}, {Osborne},
  {Ouchrif}, {Panter}, {Pelletier}, {Pita}, {P{\"u}hlhofer}, {Punch},
  {Ranchon}, {Raubenheimer}, {Raue}, {Rayner}, {Reimer}, {Ripken}, {Rob},
  {Rolland}, {Rosier-Lees}, {Rowell}, {Sahakian}, {Santangelo}, {Saug{\'e}},
  {Schlenker}, {Schlickeiser}, {Schr{\"o}der}, {Schwanke}, {Schwarzburg},
  {Schwemmer}, {Shalchi}, {Sol}, {Spangler}, {Spanier}, {Steenkamp},
  {Stegmann}, {Superina}, {Tam}, {Tavernet}, {Terrier}, {Tluczykont}, {van
  Eldik}, {Vasileiadis}, {Venter}, {Vialle}, {Vincent}, {V{\"o}lk}, {Wagner},
  \& {Ward}}]{2007ApJ...661..236A}
{Aharonian}, F., {Akhperjanian}, A.~G., {Bazer-Bachi}, A.~R., {et~al.} 2007,
  \apj, 661, 236, \dodoi{10.1086/512603}

\bibitem[{{Aharonian} {et~al.}(2008){Aharonian}, {Akhperjanian}, {Barres de
  Almeida}, {Bazer-Bachi}, {Behera}, {Beilicke}, {Benbow}, {Bernl{\"o}hr},
  {Boisson}, {Bolz}, {Borrel}, {Braun}, {Brion}, {Brown}, {B{\"u}hler},
  {Bulik}, {B{\"u}sching}, {Boutelier}, {Carrigan}, {Chadwick}, {Chounet},
  {Clapson}, {Coignet}, {Cornils}, {Costamante}, {Dalton}, {Degrange},
  {Dickinson}, {Djannati-Ata{\"\i}}, {Domainko}, {O'C. Drury}, {Dubois},
  {Dubus}, {Dyks}, {Egberts}, {Emmanoulopoulos}, {Espigat}, {Farnier},
  {Feinstein}, {Fiasson}, {F{\"o}rster}, {Fontaine}, {Funk}, {F{\"u}{\ss}ling},
  {Gallant}, {Giebels}, {Glicenstein}, {Gl{\"u}ck}, {Goret}, {Hadjichristidis},
  {Hauser}, {Hauser}, {Heinzelmann}, {Henri}, {Hermann}, {Hinton}, {Hoffmann},
  {Hofmann}, {Holleran}, {Hoppe}, {Horns}, {Jacholkowska}, {de Jager}, {Jung},
  {Katarzy{\'n}ski}, {Kendziorra}, {Kerschhaggl}, {Kh{\'e}lifi}, {Keogh},
  {Komin}, {Kosack}, {Lamanna}, {Latham}, {Lemi{\`e}re}, {Lemoine-Goumard},
  {Lenain}, {Lohse}, {Martin}, {Martineau-Huynh}, {Marcowith}, {Masterson},
  {Maurin}, {Maurin}, {McComb}, {Moderski}, {Moulin}, {de Naurois}, {Nedbal},
  {Nolan}, {Ohm}, {Olive}, {de O{\~n}a Wilhelmi}, {Orford}, {Osborne},
  {Ostrowski}, {Panter}, {Pedaletti}, {Pelletier}, {Petrucci}, {Pita},
  {P{\"u}hlhofer}, {Punch}, {Raubenheimer}, {Raue}, {Rayner}, {Reimer},
  {Renaud}, {Ripken}, {Rob}, {Rolland}, {Rosier-Lees}, {Rowell}, {Rudak},
  {Ruppel}, {Sahakian}, {Santangelo}, {Schlickeiser}, {Sch{\"o}ck},
  {Schr{\"o}der}, {Schwanke}, {Schwarzburg}, {Schwemmer}, {Shalchi}, {Sol},
  {Spangler}, {Stawarz}, {Steenkamp}, {Stegmann}, {Superina}, {Tam},
  {Tavernet}, {Terrier}, {van Eldik}, {Vasileiadis}, {Venter}, {Vialle},
  {Vincent}, {Vivier}, {V{\"o}lk}, {Volpe}, {Wagner}, {Ward}, {Zdziarski}, \&
  {Zech}}]{2008A&A...484..435A}
{Aharonian}, F., {Akhperjanian}, A.~G., {Barres de Almeida}, U., {et~al.} 2008,
  \aap, 484, 435, \dodoi{10.1051/0004-6361:20078715}

\bibitem[{{Ajello} {et~al.}(2016){Ajello}, {Baldini}, {Barbiellini},
  {Bastieri}, {Bellazzini}, {Bissaldi}, {Bloom}, {Bonino}, {Bottacini},
  {Brandt}, {Bregeon}, {Bruel}, {Buehler}, {Caliandro}, {Cameron}, {Caragiulo},
  {Cavazzuti}, {Charles}, {Chekhtman}, {Ciprini}, {Cohen-Tanugi}, {Condon},
  {Costanza}, {Cutini}, {D'Ammando}, {de Palma}, {Desiante}, {Di Lalla}, {Di
  Mauro}, {Di Venere}, {Drell}, {Dubner}, {Dumora}, {Duvidovich}, {Favuzzi},
  {Focke}, {Fusco}, {Gargano}, {Gasparrini}, {Giacani}, {Giglietto},
  {Glanzman}, {Green}, {Grenier}, {Guiriec}, {Hays}, {Hewitt}, {Hill}, {Horan},
  {Jogler}, {J{\'o}hannesson}, {Jung-Richardt}, {Kensei}, {Kuss}, {Larsson},
  {Latronico}, {Lemoine-Goumard}, {Li}, {Li}, {Longo}, {Loparco}, {Lovellette},
  {Lubrano}, {Magill}, {Maldera}, {Manfreda}, {Mayer}, {Mazziotta}, {McEnery},
  {Michelson}, {Mitthumsiri}, {Mizuno}, {Monzani}, {Morselli}, {Moskalenko},
  {Negro}, {Nuss}, {Orienti}, {Orlando}, {Ormes}, {Paneque}, {Perkins},
  {Pesce-Rollins}, {Piron}, {Pivato}, {Porter}, {Rain{\`o}}, {Rando},
  {Razzano}, {Reimer}, {Reimer}, {Reposeur}, {Schmid}, {Schulz}, {Sgr{\`o}},
  {Simone}, {Siskind}, {Spada}, {Spandre}, {Spinelli}, {Thayer}, {Tibaldo},
  {Torres}, {Tosti}, {Troja}, {Uchiyama}, {Vianello}, {Vink}, {Wood}, \&
  {Yassine}}]{2016ApJ...819...98A}
{Ajello}, M., {Baldini}, L., {Barbiellini}, G., {et~al.} 2016, \apj, 819, 98,
  \dodoi{10.3847/0004-637X/819/2/98}

\bibitem[{{Araya}(2013)}]{2013MNRAS.434.2202A}
{Araya}, M. 2013, \mnras, 434, 2202, \dodoi{10.1093/mnras/stt1162}

\bibitem[{{Araya}(2017)}]{2017ApJ...843...12A}
---. 2017, \apj, 843, 12, \dodoi{10.3847/1538-4357/aa7261}

\bibitem[{{Araya}(2020)}]{2020MNRAS.492.5980A}
---. 2020, \mnras, 492, 5980, \dodoi{10.1093/mnras/staa244}

\bibitem[{{Bamba} {et~al.}(2008){Bamba}, {Fukazawa}, {Hiraga}, {Hughes},
  {Katagiri}, {Kokubun}, {Koyama}, {Miyata}, {Mizuno}, {Mori}, {Nakajima},
  {Ozaki}, {Petre}, {Takahashi}, {Takahashi}, {Tanaka}, {Terada}, {Uchiyama},
  {Watanabe}, \& {Yamaguchi}}]{2008PASJ...60S.153B}
{Bamba}, A., {Fukazawa}, Y., {Hiraga}, J.~S., {et~al.} 2008, \pasj, 60, S153,
  \dodoi{10.1093/pasj/60.sp1.S153}

\bibitem[{{Bamba} {et~al.}(2018){Bamba}, {Ohira}, {Yamazaki}, {Sawada},
  {Terada}, {Koyama}, {Miller}, {Yamaguchi}, {Katsuda}, {Nobukawa}, \&
  {Nobukawa}}]{2018ApJ...854...71B}
{Bamba}, A., {Ohira}, Y., {Yamazaki}, R., {et~al.} 2018, \apj, 854, 71,
  \dodoi{10.3847/1538-4357/aaa5a0}

\bibitem[{{Bell} {et~al.}(2013){Bell}, {Schure}, {Reville}, \&
  {Giacinti}}]{2013MNRAS.431..415B}
{Bell}, A.~R., {Schure}, K.~M., {Reville}, B., \& {Giacinti}, G. 2013, \mnras,
  431, 415, \dodoi{10.1093/mnras/stt179}

\bibitem[{{Bocchino} {et~al.}(2000){Bocchino}, {Vink}, {Favata}, {Maggio}, \&
  {Sciortino}}]{2000AA...360..671B}
{Bocchino}, F., {Vink}, J., {Favata}, F., {Maggio}, A., \& {Sciortino}, S.
  2000, \aap, 360, 671.
\newblock \doarXiv{astro-ph/0005369}

\bibitem[{{Braun} {et~al.}(2019){Braun}, {Safi-Harb}, \&
  {Fryer}}]{2019MNRAS.489.4444B}
{Braun}, C., {Safi-Harb}, S., \& {Fryer}, C.~L. 2019, \mnras, 489, 4444,
  \dodoi{10.1093/mnras/stz2437}

\bibitem[{{Butt} {et~al.}(2008){Butt}, {Porter}, {Katz}, \&
  {Waxman}}]{2008MNRAS.386L..20B}
{Butt}, Y.~M., {Porter}, T.~A., {Katz}, B., \& {Waxman}, E. 2008, \mnras, 386,
  L20, \dodoi{10.1111/j.1745-3933.2008.00452.x}

\bibitem[{{Celli} {et~al.}(2019){Celli}, {Morlino}, {Gabici}, \&
  {Aharonian}}]{2019MNRAS.487.3199C}
{Celli}, S., {Morlino}, G., {Gabici}, S., \& {Aharonian}, F.~A. 2019, \mnras,
  487, 3199, \dodoi{10.1093/mnras/stz1425}

\bibitem[{{Chang} {et~al.}(2008){Chang}, {Konopelko}, \&
  {Cui}}]{2008ApJ...682.1177C}
{Chang}, C., {Konopelko}, A., \& {Cui}, W. 2008, \apj, 682, 1177,
  \dodoi{10.1086/589225}

\bibitem[{{Clark} {et~al.}(1975){Clark}, {Caswell}, \&
  {Green}}]{1975AuJPA..37....1C}
{Clark}, D.~H., {Caswell}, J.~L., \& {Green}, A.~J. 1975, Australian Journal of
  Physics Astrophysical Supplement, 37, 1

\bibitem[{{Cohen}(2016)}]{2016PhDT.......190C}
{Cohen}, J.~M. 2016, PhD thesis, University of Maryland, College Park

\bibitem[{{Condon} {et~al.}(2017){Condon}, {Lemoine-Goumard}, {Acero}, \&
  {Katagiri}}]{2017ApJ...851..100C}
{Condon}, B., {Lemoine-Goumard}, M., {Acero}, F., \& {Katagiri}, H. 2017, \apj,
  851, 100, \dodoi{10.3847/1538-4357/aa9be8}

\bibitem[{{DAMPE Collaboration} {et~al.}(2017){DAMPE Collaboration}, {Ambrosi},
  {An}, {Asfand iyarov}, {Azzarello}, {Bernardini}, {Bertucci}, {Cai}, {Chang},
  {Chen}, {Chen}, {Chen}, {Chen}, {Cui}, {Cui}, {D'Amone}, {de Benedittis}, {De
  Mitri}, {di Santo}, {Dong}, {Dong}, {Dong}, {Dong}, {Donvito}, {Droz},
  {Duan}, {Duan}, {Duranti}, {D'Urso}, {Fan}, {Fan}, {Fang}, {Feng}, {Feng},
  {Fusco}, {Gallo}, {Gan}, {Gao}, {Gao}, {Gargano}, {Garrappa}, {Gong}, {Gong},
  {Guo}, {Guo}, {Hu}, {Huang}, {Huang}, {Ionica}, {Jiang}, {Jiang}, {Jin},
  {Kong}, {Lei}, {Li}, {Li}, {Li}, {Li}, {Liang}, {Liang}, {Liao}, {Liu},
  {Liu}, {Liu}, {Liu}, {Liu}, {Loparco}, {Ma}, {Ma}, {Ma}, {Ma}, {Ma}, {Ma},
  {Marsella}, {Mazziotta}, {Mo}, {Niu}, {Peng}, {Peng}, {Qiao}, {Rao},
  {Salinas}, {Shang}, {H. Shen}, {Shen}, {Shen}, {Song}, {Su}, {Su}, {Sun},
  {Surdo}, {Teng}, {Tian}, {Tykhonov}, {Vagelli}, {Vitillo}, {Wang}, {Wang},
  {Wang}, {Wang}, {Wang}, {Wang}, {Wang}, {Wang}, {Wang}, {Wang}, {Wang},
  {Wang}, {Wen}, {Wang}, {Wei}, {Wei}, {Wei}, {Wu}, {Wu}, {Wu}, {Wu}, {Wu},
  {Xi}, {Xia}, {Xin}, {Xu}, {Xu}, {Xu}, {Xue}, {Yang}, {Yang}, {Yang}, {Yang},
  {Yao}, {Yu}, {Yuan}, {Yue}, {Zang}, {Zhang}, {Zhang}, {Zhang}, {Zhang},
  {Zhang}, {Zhang}, {Zhang}, {Zhang}, {Zhang}, {Zhang}, {Zhang}, {Zhang},
  {Zhang}, {Zhang}, {Zhang}, {Zhang}, {Zhang}, {Zhao}, {Zhao}, {Zhao}, {Zhou},
  {Zhou}, {Zhu}, {Zhu}, \& {Zimmer}}]{2017Natur.552...63D}
{DAMPE Collaboration}, {Ambrosi}, G., {An}, Q., {et~al.} 2017, \nat, 552, 63,
  \dodoi{10.1038/nature24475}

\bibitem[{{DAMPE Collaboration} {et~al.}(2019){DAMPE Collaboration}, An,
  Asfandiyarov, Azzarello, Bernardini, Bi, Cai, Chang, Chen, Chen, Chen, Chen,
  Cui, Cui, Dai, D{\textquoteright}Amone, De~Benedittis, De~Mitri, Di~Santo,
  Ding, Dong, Dong, Dong, Donvito, Droz, Duan, Duan, D{\textquoteright}Urso,
  Fan, Fan, Fang, Feng, Feng, Fusco, Gallo, Gan, Gao, Gargano, Gong, Gong, Guo,
  Guo, Guo, Han, Hu, Huang, Huang, Huang, Ionica, Jiang, Jin, Kong, Lei, Li,
  Li, Li, Li, Li, Liang, Liang, Liao, Liu, Liu, Liu, Liu, Liu, Liu, Loparco,
  Luo, Ma, Ma, Ma, Ma, Ma, Marsella, Mazziotta, Mo, Niu, Pan, Peng, Peng, Qiao,
  Rao, Salinas, Shang, Shen, Shen, Shen, Song, Su, Su, Sun, Surdo, Teng,
  Tykhonov, Vitillo, Wang, Wang, Wang, Wang, Wang, Wang, Wang, Wang, Wang,
  Wang, Wang, Wang, Wang, Wei, Wei, Wei, Wen, Wu, Wu, Wu, Wu, Wu, Xi, Xia, Xu,
  Xu, Xu, Xu, Xue, Yang, Yang, Yang, Yang, Yao, Yu, Yuan, Yue, Zang, Zhang,
  Zhang, Zhang, Zhang, Zhang, Zhang, Zhang, Zhang, Zhang, Zhang, Zhang, Zhang,
  Zhang, Zhao, Zhao, Zhao, Zhou, Zhou, Zhu, Zhu, \& Zimmer}]{eaax3793}
{DAMPE Collaboration}, An, Q., Asfandiyarov, R., {et~al.} 2019, Science
  Advances, 5, \dodoi{10.1126/sciadv.aax3793}

\bibitem[{{De Luca} \& {Molendi}(2004)}]{2004A&A...419..837D}
{De Luca}, A., \& {Molendi}, S. 2004, \aap, 419, 837,
  \dodoi{10.1051/0004-6361:20034421}

\bibitem[{{Devin} {et~al.}(2020){Devin}, {Lemoine-Goumard}, {Grondin},
  {Castro}, {Ballet}, {Cohen}, \& {Hewitt}}]{2020arXiv200908397D}
{Devin}, J., {Lemoine-Goumard}, M., {Grondin}, M.-H., {et~al.} 2020, arXiv
  e-prints, arXiv:2009.08397.
\newblock \doarXiv{2009.08397}

\bibitem[{{Dickel} {et~al.}(1996){Dickel}, {Green}, {Ye}, \&
  {Milne}}]{1996AJ....111..340D}
{Dickel}, J.~R., {Green}, A., {Ye}, T., \& {Milne}, D.~K. 1996, \aj, 111, 340,
  \dodoi{10.1086/117786}

\bibitem[{{Dickel} \& {Milne}(1995)}]{1995AJ....109..200D}
{Dickel}, J.~R., \& {Milne}, D.~K. 1995, \aj, 109, 200, \dodoi{10.1086/117266}

\bibitem[{{Doroshenko} {et~al.}(2017){Doroshenko}, {P{\"u}hlhofer}, {Bamba},
  {Acero}, {Tian}, {Klochkov}, \& {Santangelo}}]{2017A&A...608A..23D}
{Doroshenko}, V., {P{\"u}hlhofer}, G., {Bamba}, A., {et~al.} 2017, \aap, 608,
  A23, \dodoi{10.1051/0004-6361/201730983}

\bibitem[{{Duncan} \& {Green}(2000)}]{2000A&A...364..732D}
{Duncan}, A.~R., \& {Green}, D.~A. 2000, \aap, 364, 732.
\newblock \doarXiv{astro-ph/0009289}

\bibitem[{{Duncan} {et~al.}(1995){Duncan}, {Haynes}, {Stewart}, \&
  {Jones}}]{1995MNRAS.277..319D}
{Duncan}, A.~R., {Haynes}, R.~F., {Stewart}, R.~T., \& {Jones}, K.~L. 1995,
  \mnras, 277, 319, \dodoi{10.1093/mnras/277.1.319}

\bibitem[{{Dyer} {et~al.}(2009){Dyer}, {Cornwell}, \&
  {Maddalena}}]{2009AJ....137.2956D}
{Dyer}, K.~K., {Cornwell}, T.~J., \& {Maddalena}, R.~J. 2009, \aj, 137, 2956,
  \dodoi{10.1088/0004-6256/137/2/2956}

\bibitem[{{Fleischhack}(2019)}]{2019ICRC...36..675F}
{Fleischhack}, H. 2019, in International Cosmic Ray Conference, Vol.~36, 36th
  International Cosmic Ray Conference (ICRC2019), 675.
\newblock \doarXiv{1907.08572}

\bibitem[{{Fukuda} {et~al.}(2014){Fukuda}, {Yoshiike}, {Sano}, {Torii},
  {Yamamoto}, {Acero}, \& {Fukui}}]{2014ApJ...788...94F}
{Fukuda}, T., {Yoshiike}, S., {Sano}, H., {et~al.} 2014, \apj, 788, 94,
  \dodoi{10.1088/0004-637X/788/1/94}

\bibitem[{{Fukui} {et~al.}(2003){Fukui}, {Moriguchi}, {Tamura}, {Yamamoto},
  {Tawara}, {Mizuno}, {Onishi}, {Mizuno}, {Uchiyama}, {Hiraga}, {Takahashi},
  {Yamashita}, \& {Ikeuchi}}]{2003PASJ...55L..61F}
{Fukui}, Y., {Moriguchi}, Y., {Tamura}, K., {et~al.} 2003, \pasj, 55, L61,
  \dodoi{10.1093/pasj/55.5.L61}

\bibitem[{{Fukui} {et~al.}(2012){Fukui}, {Sano}, {Sato}, {Torii}, {Horachi},
  {Hayakawa}, {McClure-Griffiths}, {Rowell}, {Inoue}, {Inutsuka}, {Kawamura},
  {Yamamoto}, {Okuda}, {Mizuno}, {Onishi}, {Mizuno}, \&
  {Ogawa}}]{2012ApJ...746...82F}
{Fukui}, Y., {Sano}, H., {Sato}, J., {et~al.} 2012, \apj, 746, 82,
  \dodoi{10.1088/0004-637X/746/1/82}

\bibitem[{{Fukui} {et~al.}(2017){Fukui}, {Sano}, {Sato}, {Okamoto}, {Fukuda},
  {Yoshiike}, {Hayashi}, {Torii}, {Hayakawa}, \&
  {Rowell}}]{2017ApJ...850...71F}
---. 2017, \apj, 850, 71, \dodoi{10.3847/1538-4357/aa9219}

\bibitem[{{Gabici} \& {Aharonian}(2014)}]{2014MNRAS.445L..70G}
{Gabici}, S., \& {Aharonian}, F.~A. 2014, \mnras, 445, L70,
  \dodoi{10.1093/mnrasl/slu132}

\bibitem[{{Gao} {et~al.}(2011){Gao}, {Han}, {Reich}, {Reich}, {Sun}, \&
  {Xiao}}]{2011A&A...529A.159G}
{Gao}, X.~Y., {Han}, J.~L., {Reich}, W., {et~al.} 2011, \aap, 529, A159,
  \dodoi{10.1051/0004-6361/201016311}

\bibitem[{{Gerbrandt} {et~al.}(2014){Gerbrandt}, {Foster}, {Kothes},
  {Geisb{\"u}sch}, \& {Tung}}]{2014A&A...566A..76G}
{Gerbrandt}, S., {Foster}, T.~J., {Kothes}, R., {Geisb{\"u}sch}, J., \& {Tung},
  A. 2014, \aap, 566, A76, \dodoi{10.1051/0004-6361/201423679}

\bibitem[{{Giacani} {et~al.}(2000){Giacani}, {Dubner}, {Green}, {Goss}, \&
  {Gaensler}}]{2000AJ....119..281G}
{Giacani}, E.~B., {Dubner}, G.~M., {Green}, A.~J., {Goss}, W.~M., \&
  {Gaensler}, B.~M. 2000, \aj, 119, 281, \dodoi{10.1086/301173}

\bibitem[{Giuliani {et~al.}(2011)Giuliani, Cardillo, Tavani, Fukui, Yoshiike,
  Torii, Dubner, Castelletti, Barbiellini, Bulgarelli, Caraveo, Costa,
  Cattaneo, Chen, Contessi, Monte, Donnarumma, Evangelista, Feroci, Gianotti,
  Lazzarotto, Lucarelli, Longo, Marisaldi, Mereghetti, Pacciani, Pellizzoni,
  Piano, Picozza, Pittori, Pucella, Rapisarda, Rappoldi, Sabatini, Soffitta,
  Striani, Trifoglio, Trois, Vercellone, Verrecchia, Vittorini, Colafrancesco,
  Giommi, \& Bignami}]{Giuliani_2011}
Giuliani, A., Cardillo, M., Tavani, M., {et~al.} 2011, The Astrophysical
  Journal, 742, L30, \dodoi{10.1088/2041-8205/742/2/l30}

\bibitem[{{Guo} {et~al.}(2018){Guo}, {Xin}, {Liao}, {Yuan}, {Gao}, \&
  {Fan}}]{2018ApJ...853....2G}
{Guo}, X.-L., {Xin}, Y.-L., {Liao}, N.-H., {et~al.} 2018, \apj, 853, 2,
  \dodoi{10.3847/1538-4357/aaa3f8}

\bibitem[{{H.~E.~S.~S. Collaboration} {et~al.}(2011{\natexlab{a}}){H.~E.~S.~S.
  Collaboration}, {Abramowski}, {Acero}, {Aharonian}, {Akhperjanian}, {Anton},
  {Balzer}, {Barnacka}, {Barres de Almeida}, {Becherini}, {Becker}, {Behera},
  {Bernl{\"o}hr}, {Bochow}, {Boisson}, {Bolmont}, {Bordas}, {Brucker}, {Brun},
  {Brun}, {Bulik}, {B{\"u}sching}, {Carrigan}, {Casanova}, {Cerruti},
  {Chadwick}, {Charbonnier}, {Chaves}, {Cheesebrough}, {Chounet}, {Clapson},
  {Coignet}, {Cologna}, {Conrad}, {Dalton}, {Daniel}, {Davids}, {Degrange},
  {Deil}, {Dickinson}, {Djannati-Ata{\"\i}}, {Domainko}, {Drury}, {Dubois},
  {Dubus}, {Dutson}, {Dyks}, {Dyrda}, {Egberts}, {Eger}, {Espigat}, {Fallon},
  {Farnier}, {Fegan}, {Feinstein}, {Fernandes}, {Fiasson}, {Fontaine},
  {F{\"o}rster}, {F{\"u}{\ss}ling}, {Gallant}, {Gast}, {G{\'e}rard}, {Gerbig},
  {Giebels}, {Glicenstein}, {Gl{\"u}ck}, {Goret}, {G{\"o}ring}, {H{\"a}ffner},
  {Hague}, {Hampf}, {Hauser}, {Heinz}, {Heinzelmann}, {Henri}, {Hermann},
  {Hinton}, {Hoffmann}, {Hofmann}, {Hofverberg}, {Holler}, {Horns},
  {Jacholkowska}, {de Jager}, {Jahn}, {Jamrozy}, {Jung}, {Kastendieck},
  {Katarzy{\'n}ski}, {Katz}, {Kaufmann}, {Keogh}, {Khangulyan}, {Kh{\'e}lifi},
  {Klochkov}, {Klu{\'z}niak}, {Kneiske}, {Komin}, {Kosack}, {Kossakowski},
  {Laffon}, {Lamanna}, {Lennarz}, {Lohse}, {Lopatin}, {Lu}, {Marandon},
  {Marcowith}, {Masbou}, {Maurin}, {Maxted}, {McComb}, {Medina}, {M{\'e}hault},
  {Moderski}, {Moulin}, {Naumann}, {Naumann-Godo}, {de Naurois}, {Nedbal},
  {Nekrassov}, {Nguyen}, {Nicholas}, {Niemiec}, {Nolan}, {Ohm}, {de O{\~n}a
  Wilhelmi}, {Opitz}, {Ostrowski}, {Oya}, {Panter}, {Paz Arribas}, {Pedaletti},
  {Pelletier}, {Petrucci}, {Pita}, {P{\"u}hlhofer}, {Punch}, {Quirrenbach},
  {Raue}, {Rayner}, {Reimer}, {Reimer}, {Renaud}, {de los Reyes}, {Rieger},
  {Ripken}, {Rob}, {Rosier-Lees}, {Rowell}, {Rudak}, {Rulten}, {Ruppel},
  {Ryde}, {Sahakian}, {Santangelo}, {Schlickeiser}, {Sch{\"o}ck}, {Schulz},
  {Schwanke}, {Schwarzburg}, {Schwemmer}, {Sikora}, {Skilton}, {Sol},
  {Spengler}, {Stawarz}, {Steenkamp}, {Stegmann}, {Stinzing}, {Stycz},
  {Sushch}, {Szostek}, {Tavernet}, {Terrier}, {Tluczykont}, {Valerius}, {van
  Eldik}, {Vasileiadis}, {Venter}, {Vialle}, {Viana}, {Vincent}, {V{\"o}lk},
  {Volpe}, {Vorobiov}, {Vorster}, {Wagner}, {Ward}, {White}, {Wierzcholska},
  {Zacharias}, {Zajczyk}, {Zdziarski}, {Zech}, \&
  {Zechlin}}]{2011A&A...531A..81H}
{H.~E.~S.~S. Collaboration}, {Abramowski}, A., {Acero}, F., {et~al.}
  2011{\natexlab{a}}, \aap, 531, A81, \dodoi{10.1051/0004-6361/201016425}

\bibitem[{{H.~E.~S.~S. Collaboration} {et~al.}(2011{\natexlab{b}}){H.~E.~S.~S.
  Collaboration}, {Abramowski}, {Acero}, {Aharonian}, {Akhperjanian}, {Anton},
  {Balzer}, {Barnacka}, {Barres de Almeida}, {Becherini}, {Becker}, {Behera},
  {Bernl{\"o}hr}, {Bochow}, {Boisson}, {Bolmont}, {Bordas}, {Brucker}, {Brun},
  {Brun}, {Bulik}, {B{\"u}sching}, {Carrigan}, {Casanova}, {Cerruti},
  {Chadwick}, {Charbonnier}, {Chaves}, {Cheesebrough}, {Chounet}, {Clapson},
  {Coignet}, {Cologna}, {Conrad}, {Dalton}, {Daniel}, {Davids}, {Degrange},
  {Deil}, {Dickinson}, {Djannati-Ata{\"\i}}, {Domainko}, {Drury}, {Dubois},
  {Dubus}, {Dutson}, {Dyks}, {Dyrda}, {Egberts}, {Eger}, {Espigat}, {Fallon},
  {Farnier}, {Fegan}, {Feinstein}, {Fernandes}, {Fiasson}, {Fontaine},
  {F{\"o}rster}, {F{\"u}{\ss}ling}, {Gallant}, {Gast}, {G{\'e}rard}, {Gerbig},
  {Giebels}, {Glicenstein}, {Gl{\"u}ck}, {Goret}, {G{\"o}ring}, {H{\"a}ffner},
  {Hague}, {Hampf}, {Hauser}, {Heinz}, {Heinzelmann}, {Henri}, {Hermann},
  {Hinton}, {Hoffmann}, {Hofmann}, {Hofverberg}, {Holler}, {Horns},
  {Jacholkowska}, {de Jager}, {Jahn}, {Jamrozy}, {Jung}, {Kastendieck},
  {Katarzy{\'n}ski}, {Katz}, {Kaufmann}, {Keogh}, {Khangulyan}, {Kh{\'e}lifi},
  {Klochkov}, {Klu{\'z}niak}, {Kneiske}, {Komin}, {Kosack}, {Kossakowski},
  {Laffon}, {Lamanna}, {Lennarz}, {Lohse}, {Lopatin}, {Lu}, {Marandon},
  {Marcowith}, {Masbou}, {Maurin}, {Maxted}, {McComb}, {Medina}, {M{\'e}hault},
  {Moderski}, {Moulin}, {Naumann}, {Naumann-Godo}, {de Naurois}, {Nedbal},
  {Nekrassov}, {Nguyen}, {Nicholas}, {Niemiec}, {Nolan}, {Ohm}, {de O{\~n}a
  Wilhelmi}, {Opitz}, {Ostrowski}, {Oya}, {Panter}, {Paz Arribas}, {Pedaletti},
  {Pelletier}, {Petrucci}, {Pita}, {P{\"u}hlhofer}, {Punch}, {Quirrenbach},
  {Raue}, {Rayner}, {Reimer}, {Reimer}, {Renaud}, {de los Reyes}, {Rieger},
  {Ripken}, {Rob}, {Rosier-Lees}, {Rowell}, {Rudak}, {Rulten}, {Ruppel},
  {Ryde}, {Sahakian}, {Santangelo}, {Schlickeiser}, {Sch{\"o}ck}, {Schulz},
  {Schwanke}, {Schwarzburg}, {Schwemmer}, {Sikora}, {Skilton}, {Sol},
  {Spengler}, {Stawarz}, {Steenkamp}, {Stegmann}, {Stinzing}, {Stycz},
  {Sushch}, {Szostek}, {Tavernet}, {Terrier}, {Tluczykont}, {Valerius}, {van
  Eldik}, {Vasileiadis}, {Venter}, {Vialle}, {Viana}, {Vincent}, {V{\"o}lk},
  {Volpe}, {Vorobiov}, {Vorster}, {Wagner}, {Ward}, {White}, {Wierzcholska},
  {Zacharias}, {Zajczyk}, {Zdziarski}, {Zech}, \&
  {Zechlin}}]{2011AA...531A..81H}
---. 2011{\natexlab{b}}, \aap, 531, A81, \dodoi{10.1051/0004-6361/201016425}

\bibitem[{{H.~E.~S.~S. Collaboration} {et~al.}(2015){H.~E.~S.~S.
  Collaboration}, {Abramowski}, {Aharonian}, {Ait Benkhali}, {Akhperjanian},
  {Ang{\"u}ner}, {Backes}, {Balenderan}, {Balzer}, {Barnacka}, {Becherini},
  {Becker-Tjus}, {Berge}, {Bernhard}, {Bernl{\"o}hr}, {Birsin}, {Biteau},
  {B{\"o}ttcher}, {Boisson}, {Bolmont}, {Bordas}, {Bregeon}, {Brun}, {Brun},
  {Bryan}, {Bulik}, {Carrigan}, {Casanova}, {Chadwick}, {Chakraborty},
  {Chalme-Calvet}, {Chaves}, {Chr{\'e}tien}, {Colafrancesco}, {Cologna},
  {Conrad}, {Couturier}, {Cui}, {Dalton}, {Davids}, {Degrange}, {Deil}, {de
  Wilt}, {Djannati-Ata{\"\i}}, {Domainko}, {Donath}, {Drury}, {Dubus},
  {Dutson}, {Dyks}, {Dyrda}, {Edwards}, {Egberts}, {Eger}, {Espigat},
  {Farnier}, {Fegan}, {Feinstein}, {Fernandes}, {Fernandez}, {Fiasson},
  {Fontaine}, {F{\"o}rster}, {F{\"u}{\ss}ling}, {Gabici}, {Gajdus}, {Gallant},
  {Garrigoux}, {Giavitto}, {Giebels}, {Glicenstein}, {Gottschall}, {Grondin},
  {Grudzi{\'n}ska}, {Hadasch}, {H{\"a}ffner}, {Hahn}, {Harris}, {Heinzelmann},
  {Henri}, {Hermann}, {Hervet}, {Hillert}, {Hinton}, {Hofmann}, {Hofverberg},
  {Holler}, {Horns}, {Ivascenko}, {Jacholkowska}, {Jahn}, {Jamrozy}, {Janiak},
  {Jankowsky}, {Jung}, {Kastendieck}, {Katarzy{\'n}ski}, {Katz}, {Kaufmann},
  {Kh{\'e}lifi}, {Kieffer}, {Klepser}, {Klochkov}, {Klu{\'z}niak}, {Kolitzus},
  {Komin}, {Kosack}, {Krakau}, {Krayzel}, {Kr{\"u}ger}, {Laffon}, {Lamanna},
  {Lefaucheur}, {Lefranc}, {Lemi{\`e}re}, {Lemoine-Goumard}, {Lenain}, {Lohse},
  {Lopatin}, {Lu}, {Marandon}, {Marcowith}, {Marx}, {Maurin}, {Maxted},
  {Mayer}, {McComb}, {M{\'e}hault}, {Meintjes}, {Menzler}, {Meyer}, {Mitchell},
  {Moderski}, {Mohamed}, {Mor{\r{a}}}, {Moulin}, {Murach}, {de Naurois},
  {Niemiec}, {Nolan}, {Oakes}, {Odaka}, {Ohm}, {Opitz}, {Ostrowski}, {Oya},
  {Panter}, {Parsons}, {Paz Arribas}, {Pekeur}, {Pelletier}, {Perez},
  {Petrucci}, {Peyaud}, {Pita}, {Poon}, {P{\"u}hlhofer}, {Punch},
  {Quirrenbach}, {Raab}, {Reichardt}, {Reimer}, {Reimer}, {Renaud}, {de los
  Reyes}, {Rieger}, {Rob}, {Romoli}, {Rosier-Lees}, {Rowell}, {Rudak},
  {Rulten}, {Sahakian}, {Salek}, {Sanchez}, {Santangelo}, {Schlickeiser},
  {Sch{\"u}ssler}, {Schulz}, {Schwanke}, {Schwarzburg}, {Schwemmer}, {Sol},
  {Spanier}, {Spengler}, {Spies}, {Stawarz}, {Steenkamp}, {Stegmann},
  {Stinzing}, {Stycz}, {Sushch}, {Tavernet}, {Tavernier}, {Taylor}, {Terrier},
  {Tluczykont}, {Trichard}, {Valerius}, {van Eldik}, {van Soelen},
  {Vasileiadis}, {Veh}, {Venter}, {Viana}, {Vincent}, {Vink}, {V{\"o}lk},
  {Volpe}, {Vorster}, {Vuillaume}, {Wagner}, {Wagner}, {Wagner}, {Ward},
  {Weidinger}, {Weitzel}, {White}, {Wierzcholska}, {Willmann}, {W{\"o}rnlein},
  {Wouters}, {Yang}, {Zabalza}, {Zaborov}, {Zacharias}, {Zdziarski}, {Zech}, \&
  {Zechlin}}]{2015Sci...347..406H}
{H.~E.~S.~S. Collaboration}, {Abramowski}, A., {Aharonian}, F., {et~al.} 2015,
  Science, 347, 406, \dodoi{10.1126/science.1261313}

\bibitem[{{H.~E.~S.~S. Collaboration} {et~al.}(2018{\natexlab{a}}){H.~E.~S.~S.
  Collaboration}, {Abdalla}, {Abramowski}, {Aharonian}, {Ait Benkhali},
  {Akhperjanian}, {Andersson}, {Ang{\"u}ner}, {Arrieta}, {Aubert}, {Backes},
  {Balzer}, {Barnard}, {Becherini}, {Becker Tjus}, {Berge}, {Bernhard},
  {Bernl{\"o}hr}, {Blackwell}, {B{\"o}ttcher}, {Boisson}, {Bolmont}, {Bordas},
  {Bregeon}, {Brun}, {Brun}, {Bryan}, {Bulik}, {Capasso}, {Carr}, {Casanova},
  {Cerruti}, {Chakraborty}, {Chalme-Calvet}, {Chaves}, {Chen}, {Chevalier},
  {Chr{\'e}tien}, {Colafrancesco}, {Cologna}, {Condon}, {Conrad}, {Cui},
  {Davids}, {Decock}, {Degrange}, {Deil}, {Devin}, {deWilt}, {Dirson},
  {Djannati-Ata{\"\i}}, {Domainko}, {Donath}, {Drury}, {Dubus}, {Dutson},
  {Dyks}, {Edwards}, {Egberts}, {Eger}, {Ernenwein}, {Eschbach}, {Farnier},
  {Fegan}, {Fernand es}, {Fiasson}, {Fontaine}, {F{\"o}rster}, {Fukuyama},
  {Funk}, {F{\"u}{\ss}ling}, {Gabici}, {Gajdus}, {Gallant}, {Garrigoux},
  {Giavitto}, {Giebels}, {Glicenstein}, {Gottschall}, {Goyal}, {Grondin},
  {Hadasch}, {Hahn}, {Haupt}, {Hawkes}, {Heinzelmann}, {Henri}, {Hermann},
  {Hervet}, {Hinton}, {Hofmann}, {Hoischen}, {Holler}, {Horns}, {Ivascenko},
  {Jacholkowska}, {Jamrozy}, {Janiak}, {Jankowsky}, {Jankowsky}, {Jingo},
  {Jogler}, {Jouvin}, {Jung-Richardt}, {Kastendieck}, {Katarzy{\'n}ski},
  {Katz}, {Kerszberg}, {Kh{\'e}lifi}, {Kieffer}, {King}, {Klepser}, {Klochkov},
  {Klu{\'z}niak}, {Kolitzus}, {Komin}, {Kosack}, {Krakau}, {Kraus}, {Krayzel},
  {Kr{\"u}ger}, {Laffon}, {Lamanna}, {Lau}, {Lees}, {Lefaucheur}, {Lefranc},
  {Lemi{\`e}re}, {Lemoine-Goumard}, {Lenain}, {Leser}, {Lohse}, {Lorentz},
  {Liu}, {L{\'o}pez-Coto}, {Lypova}, {Marandon}, {Marcowith}, {Mariaud},
  {Marx}, {Maurin}, {Maxted}, {Mayer}, {Meintjes}, {Meyer}, {Mitchell},
  {Moderski}, {Mohamed}, {Mohrmann}, {Mor{\r{a}}}, {Moulin}, {Murach}, {de
  Naurois}, {Niederwanger}, {Niemiec}, {Oakes}, {O'Brien}, {Odaka}, {{\"O}ttl},
  {Ohm}, {Ostrowski}, {Oya}, {Padovani}, {Panter}, {Parsons}, {Pekeur},
  {Pelletier}, {Perennes}, {Petrucci}, {Peyaud}, {Piel}, {Pita}, {Poon},
  {Prokhorov}, {Prokoph}, {P{\"u}hlhofer}, {Punch}, {Quirrenbach}, {Raab},
  {Reimer}, {Reimer}, {Renaud}, {de los Reyes}, {Rieger}, {Romoli},
  {Rosier-Lees}, {Rowell}, {Rudak}, {Rulten}, {Sahakian}, {Salek}, {Sanchez},
  {Santangelo}, {Sasaki}, {Schlickeiser}, {Sch{\"u}ssler}, {Schulz},
  {Schwanke}, {Schwemmer}, {Settimo}, {Seyffert}, {Shafi}, {Shilon}, {Simoni},
  {Sol}, {Spanier}, {Spengler}, {Spies}, {Stawarz}, {Steenkamp}, {Stegmann},
  {Stinzing}, {Stycz}, {Sushch}, {Takahashi}, {Tavernet}, {Tavernier},
  {Taylor}, {Terrier}, {Tibaldo}, {Tiziani}, {Tluczykont}, {Trichard}, {Tuffs},
  {Uchiyama}, {van der Walt}, {van Eldik}, {van Rensburg}, {van Soelen},
  {Vasileiadis}, {Veh}, {Venter}, {Viana}, {Vincent}, {Vink}, {Voisin},
  {V{\"o}lk}, {Volpe}, {Vuillaume}, {Wadiasingh}, {Wagner}, {Wagner}, {Wagner},
  {White}, {Wierzcholska}, {Willmann}, {W{\"o}rnlein}, {Wouters}, {Yang},
  {Zabalza}, {Zaborov}, {Zacharias}, {Zdziarski}, {Zech}, {Zefi}, {Ziegler}, \&
  {{\.Z}ywucka}}]{2018A&A...612A...6H}
{H.~E.~S.~S. Collaboration}, {Abdalla}, H., {Abramowski}, A., {et~al.}
  2018{\natexlab{a}}, \aap, 612, A6, \dodoi{10.1051/0004-6361/201629790}

\bibitem[{{H.~E.~S.~S. Collaboration} {et~al.}(2018{\natexlab{b}}){H.~E.~S.~S.
  Collaboration}, {Abdalla}, {Abramowski}, {Aharonian}, {Ait Benkhali},
  {Akhperjanian}, {Andersson}, {Ang{\"u}ner}, {Arakawa}, {Arrieta}, {Aubert},
  {Backes}, {Balzer}, {Barnard}, {Becherini}, {Becker Tjus}, {Berge},
  {Bernhard}, {Bernl{\"o}hr}, {Blackwell}, {B{\"o}ttcher}, {Boisson},
  {Bolmont}, {Bonnefoy}, {Bordas}, {Bregeon}, {Brun}, {Brun}, {Bryan},
  {B{\"u}chele}, {Bulik}, {Capasso}, {Carr}, {Casanova}, {Cerruti},
  {Chakraborty}, {Chaves}, {Chen}, {Chevalier}, {Coffaro}, {Colafrancesco},
  {Cologna}, {Condon}, {Conrad}, {Cui}, {Davids}, {Decock}, {Degrange}, {Deil},
  {Devin}, {deWilt}, {Dirson}, {Djannati-Ata{\"\i}}, {Domainko}, {Donath},
  {Drury}, {Dutson}, {Dyks}, {Edwards}, {Egberts}, {Eger}, {Ernenwein},
  {Eschbach}, {Farnier}, {Fegan}, {Fernand es}, {Fiasson}, {Fontaine},
  {F{\"o}rster}, {Funk}, {F{\"u}{\ss}ling}, {Gabici}, {Gajdus}, {Gallant},
  {Garrigoux}, {Giavitto}, {Giebels}, {Glicenstein}, {Gottschall}, {Goyal},
  {Grondin}, {Hahn}, {Haupt}, {Hawkes}, {Heinzelmann}, {Henri}, {Hermann},
  {Hervet}, {Hinton}, {Hofmann}, {Hoischen}, {Holch}, {Holler}, {Horns},
  {Ivascenko}, {Iwasaki}, {Jacholkowska}, {Jamrozy}, {Janiak}, {Jankowsky},
  {Jankowsky}, {Jingo}, {Jogler}, {Jouvin}, {Jung-Richardt}, {Kastendieck},
  {Katarzy{\'n}ski}, {Katsuragawa}, {Katz}, {Kerszberg}, {Khangulyan},
  {Kh{\'e}lifi}, {King}, {Klepser}, {Klochkov}, {Klu{\'z}niak}, {Kolitzus},
  {Komin}, {Kosack}, {Krakau}, {Kraus}, {Kr{\"u}ger}, {Laffon}, {Lamanna},
  {Lau}, {Lees}, {Lefaucheur}, {Lefranc}, {Lemi{\`e}re}, {Lemoine-Goumard},
  {Lenain}, {Leser}, {Lohse}, {Lorentz}, {Liu}, {L{\'o}pez-Coto}, {Lypova},
  {Marandon}, {Marcowith}, {Mariaud}, {Marx}, {Maurin}, {Maxted}, {Mayer},
  {Meintjes}, {Meyer}, {Mitchell}, {Moderski}, {Mohamed}, {Mohrmann},
  {Mor{\r{a}}}, {Moulin}, {Murach}, {Nakashima}, {de Naurois}, {Niederwanger},
  {Niemiec}, {Oakes}, {O'Brien}, {Odaka}, {{\"O}ttl}, {Ohm}, {Ostrowski},
  {Oya}, {Padovani}, {Panter}, {Parsons}, {Pekeur}, {Pelletier}, {Perennes},
  {Petrucci}, {Peyaud}, {Piel}, {Pita}, {Poon}, {Prokhorov}, {Prokoph},
  {P{\"u}hlhofer}, {Punch}, {Quirrenbach}, {Raab}, {Reimer}, {Reimer},
  {Renaud}, {de los Reyes}, {Richter}, {Rieger}, {Romoli}, {Rowell}, {Rudak},
  {Rulten}, {Sahakian}, {Saito}, {Salek}, {Sanchez}, {Santangelo}, {Sasaki},
  {Schlickeiser}, {Sch{\"u}ssler}, {Schulz}, {Schwanke}, {Schwemmer},
  {Seglar-Arroyo}, {Settimo}, {Seyffert}, {Shafi}, {Shilon}, {Simoni}, {Sol},
  {Spanier}, {Spengler}, {Spies}, {Stawarz}, {Steenkamp}, {Stegmann}, {Stycz},
  {Sushch}, {Takahashi}, {Tavernet}, {Tavernier}, {Taylor}, {Terrier},
  {Tibaldo}, {Tiziani}, {Tluczykont}, {Trichard}, {Tsuji}, {Tuffs}, {Uchiyama},
  {van der Walt}, {van Eldik}, {van Rensburg}, {van Soelen}, {Vasileiadis},
  {Veh}, {Venter}, {Viana}, {Vincent}, {Vink}, {Voisin}, {V{\"o}lk},
  {Vuillaume}, {Wadiasingh}, {Wagner}, {Wagner}, {Wagner}, {White},
  {Wierzcholska}, {Willmann}, {W{\"o}rnlein}, {Wouters}, {Yang}, {Zabalza},
  {Zaborov}, {Zacharias}, {Zanin}, {Zdziarski}, {Zech}, {Zefi}, {Ziegler},
  {{\.Z}ywucka}, {Bamba}, {Fukui}, {Sano}, \& {Yoshiike}}]{2018A&A...612A...8H}
---. 2018{\natexlab{b}}, \aap, 612, A8, \dodoi{10.1051/0004-6361/201730737}

\bibitem[{{H.~E.~S.~S. Collaboration} {et~al.}(2018{\natexlab{c}}){H.~E.~S.~S.
  Collaboration}, {Abdalla}, {Abramowski}, {Aharonian}, {Ait Benkhali},
  {Akhperjanian}, {Ang{\"u}ner}, {Arakawa}, {Arrieta}, {Aubert}, {Backes},
  {Balzer}, {Barnard}, {Becherini}, {Becker Tjus}, {Berge}, {Bernhard},
  {Bernl{\"o}hr}, {Blackwell}, {B{\"o}ttcher}, {Boisson}, {Bolmont}, {Bordas},
  {Bregeon}, {Brun}, {Brun}, {Bryan}, {B{\"u}chele}, {Bulik}, {Capasso},
  {Carr}, {Casanova}, {Cerruti}, {Chakraborty}, {Chalme-Calvet}, {Chaves},
  {Chen}, {Chevalier}, {Chr{\'e}tien}, {Coffaro}, {Colafrancesco}, {Cologna},
  {Condon}, {Conrad}, {Cui}, {Davids}, {Decock}, {Degrange}, {Deil}, {Devin},
  {deWilt}, {Dirson}, {Djannati-Ata{\"\i}}, {Domainko}, {Donath}, {Drury},
  {Dutson}, {Dyks}, {Edwards}, {Egberts}, {Eger}, {Ernenwein}, {Eschbach},
  {Farnier}, {Fegan}, {Fernand es}, {Fiasson}, {Fontaine}, {F{\"o}rster},
  {Funk}, {F{\"u}{\ss}ling}, {Gabici}, {Gajdus}, {Gallant}, {Garrigoux},
  {Giavitto}, {Giebels}, {Glicenstein}, {Gottschall}, {Goyal}, {Grondin},
  {Hahn}, {Haupt}, {Hawkes}, {Heinzelmann}, {Henri}, {Hermann}, {Hervet},
  {Hinton}, {Hofmann}, {Hoischen}, {Holler}, {Horns}, {Ivascenko}, {Iwasaki},
  {Jacholkowska}, {Jamrozy}, {Janiak}, {Jankowsky}, {Jankowsky}, {Jingo},
  {Jogler}, {Jouvin}, {Jung-Richardt}, {Kastendieck}, {Katarzy{\'n}ski},
  {Katsuragawa}, {Katz}, {Kerszberg}, {Khangulyan}, {Kh{\'e}lifi}, {Kieffer},
  {King}, {Klepser}, {Klochkov}, {Klu{\'z}niak}, {Kolitzus}, {Komin}, {Krakau},
  {Kraus}, {Kr{\"u}ger}, {Laffon}, {Lamanna}, {Lau}, {Lees}, {Lefaucheur},
  {Lefranc}, {Lemi{\`e}re}, {Lemoine-Goumard}, {Lenain}, {Leser}, {Lohse},
  {Lorentz}, {Liu}, {L{\'o}pez-Coto}, {Lypova}, {Marandon}, {Marcowith},
  {Mariaud}, {Marx}, {Maurin}, {Maxted}, {Mayer}, {Meintjes}, {Meyer},
  {Mitchell}, {Moderski}, {Mohamed}, {Mohrmann}, {Mor{\r{a}}}, {Moulin},
  {Murach}, {Nakashima}, {de Naurois}, {Niederwanger}, {Niemiec}, {Oakes},
  {O'Brien}, {Odaka}, {{\"O}ttl}, {Ohm}, {Ostrowski}, {Oya}, {Padovani},
  {Panter}, {Parsons}, {Paz Arribas}, {Pekeur}, {Pelletier}, {Perennes},
  {Petrucci}, {Peyaud}, {Piel}, {Pita}, {Poon}, {Prokhorov}, {Prokoph},
  {P{\"u}hlhofer}, {Punch}, {Quirrenbach}, {Raab}, {Reimer}, {Reimer},
  {Renaud}, {de los Reyes}, {Richter}, {Rieger}, {Romoli}, {Rowell}, {Rudak},
  {Rulten}, {Sahakian}, {Saito}, {Salek}, {Sanchez}, {Santangelo}, {Sasaki},
  {Schlickeiser}, {Sch{\"u}ssler}, {Schulz}, {Schwanke}, {Schwemmer},
  {Seglar-Arroyo}, {Settimo}, {Seyffert}, {Shafi}, {Shilon}, {Simoni}, {Sol},
  {Spanier}, {Spengler}, {Spies}, {Stawarz}, {Steenkamp}, {Stegmann}, {Stycz},
  {Sushch}, {Takahashi}, {Tavernet}, {Tavernier}, {Taylor}, {Terrier},
  {Tibaldo}, {Tiziani}, {Tluczykont}, {Trichard}, {Tsuji}, {Tuffs}, {Uchiyama},
  {van der Walt}, {van Eldik}, {van Rensburg}, {van Soelen}, {Vasileiadis},
  {Veh}, {Venter}, {Viana}, {Vincent}, {Vink}, {Voisin}, {V{\"o}lk},
  {Vuillaume}, {Wadiasingh}, {Wagner}, {Wagner}, {Wagner}, {White},
  {Wierzcholska}, {Willmann}, {W{\"o}rnlein}, {Wouters}, {Yang}, {Zabalza},
  {Zaborov}, {Zacharias}, {Zanin}, {Zdziarski}, {Zech}, {Zefi}, {Ziegler}, \&
  {{\.Z}ywucka}}]{2018A&A...612A...7H}
---. 2018{\natexlab{c}}, \aap, 612, A7, \dodoi{10.1051/0004-6361/201630002}

\bibitem[{{H.~E.~S.~S. Collaboration} {et~al.}(2018{\natexlab{d}}){H.~E.~S.~S.
  Collaboration}, {Abramowski}, {Aharonian}, {Ait Benkhali}, {Akhperjanian},
  {Ang{\"u}ner}, {Backes}, {Balzer}, {Becherini}, {Becker Tjus}, {Berge},
  {Bernhard}, {Bernl{\"o}hr}, {Birsin}, {Blackwell}, {B{\"o}ttcher}, {Boisson},
  {Bolmont}, {Bordas}, {Bregeon}, {Brun}, {Brun}, {Bryan}, {Bulik}, {Carr},
  {Casanova}, {Chakraborty}, {Chalme-Calvet}, {Chaves}, {Chen}, {Chevalier},
  {Chr{\'e}tien}, {Colafrancesco}, {Cologna}, {Condon}, {Conrad}, {Couturier},
  {Cui}, {Davids}, {Degrange}, {Deil}, {deWilt}, {Djannati-Ata{\"\i}},
  {Domainko}, {Donath}, {Drury}, {Dubus}, {Dutson}, {Dyks}, {Dyrda}, {Edwards},
  {Egberts}, {Eger}, {Ernenwein}, {Espigat}, {Farnier}, {Fegan}, {Feinstein},
  {Fernandes}, {Fernand ez}, {Fiasson}, {Fontaine}, {F{\"o}rster},
  {F{\"u}{\ss}ling}, {Gabici}, {Gajdus}, {Gallant}, {Garrigoux}, {Giavitto},
  {Giebels}, {Glicenstein}, {Gottschall}, {Goyal}, {Grondin}, {Grudzi{\'n}ska},
  {Hadasch}, {H{\"a}ffner}, {Hahn}, {Hawkes}, {Heinzelmann}, {Henri},
  {Hermann}, {Hervet}, {Hillert}, {Hinton}, {Hofmann}, {Hofverberg},
  {Hoischen}, {Holler}, {Horns}, {Ivascenko}, {Jacholkowska}, {Jamrozy},
  {Janiak}, {Jankowsky}, {Jung-Richardt}, {Kastendieck}, {Katarzy{\'n}ski},
  {Katz}, {Kerszberg}, {Kh{\'e}lifi}, {Kieffer}, {Klepser}, {Klochkov},
  {Klu{\'z}niak}, {Kolitzus}, {Komin}, {Kosack}, {Krakau}, {Krayzel},
  {Kr{\"u}ger}, {Laffon}, {Lamanna}, {Lau}, {Lefaucheur}, {Lefranc},
  {Lemi{\`e}re}, {Lemoine-Goumard}, {Lenain}, {Lohse}, {Lopatin}, {Lorentz},
  {Lu}, {Lui}, {Marandon}, {Marcowith}, {Mariaud}, {Marx}, {Maurin}, {Maxted},
  {Mayer}, {Meintjes}, {Menzler}, {Meyer}, {Mitchell}, {Moderski}, {Mohamed},
  {Mor{\r{a}}}, {Moulin}, {Murach}, {de Naurois}, {Niemiec}, {Oakes}, {Odaka},
  {{\"O}ttl}, {Ohm}, {Opitz}, {Ostrowski}, {Oya}, {Panter}, {Parsons}, {Paz
  Arribas}, {Pekeur}, {Pelletier}, {Petrucci}, {Peyaud}, {Pita}, {Poon},
  {Prokhorov}, {Prokoph}, {P{\"u}hlhofer}, {Punch}, {Quirrenbach}, {Raab},
  {Reichardt}, {Reimer}, {Reimer}, {Renaud}, {de los Reyes}, {Rieger},
  {Romoli}, {Rosier-Lees}, {Rowell}, {Rudak}, {Rulten}, {Sahakian}, {Salek},
  {Sanchez}, {Santangelo}, {Sasaki}, {Schlickeiser}, {Sch{\"u}ssler}, {Schulz},
  {Schwanke}, {Schwemmer}, {Seyffert}, {Simoni}, {Sol}, {Spanier}, {Spengler},
  {Spies}, {Stawarz}, {Steenkamp}, {Stegmann}, {Stinzing}, {Stycz}, {Sushch},
  {Tavernet}, {Tavernier}, {Taylor}, {Terrier}, {Tluczykont}, {Trichard},
  {Tuffs}, {Valerius}, {van der Walt}, {van Eldik}, {van Soelen},
  {Vasileiadis}, {Veh}, {Venter}, {Viana}, {Vincent}, {Vink}, {Voisin},
  {V{\"o}lk}, {Vuillaume}, {Wagner}, {Wagner}, {Wagner}, {Weidinger}, {White},
  {Wierzcholska}, {Willmann}, {W{\"o}rnlein}, {Wouters}, {Yang}, {Zabalza},
  {Zaborov}, {Zacharias}, {Zdziarski}, {Zech}, {Zefi}, \&
  {{\.Z}ywucka}}]{2018A&A...612A...4H}
{H.~E.~S.~S. Collaboration}, {Abramowski}, A., {Aharonian}, F., {et~al.}
  2018{\natexlab{d}}, \aap, 612, A4, \dodoi{10.1051/0004-6361/201526545}

\bibitem[{{Helder} {et~al.}(2013){Helder}, {Vink}, {Bamba}, {Bleeker},
  {Burrows}, {Ghavamian}, \& {Yamazaki}}]{2013MNRAS.435..910H}
{Helder}, E.~A., {Vink}, J., {Bamba}, A., {et~al.} 2013, \mnras, 435, 910,
  \dodoi{10.1093/mnras/stt993}

\bibitem[{{Helfand} \& {Becker}(1984)}]{1984Natur.307..215H}
{Helfand}, D.~J., \& {Becker}, R.~H. 1984, \nat, 307, 215,
  \dodoi{10.1038/307215a0}

\bibitem[{{HI4PI Collaboration} {et~al.}(2016){HI4PI Collaboration}, {Ben
  Bekhti}, {Fl{\"o}er}, {Keller}, {Kerp}, {Lenz}, {Winkel}, {Bailin},
  {Calabretta}, {Dedes}, {Ford}, {Gibson}, {Haud}, {Janowiecki}, {Kalberla},
  {Lockman}, {McClure-Griffiths}, {Murphy}, {Nakanishi}, {Pisano}, \&
  {Staveley-Smith}}]{2016A&A...594A.116H}
{HI4PI Collaboration}, {Ben Bekhti}, N., {Fl{\"o}er}, L., {et~al.} 2016, \aap,
  594, A116, \dodoi{10.1051/0004-6361/201629178}

\bibitem[{{Higgs}(1977)}]{1977AJ.....82..329H}
{Higgs}, L.~A. 1977, \aj, 82, 329, \dodoi{10.1086/112054}

\bibitem[{{Hughes} {et~al.}(1998){Hughes}, {Hayashi}, \&
  {Koyama}}]{1998ApJ...505..732H}
{Hughes}, J.~P., {Hayashi}, I., \& {Koyama}, K. 1998, \apj, 505, 732,
  \dodoi{10.1086/306202}

\bibitem[{{Inoue}(2019)}]{2019ApJ...872...46I}
{Inoue}, T. 2019, \apj, 872, 46, \dodoi{10.3847/1538-4357/aafb70}

\bibitem[{{Inoue} {et~al.}(2012){Inoue}, {Yamazaki}, {Inutsuka}, \&
  {Fukui}}]{2012ApJ...744...71I}
{Inoue}, T., {Yamazaki}, R., {Inutsuka}, S.-i., \& {Fukui}, Y. 2012, \apj, 744,
  71, \dodoi{10.1088/0004-637X/744/1/71}

\bibitem[{{Jones}(1968)}]{1968PhRv..167.1159J}
{Jones}, F.~C. 1968, Physical Review, 167, 1159,
  \dodoi{10.1103/PhysRev.167.1159}

\bibitem[{{Katsuda} {et~al.}(2009){Katsuda}, {Petre}, {Long}, {Reynolds},
  {Winkler}, {Mori}, \& {Tsunemi}}]{2009ApJ...692L.105K}
{Katsuda}, S., {Petre}, R., {Long}, K.~S., {et~al.} 2009, \apjl, 692, L105,
  \dodoi{10.1088/0004-637X/692/2/L105}

\bibitem[{{Katsuda} {et~al.}(2008){Katsuda}, {Tsunemi}, \&
  {Mori}}]{2008ApJ...678L..35K}
{Katsuda}, S., {Tsunemi}, H., \& {Mori}, K. 2008, \apjl, 678, L35,
  \dodoi{10.1086/588499}

\bibitem[{{Kellett} {et~al.}(1987){Kellett}, {Branduardi-Raymont}, {Culhane},
  {Mason}, {Mason}, \& {Whitehouse}}]{1987MNRAS.225..199K}
{Kellett}, B.~J., {Branduardi-Raymont}, G., {Culhane}, J.~L., {et~al.} 1987,
  \mnras, 225, 199, \dodoi{10.1093/mnras/225.2.199}

\bibitem[{{Kothes} {et~al.}(2006){Kothes}, {Fedotov}, {Foster}, \&
  {Uyan{\i}ker}}]{2006A&A...457.1081K}
{Kothes}, R., {Fedotov}, K., {Foster}, T.~J., \& {Uyan{\i}ker}, B. 2006, \aap,
  457, 1081, \dodoi{10.1051/0004-6361:20065062}

\bibitem[{{Kuntz} \& {Snowden}(2008)}]{2008A&A...478..575K}
{Kuntz}, K.~D., \& {Snowden}, S.~L. 2008, \aap, 478, 575,
  \dodoi{10.1051/0004-6361:20077912}

\bibitem[{{Lagage} \& {Cesarsky}(1983)}]{1983A&A...125..249L}
{Lagage}, P.~O., \& {Cesarsky}, C.~J. 1983, \aap, 125, 249

\bibitem[{{Lazendic} {et~al.}(2004){Lazendic}, {Slane}, {Gaensler}, {Reynolds},
  {Plucinsky}, \& {Hughes}}]{2004ApJ...602..271L}
{Lazendic}, J.~S., {Slane}, P.~O., {Gaensler}, B.~M., {et~al.} 2004, \apj, 602,
  271, \dodoi{10.1086/380956}

\bibitem[{{Leahy} {et~al.}(2013){Leahy}, {Green}, \&
  {Ranasinghe}}]{2013MNRAS.436..968L}
{Leahy}, D.~A., {Green}, K., \& {Ranasinghe}, S. 2013, \mnras, 436, 968,
  \dodoi{10.1093/mnras/stt1596}

\bibitem[{{Lemoine-Goumard} {et~al.}(2012){Lemoine-Goumard}, {Renaud}, {Vink},
  {Allen}, {Bamba}, {Giordano}, \& {Uchiyama}}]{2012AA...545A..28L}
{Lemoine-Goumard}, M., {Renaud}, M., {Vink}, J., {et~al.} 2012, \aap, 545, A28,
  \dodoi{10.1051/0004-6361/201219896}

\bibitem[{{MAGIC Collaboration} {et~al.}(2020){MAGIC Collaboration}, {Acciari},
  {Ansoldi}, {Antonelli}, {Arbet Engels}, {Baack}, {Babi{\'c}}, {Banerjee},
  {Barres de Almeida}, {Barrio}, {Becerra Gonz{\'a}lez}, {Bednarek},
  {Bellizzi}, {Bernardini}, {Berti}, {Besenrieder}, {Bhattacharyya},
  {Bigongiari}, {Biland}, {Blanch}, {Bonnoli}, {Bo{\v{s}}njak}, {Busetto},
  {Carosi}, {Ceribella}, {Cerruti}, {Chai}, {Chilingarian}, {Cikota}, {Colak},
  {Colin}, {Colombo}, {Contreras}, {Cortina}, {Covino}, {D'Elia}, {Da Vela},
  {Dazzi}, {De Angelis}, {De Lotto}, {Delfino}, {Delgado}, {Depaoli}, {Di
  Pierro}, {Di Venere}, {Do Souto Espi{\~n}eira}, {Dominis Prester}, {Donini},
  {Dorner}, {Doro}, {Elsaesser}, {Fallah Ramazani}, {Fattorini}, {Ferrara},
  {Foffano}, {Fonseca}, {Font}, {Fruck}, {Fukami}, {Garc{\'\i}a L{\'o}pez},
  {Garczarczyk}, {Gasparyan}, {Gaug}, {Giglietto}, {Giordano}, {Gliwny},
  {Godinovi{\'c}}, {Green}, {Hadasch}, {Hahn}, {Herrera}, {Hoang}, {Hrupec},
  {H{\"u}tten}, {Inada}, {Inoue}, {Ishio}, {Iwamura}, {Jouvin}, {Kajiwara},
  {Karjalainen}, {Kerszberg}, {Kobayashi}, {Kubo}, {Kushida}, {Lamastra},
  {Lelas}, {Leone}, {Lindfors}, {Lombardi}, {Longo}, {L{\'o}pez},
  {L{\'o}pez-Coto}, {L{\'o}pez-Oramas}, {Loporchio}, {Machado de Oliveira
  Fraga}, {Masuda}, {Maggio}, {Majumdar}, {Makariev}, {Mallamaci}, {Maneva},
  {Manganaro}, {Mannheim}, {Maraschi}, {Mariotti}, {Mart{\'\i}nez}, {Mazin},
  {Mender}, {Mi{\'c}anovi{\'c}}, {Miceli}, {Miener}, {Minev}, {Miranda},
  {Mirzoyan}, {Molina}, {Moralejo}, {Morcuende}, {Moreno}, {Moretti},
  {Munar-Adrover}, {Neustroev}, {Nigro}, {Nilsson}, {Ninci}, {Nishijima},
  {Noda}, {Nogu{\'e}s}, {Nozaki}, {Ohtani}, {Oka}, {Otero-Santos},
  {Palatiello}, {Paneque}, {Paoletti}, {Paredes}, {Pavleti{\'c}}, {Pe{\~n}il},
  {Peresano}, {Persic}, {Prada Moroni}, {Prandini}, {Puljak}, {Rhode},
  {Rib{\'o}}, {Rico}, {Righi}, {Rugliancich}, {Saha}, {Sahakyan}, {Saito},
  {Sakurai}, {Satalecka}, {Schleicher}, {Schmidt}, {Schweizer}, {Sitarek},
  {{\v{S}}nidari{\'c}}, {Sobczynska}, {Spolon}, {Stamerra}, {Strom}, {Strzys},
  {Suda}, {Suri{\'c}}, {Takahashi}, {Tavecchio}, {Temnikov}, {Terzi{\'c}},
  {Teshima}, {Torres-Alb{\`a}}, {Tosti}, {van Scherpenberg}, {Vanzo}, {Vazquez
  Acosta}, {Ventura}, {Verguilov}, {Vigorito}, {Vitale}, {Vovk}, {Will},
  {Zari{\'c}}, {authors}, {:}, {Celli}, \& {Morlino}}]{2020arXiv201015854M}
{MAGIC Collaboration}, {Acciari}, V.~A., {Ansoldi}, S., {et~al.} 2020, arXiv
  e-prints, arXiv:2010.15854.
\newblock \doarXiv{2010.15854}

\bibitem[{{Maxted} {et~al.}(2012){Maxted}, {Rowell}, {Dawson}, {Burton},
  {Nicholas}, {Fukui}, {Walsh}, {Kawamura}, {Horachi}, \&
  {Sano}}]{2012MNRAS.422.2230M}
{Maxted}, N.~I., {Rowell}, G.~P., {Dawson}, B.~R., {et~al.} 2012, \mnras, 422,
  2230, \dodoi{10.1111/j.1365-2966.2012.20766.x}

\bibitem[{{Maxted} {et~al.}(2018){Maxted}, {Braiding}, {Wong}, {Rowell},
  {Burton}, {Filipovi{\'c}}, {Voisin}, {Uro{\v{s}}evi{\'c}}, {Vukoti{\'c}}, \&
  {Pavlovi{\'c}}}]{2018MNRAS.480..134M}
{Maxted}, N.~I., {Braiding}, C., {Wong}, G.~F., {et~al.} 2018, \mnras, 480,
  134, \dodoi{10.1093/mnras/sty1797}

\bibitem[{{Milne} \& {Haynes}(1994)}]{1994MNRAS.270..106M}
{Milne}, D.~K., \& {Haynes}, R.~F. 1994, \mnras, 270, 106,
  \dodoi{10.1093/mnras/270.1.106}

\bibitem[{{Mori}(2009)}]{2009APh....31..341M}
{Mori}, M. 2009, Astroparticle Physics, 31, 341,
  \dodoi{10.1016/j.astropartphys.2009.03.004}

\bibitem[{{Morris} {et~al.}(2002){Morris}, {Hobbs}, {Lyne}, {Stairs}, {Camilo},
  {Manchester}, {Possenti}, {Bell}, {Kaspi}, {Amico}, {McKay}, {Crawford}, \&
  {Kramer}}]{2002MNRAS.335..275M}
{Morris}, D.~J., {Hobbs}, G., {Lyne}, A.~G., {et~al.} 2002, \mnras, 335, 275,
  \dodoi{10.1046/j.1365-8711.2002.05551.x}

\bibitem[{{Porter} {et~al.}(2006){Porter}, {Moskalenko}, \&
  {Strong}}]{2006ApJ...648L..29P}
{Porter}, T.~A., {Moskalenko}, I.~V., \& {Strong}, A.~W. 2006, \apjl, 648, L29,
  \dodoi{10.1086/507770}

\bibitem[{{Profumo} {et~al.}(2018){Profumo}, {Reynoso-Cordova}, {Kaaz}, \&
  {Silverman}}]{2018PhRvD..97l3008P}
{Profumo}, S., {Reynoso-Cordova}, J., {Kaaz}, N., \& {Silverman}, M. 2018,
  \prd, 97, 123008, \dodoi{10.1103/PhysRevD.97.123008}

\bibitem[{{Qiao} {et~al.}(2019){Qiao}, {Liu}, {Guo}, \&
  {Yuan}}]{2019JCAP...12..007Q}
{Qiao}, B.-Q., {Liu}, W., {Guo}, Y.-Q., \& {Yuan}, Q. 2019, \jcap, 2019, 007,
  \dodoi{10.1088/1475-7516/2019/12/007}

\bibitem[{{Reich} \& {Sun}(2019)}]{2019RAA....19...45R}
{Reich}, W., \& {Sun}, X.-H. 2019, Research in Astronomy and Astrophysics, 19,
  045, \dodoi{10.1088/1674-4527/19/3/45}

\bibitem[{{Reynoso} {et~al.}(2004){Reynoso}, {Green}, {Johnston}, {Goss},
  {Dubner}, \& {Giacani}}]{2004PASA...21...82R}
{Reynoso}, E.~M., {Green}, A.~J., {Johnston}, S., {et~al.} 2004, \pasa, 21, 82,
  \dodoi{10.1071/AS03053}

\bibitem[{{Sano} {et~al.}(2010){Sano}, {Sato}, {Horachi}, {Moribe}, {Yamamoto},
  {Hayakawa}, {Torii}, {Kawamura}, {Okuda}, {Mizuno}, {Onishi}, {Maezawa},
  {Inoue}, {Inutsuka}, {Tanaka}, {Matsumoto}, {Mizuno}, {Ogawa}, {Stutzki},
  {Bertoldi}, {Anderl}, {Bronfman}, {Koo}, {Burton}, {Benz}, \&
  {Fukui}}]{2010ApJ...724...59S}
{Sano}, H., {Sato}, J., {Horachi}, H., {et~al.} 2010, \apj, 724, 59,
  \dodoi{10.1088/0004-637X/724/1/59}

\bibitem[{{Sano} {et~al.}(2019){Sano}, {Rowell}, {Reynoso}, {Jung-Richardt},
  {Yamane}, {Nagaya}, {Yoshiike}, {Hayashi}, {Torii}, \&
  {Maxted}}]{2019ApJ...876...37S}
{Sano}, H., {Rowell}, G., {Reynoso}, E.~M., {et~al.} 2019, \apj, 876, 37,
  \dodoi{10.3847/1538-4357/ab108f}

\bibitem[{{Sano} {et~al.}(2020){Sano}, {Plucinsky}, {Bamba}, {Sharda},
  {Filipovi{\'c}}, {Law}, {Alsaberi}, {Yamane}, {Tokuda}, {Acero}, {Sasaki},
  {Vink}, {Inoue}, {Inutsuka}, {Shimoda}, {Tsuge}, {Fujii}, {Voisin}, {Maxted},
  {Rowell}, {Onishi}, {Kawamura}, {Mizuno}, {Yamamoto}, {Tachihara}, \&
  {Fukui}}]{2020ApJ...902...53S}
{Sano}, H., {Plucinsky}, P.~P., {Bamba}, A., {et~al.} 2020, \apj, 902, 53,
  \dodoi{10.3847/1538-4357/abb469}

\bibitem[{{Smith} {et~al.}(2019){Smith}, {Bruel}, {Cognard}, {Cameron},
  {Camilo}, {Dai}, {Guillemot}, {Johnson}, {Johnston}, {Keith}, {Kerr},
  {Kramer}, {Lyne}, {Manchester}, {Shannon}, {Sobey}, {Stappers}, \&
  {Weltevrede}}]{2019ApJ...871...78S}
{Smith}, D.~A., {Bruel}, P., {Cognard}, I., {et~al.} 2019, \apj, 871, 78,
  \dodoi{10.3847/1538-4357/aaf57d}

\bibitem[{Su {et~al.}(2017)Su, Zhou, Yang, Chen, Chen, Gong, \&
  Zhang}]{Su_2017}
Su, Y., Zhou, X., Yang, J., {et~al.} 2017, The Astrophysical Journal, 845, 48,
  \dodoi{10.3847/1538-4357/aa7f2a}

\bibitem[{{Tanaka} {et~al.}(2008){Tanaka}, {Uchiyama}, {Aharonian},
  {Takahashi}, {Bamba}, {Hiraga}, {Kataoka}, {Kishishita}, {Kokubun}, {Mori},
  {Nakazawa}, {Petre}, {Tajima}, \& {Watanabe}}]{2008ApJ...685..988T}
{Tanaka}, T., {Uchiyama}, Y., {Aharonian}, F.~A., {et~al.} 2008, \apj, 685,
  988, \dodoi{10.1086/591020}

\bibitem[{{Tanaka} {et~al.}(2011){Tanaka}, {Allafort}, {Ballet}, {Funk},
  {Giordano}, {Hewitt}, {Lemoine-Goumard}, {Tajima}, {Tibolla}, \&
  {Uchiyama}}]{2011ApJ...740L..51T}
{Tanaka}, T., {Allafort}, A., {Ballet}, J., {et~al.} 2011, \apjl, 740, L51,
  \dodoi{10.1088/2041-8205/740/2/L51}

\bibitem[{{Tang} \& {Chevalier}(2014)}]{2014ApJ...784L..35T}
{Tang}, X., \& {Chevalier}, R.~A. 2014, \apjl, 784, L35,
  \dodoi{10.1088/2041-8205/784/2/L35}

\bibitem[{{Tian} {et~al.}(2008){Tian}, {Leahy}, {Haverkorn}, \&
  {Jiang}}]{2008ApJ...679L..85T}
{Tian}, W.~W., {Leahy}, D.~A., {Haverkorn}, M., \& {Jiang}, B. 2008, \apjl,
  679, L85, \dodoi{10.1086/589506}

\bibitem[{{Tsuji} \& {Uchiyama}(2016)}]{2016PASJ...68..108T}
{Tsuji}, N., \& {Uchiyama}, Y. 2016, \pasj, 68, 108,
  \dodoi{10.1093/pasj/psw102}

\bibitem[{{Vasisht} {et~al.}(1997){Vasisht}, {Kulkarni}, {Anderson},
  {Hamilton}, \& {Kawai}}]{1997ApJ...476L..43V}
{Vasisht}, G., {Kulkarni}, S.~R., {Anderson}, S.~B., {Hamilton}, T.~T., \&
  {Kawai}, N. 1997, \apjl, 476, L43, \dodoi{10.1086/310493}

\bibitem[{{Vogt} \& {Dopita}(2011)}]{2011ApSS.331..521V}
{Vogt}, F., \& {Dopita}, M.~A. 2011, \apss, 331, 521,
  \dodoi{10.1007/s10509-010-0479-7}

\bibitem[{{Wendker} {et~al.}(1991){Wendker}, {Higgs}, \&
  {Landecker}}]{1991A&A...241..551W}
{Wendker}, H.~J., {Higgs}, L.~A., \& {Landecker}, T.~L. 1991, \aap, 241, 551

\bibitem[{{Winkler} {et~al.}(2003){Winkler}, {Gupta}, \&
  {Long}}]{2003ApJ...585..324W}
{Winkler}, P.~F., {Gupta}, G., \& {Long}, K.~S. 2003, \apj, 585, 324,
  \dodoi{10.1086/345985}

\bibitem[{{Woermann} \& {Jonas}(1988)}]{1988MNRAS.234..971W}
{Woermann}, B., \& {Jonas}, J.~L. 1988, \mnras, 234, 971,
  \dodoi{10.1093/mnras/234.4.971}

\bibitem[{{Wood} {et~al.}(2017){Wood}, {Caputo}, {Charles}, {Di Mauro},
  {Magill}, {Perkins}, \& {Fermi-LAT Collaboration}}]{2017ICRC...35..824W}
{Wood}, M., {Caputo}, R., {Charles}, E., {et~al.} 2017, in International Cosmic
  Ray Conference, Vol. 301, 35th International Cosmic Ray Conference
  (ICRC2017), 824.
\newblock \doarXiv{1707.09551}

\bibitem[{{Xing} {et~al.}(2014){Xing}, {Wang}, {Zhang}, \&
  {Chen}}]{2014ApJ...781...64X}
{Xing}, Y., {Wang}, Z., {Zhang}, X., \& {Chen}, Y. 2014, \apj, 781, 64,
  \dodoi{10.1088/0004-637X/781/2/64}

\bibitem[{{Yuan} {et~al.}(2012){Yuan}, {Liu}, \& {Bi}}]{2012ApJ...761..133Y}
{Yuan}, Q., {Liu}, S., \& {Bi}, X. 2012, \apj, 761, 133,
  \dodoi{10.1088/0004-637X/761/2/133}

\bibitem[{{Zavlin} {et~al.}(2000){Zavlin}, {Pavlov}, {Sanwal}, \&
  {Tr{\"u}mper}}]{2000ApJ...540L..25Z}
{Zavlin}, V.~E., {Pavlov}, G.~G., {Sanwal}, D., \& {Tr{\"u}mper}, J. 2000,
  \apjl, 540, L25, \dodoi{10.1086/312866}

\bibitem[{{Zeng} {et~al.}(2019){Zeng}, {Xin}, \& {Liu}}]{2019ApJ...874...50Z}
{Zeng}, H., {Xin}, Y., \& {Liu}, S. 2019, \apj, 874, 50,
  \dodoi{10.3847/1538-4357/aaf392}

\bibitem[{{Zhang} {et~al.}(2020){Zhang}, {Xi}, {Liu}, {Xin}, {Liu}, \&
  {Wang}}]{2020ApJ...889...12Z}
{Zhang}, H.-M., {Xi}, S.-Q., {Liu}, R.-Y., {et~al.} 2020, \apj, 889, 12,
  \dodoi{10.3847/1538-4357/ab5af6}

\bibitem[{{Zhang} \& {Chen}(2016)}]{2016ApJ...821...43Z}
{Zhang}, X., \& {Chen}, Y. 2016, \apj, 821, 43,
  \dodoi{10.3847/0004-637X/821/1/43}

\bibitem[{{Zhang} \& {Liu}(2019{\natexlab{a}})}]{2019ApJ...876...24Z}
{Zhang}, X., \& {Liu}, S. 2019{\natexlab{a}}, \apj, 876, 24,
  \dodoi{10.3847/1538-4357/ab14df}

\bibitem[{{Zhang} \& {Liu}(2019{\natexlab{b}})}]{2019ApJ...874...98Z}
---. 2019{\natexlab{b}}, \apj, 874, 98, \dodoi{10.3847/1538-4357/ab09fe}

\bibitem[{{Zhang} {et~al.}(1997){Zhang}, {Zheng}, {Landecker}, \&
  {Higgs}}]{1997A&A...324..641Z}
{Zhang}, X., {Zheng}, Y., {Landecker}, T.~L., \& {Higgs}, L.~A. 1997, \aap,
  324, 641

\bibitem[{{Zhang} \& {Liu}(2019{\natexlab{c}})}]{2019MNRAS.482.5268Z}
{Zhang}, Y., \& {Liu}, S. 2019{\natexlab{c}}, \mnras, 482, 5268,
  \dodoi{10.1093/mnras/sty3136}

\bibitem[{{Zhang} {et~al.}(2017){Zhang}, {Liu}, \&
  {Yuan}}]{2017ApJ...844L...3Z}
{Zhang}, Y., {Liu}, S., \& {Yuan}, Q. 2017, \apjl, 844, L3,
  \dodoi{10.3847/2041-8213/aa7de1}

\bibitem[{{Zhang} \& {Liu}(2020)}]{2020ChA&A..44....1Z}
{Zhang}, Y.-R., \& {Liu}, S.-M. 2020, \caa, 44, 1,
  \dodoi{10.1016/j.chinastron.2020.04.001}

\bibitem[{{Zirakashvili} \& {Aharonian}(2010)}]{2010ApJ...708..965Z}
{Zirakashvili}, V.~N., \& {Aharonian}, F.~A. 2010, \apj, 708, 965,
  \dodoi{10.1088/0004-637X/708/2/965}

\end{thebibliography}
\bibliographystyle{aasjournal}



\end{document}